\pdfoutput=1
\documentclass[11pt,twoside,a4paper,cmspaper,final,collab]{cms-tdr}
\def\svnVersion{9d70032}\def\svnDate{2023/12/30}\def\cmsCernNoTag{CERN-EP-2023-124}\def\cmsCernDate{\today}\def\cmsMessage{Published in the Journal of High Energy Physics as \href{http://dx.doi.org/10.1007/JHEP12(2023)068}{\doi{10.1007/JHEP12(2023)068}.}}
\begin{document}\cmsNoteHeader{TOP-22-006}

\cmsNoteHeader{TOP-22-006}

\newcommand{\ttX}{\ensuremath{\ttbar\PX}\xspace}
\newcommand{\ttH}{\ensuremath{\ttbar\PH}\xspace}
\newcommand{\ttZ}{\ensuremath{\ttbar\PZ}\xspace}
\newcommand{\ttW}{\ensuremath{\ttbar\PW}\xspace}
\newcommand{\ttWp}{\ensuremath{\ttbar\PWp}\xspace}
\newcommand{\ttWm}{\ensuremath{\ttbar\PWm}\xspace}
\newcommand{\ttll}{\ensuremath{\ttbar\Pell\overline{\Pell}}\xspace}
\newcommand{\ttlnu}{\ensuremath{\ttbar\Pell\PGn}\xspace}
\newcommand{\ttgamma}{\ensuremath{\ttbar\PGg}\xspace}
\newcommand{\tllq}{\ensuremath{\PQt\Pell\overline{\Pell}\PQq}\xspace}
\newcommand{\tXq}{\ensuremath{\PQt\PX\PQq}\xspace}
\newcommand{\tHq}{\ensuremath{\PQt\PH\PQq}\xspace}
\newcommand{\tttt}{\ensuremath{\ttbar\ttbar}\xspace}
\newcommand{\tZq}{\ensuremath{\PQt\PZ\PQq}\xspace}
\newcommand{\tWZ}{\ensuremath{\PQt\PW\PZ}\xspace}
\newcommand{\WZ}{\ensuremath{\PW\PZ}\xspace}
\newcommand{\ZZ}{\ensuremath{\PZ\PZ}\xspace}

\newcommand{\PQj}{{\HepParticle{j}{}{}}\xspace}
\newcommand{\ptljz}{\ensuremath{\pt(\Pell\PQj)_{\text{max}}}\xspace}
\newcommand{\ptZ}{\ensuremath{\pt(\PZ)}\xspace}

\newcommand*{\eftOp}[4]{\ensuremath{
    {#4
    \ifx\empty#3\empty\ifx\empty#1\empty\else^{#1}\fi\else^{#1(#3)}\fi
    \ifx\empty#2\empty\else_{#2}\fi}
}}

\newcommand{\ctp}  {\eftOp{}{\PQt\varphi}{}{c}\xspace}
\newcommand{\ctpI} {\eftOp{I}{\PQt\varphi}{}{c}\xspace}
\newcommand{\cpQM} {\eftOp{-}{\varphi Q}{}{c}\xspace}
\newcommand{\cpQa} {\eftOp{3}{\varphi Q}{}{c}\xspace}
\newcommand{\cpt}  {\eftOp{}{\varphi\PQt}{}{c}\xspace}
\newcommand{\cptb} {\eftOp{}{\varphi\PQt\PQb}{}{c}\xspace}
\newcommand{\cptbI}{\eftOp{I}{\varphi\PQt\PQb}{}{c}\xspace}
\newcommand{\ctW}  {\eftOp{}{\PQt\PW}{}{c}\xspace}
\newcommand{\ctWI} {\eftOp{I}{\PQt\PW}{}{c}\xspace}
\newcommand{\ctZ}  {\eftOp{}{\PQt\PZ}{}{c}\xspace}
\newcommand{\ctZI} {\eftOp{I}{\PQt\PZ}{}{c}\xspace}
\newcommand{\cbW}  {\eftOp{}{\PQb\PW}{}{c}\xspace}
\newcommand{\cbWI} {\eftOp{I}{\PQb\PW}{}{c}\xspace}
\newcommand{\ctG}  {\eftOp{}{\PQt G}{}{c}\xspace}
\newcommand{\ctGI} {\eftOp{I}{\PQt G}{}{c}\xspace}

\newcommand{\cQla} {\eftOp{3}{Q\Pell}{\Pell}{c}\xspace}
\newcommand{\cQlM} {\eftOp{-}{Q\Pell}{\Pell}{c}\xspace}
\newcommand{\cQe}  {\eftOp{}{Q\Pe}{\Pell}{c}\xspace}
\newcommand{\ctl}  {\eftOp{}{\PQt \Pell}{\Pell}{c}\xspace}
\newcommand{\cte}  {\eftOp{}{\PQt \Pe}{\Pell}{c}\xspace}
\newcommand{\ctlS} {\eftOp{S}{\PQt}{\Pell}{c}\xspace}
\newcommand{\ctlSI}{\eftOp{SI}{\PQt}{\Pell}{c}\xspace}
\newcommand{\ctlT} {\eftOp{T}{\PQt}{\Pell}{c}\xspace}
\newcommand{\ctlTI}{\eftOp{TI}{\PQt}{\Pell}{c}\xspace}

\newcommand{\cttOne}   {\eftOp{1}{\PQt\PQt}{}{c}\xspace}
\newcommand{\cQQOne}   {\eftOp{1}{\PQQ\PQQ}{}{c}\xspace}
\newcommand{\cQtOne}   {\eftOp{1}{\PQQ\PQt}{}{c}\xspace}
\newcommand{\cQtEight} {\eftOp{8}{\PQQ\PQt}{}{c}\xspace}

\newcommand{\cQqOneThree}   {\eftOp{31}{\PQQ\PQq}{}{c}\xspace}
\newcommand{\cQqEightThree} {\eftOp{38}{\PQQ\PQq}{}{c}\xspace}
\newcommand{\cQqOneOne}     {\eftOp{11}{\PQQ\PQq}{}{c}\xspace}
\newcommand{\ctqOne}        {\eftOp{1}{\PQt\PQq}{}{c}\xspace}
\newcommand{\cQqEightOne}   {\eftOp{18}{\PQQ\PQq}{}{c}\xspace}
\newcommand{\ctqEight}      {\eftOp{8}{\PQt\PQq}{}{c}\xspace}

\newcommand{\nwc}{\ensuremath{26}\xspace}
\newcommand{\nwcone}{\ensuremath{25}\xspace}
\newcommand{\nwctwo}{\ensuremath{24}\xspace}

\newcommand{\twolss}{\ensuremath{2\Pell\text{ss}}\xspace}
\newcommand{\twolsstwob}{\ensuremath{2\Pell\text{ss }2\PQb}\xspace}
\newcommand{\twolssthreeb}{\ensuremath{2\Pell\text{ss }3\PQb}\xspace}
\newcommand{\threel}{\ensuremath{3\Pell}\xspace}
\newcommand{\threeloffZ}{\ensuremath{3\Pell\text{ off-}\PZ}\xspace}
\newcommand{\threeloffZoneb}{\ensuremath{3\Pell\text{ off-}\PZ\text{ }1\PQb}\xspace}
\newcommand{\threeloffZtwob}{\ensuremath{3\Pell\text{ off-}\PZ\text{ }2\PQb}\xspace}
\newcommand{\threelonZ}{\ensuremath{3\Pell\text{ on-}\PZ}\xspace}
\newcommand{\threelonZoneb}{\ensuremath{3\Pell\text{ on-}\PZ\text{ }1\PQb}\xspace}
\newcommand{\threelonZtwob}{\ensuremath{3\Pell\text{ on-}\PZ\text{ }2\PQb}\xspace}
\newcommand{\fourl}{\ensuremath{4\Pell}\xspace}
\newcommand{\mll}{\ensuremath{m_{\Pell\Pell}\xspace}}
\newcommand{\mZ}{\ensuremath{m_{\PZ}}\xspace}
\newcommand{\mZmll}{\ensuremath{\abs{\mZ-\mll}}\xspace}

\providecommand{\cmsTable}[1]{\resizebox{\textwidth}{!}{#1}}

\newcommand{\mur}{\ensuremath{\mu_{\mathrm{R}}}\xspace}
\newcommand{\muf}{\ensuremath{\mu_{\mathrm{F}}}\xspace}

\newcommand{\twohqV}{\ensuremath{\textrm{2hqV}}\xspace}
\newcommand{\fourhq}{\ensuremath{\textrm{4hq}}\xspace}
\newcommand{\twohqtwolq}{\ensuremath{\textrm{2hq2lq}}\xspace}
\newcommand{\twohqtwolep}{\ensuremath{\textrm{2hq2\Pell}}\xspace}

\newcommand{\PDF}{\ensuremath{\,\text{(PDF)}}\xspace}
\newcommand{\QCD}{\ensuremath{\,\text{(QCD)}}\xspace}
\newcommand{\abseta}{\ensuremath{\abs{\eta}}\xspace}
\newcommand{\DeepJet}{\ensuremath{\textsc{DeepJet}}\xspace}
\newcommand{\perTeVsqns}{\ensuremath{\text{Te\hspace{-.08em}V}^{-2}}\xspace}

\title{Search for physics beyond the standard model in top quark production with additional leptons in the context of effective field theory}
\date{\today}

\abstract{A search for new physics in top quark production with additional final-state leptons is performed using data collected by the CMS experiment in proton-proton collisions at $\sqrt{s}=13\TeV$ at the LHC during 2016--2018. The data set corresponds to an integrated luminosity of 138\fbinv. Using the framework of effective field theory (EFT), potential new physics effects are parametrized in terms of \nwc dimension-six EFT operators. The impacts of EFT operators are incorporated through the event-level reweighting of Monte Carlo simulations, which allows for detector-level predictions. The events are divided into several categories based on lepton multiplicity, total lepton charge, jet multiplicity, and \PQb-tagged jet multiplicity. Kinematic variables corresponding to the transverse momentum (\pt) of the leading pair of leptons and/or jets as well as the \pt of on-shell \PZ bosons are used to extract the 95\% confidence intervals of the \nwc Wilson coefficients corresponding to these EFT operators. No significant deviation with respect to the standard model prediction is found.}

\hypersetup{
pdfauthor={CMS Collaboration},
pdftitle={Search for physics beyond the standard model in top quark production with additional leptons in the context of effective field theory},
pdfsubject={CMS},
pdfkeywords={CMS, top quark, EFT}}

\maketitle

\section{Introduction}

Searches for new fundamental particles and interactions are motivated by the strong evidence for phenomena (such as dark matter~\cite{Feng:2010gw,Porter:2011nv}) that are not described by the standard model (SM) of particle physics.
However, there is no a priori reason to assume that particles will be light enough to be produced on-shell at the CERN LHC. Indirect methods of probing higher energy scales are thus an important part of searches for new physics at the energy frontier.
One example of this type of approach is effective field theory (EFT), a flexible framework that comprehensively describes the off-shell effects of new physics phenomena at a mass scale $\Lambda$.
The EFT treats the SM Lagrangian as the lowest order term in an expansion of a more complete Lagrangian at a mass scale $\Lambda$ in the form of a series of higher-dimensional operators, which are built from products of SM fields that respect the SM symmetries.
The EFT Lagrangian is written as
\begin{linenomath}
\begin{equation}
\label{eq:Left}
    \mathcal{L}_{\mathrm{EFT}} = \mathcal{L}_{\mathrm{SM}} +
    \sum\limits_{d,i} \frac{c^{d}_i}{\Lambda^{d-4}}
    \mathcal{O}^{d}_i ,
\end{equation}
\end{linenomath}
where $\mathcal{L}_{\mathrm{SM}}$ is the SM Lagrangian, $\mathcal{O}^{d}_i$ are the EFT operators of dimension $d$, and $c^{d}_i$ are the Wilson coefficients (WCs) which control the strength of the EFT effects.
Since each higher order term in Eq.~\eqref{eq:Left} is suppressed by powers of $\Lambda^{d-4}$,
the smallest dimension operators tend to produce the largest expected deviations from the SM ($d \leq 4$) processes.
We do not consider operators that violate baryon or lepton number, so all operators of odd dimension are excluded, making the dimension-six operators the leading new physics contributions~\cite{Degrande:2012wf}. The next contributions would arise from dimension-eight operators, which are not considered here.

This paper focuses specifically on operators that couple the top quark to leptons, bosons, and other heavy (top or bottom) quarks.
Searching for new physics in the top quark sector is motivated by the uniquely large mass of the top quark~\cite{CDF:1995wbb,D0:1995jca} and the resulting Yukawa coupling to the Higgs field~\cite{Workman:2022ynf} of roughly unity.
The LHC provides a rich environment of top quarks produced with additional leptons. Furthermore, the leptonic final-state decays of the top quark provide experimentally clean signatures with relatively low background contributions.
The dominant SM contributions to these signatures arise from processes in which one or more top quarks are produced in association with a heavy boson or other top quarks.
Referred to as associated top quark production, these processes include \ttH~\cite{CMS:2020mpn}, \ttW~\cite{CMS:2022tkv}, \ttZ~\cite{CMS:2019too}, \tZq~\cite{CMS:2021ugv}, \tHq~\cite{CMS:2020mpn}, \tttt~\cite{CMS:2023zdh}.
While each of these processes have been studied individually, the analysis presented in this paper takes a more global approach, using the EFT framework to probe the potential effects of heavy new physics impacting these associated top quark processes simultaneously.
In addition to these SM contributions, the analysis also aims to probe new physics effects that may impact these final state signatures without an intermediate boson (via a four-fermion EFT vertex).

The analysis described in this paper builds on the approach developed in Ref.~\cite{CMS:2020lrr}, which studied 16 dimension-six EFT operators with data collected in 2017, corresponding to an integrated luminosity of 41.5\fbinv.
With this approach, EFT effects are incorporated into the event weights of the simulated samples, allowing detector-level predictions that account for all relevant interference effects (not only between new physics and the SM, but also among new physics operators) and correlations among WCs.
This approach has been subsequently utilized to study \ttZ, \tZq, and \tWZ~\cite{CMS:2021aly} and to study \ttH and \ttZ processes in which the Higgs or \PZ boson is boosted~\cite{CMS:2022hjj}; the former simultaneously probes five WCs, while the latter simultaneously probes eight WCs.
Expanding on these previous analyses, we study the effects of \nwc operators, incorporate additional signatures, and improve the sensitivity by fitting differential kinematical distributions and making use of all data collected by the CMS experiment in 2016--2018, corresponding to a total integrated luminosity of 138\fbinv.
This paper thus represents the most global detector-level EFT analysis to date.

The sections in this paper are organized as follows.
The CMS detector is introduced in Section~\ref{sec:detector}.
In Section~\ref{sec:samples}, the data and simulated samples are discussed.
The object reconstruction and event selection are covered in Sections~\ref{sec:reconstruction} and \ref{sec:selection}.
Section~\ref{sec:backgrounds} describes the background estimation.
The statistical methods are explained in Section~\ref{sec:fitting}, and the systematic uncertainties are detailed in Section~\ref{sec:systematics}.
Section~\ref{sec:results} presents the results, which are summarized in Section~\ref{sec:summary}.
The tabulated results are provided in the HEPData record for this analysis~\cite{hepdata}.

\section{The CMS detector}
\label{sec:detector}

The central feature of the CMS apparatus is a superconducting solenoid of 6\unit{m} internal diameter,
providing a magnetic field of 3.8\unit{T}. Within the solenoid volume are a silicon pixel and strip tracker,
a lead tungstate crystal electromagnetic calorimeter (ECAL), and a brass and scintillator
hadron calorimeter (HCAL), each composed of a barrel and two endcap sections. Forward calorimeters extend
the pseudorapidity ($\eta$) coverage provided by the barrel and endcap detectors. Muons are measured in
gas-ionization detectors embedded in the steel flux-return yoke outside the solenoid. A more detailed
description of the CMS detector, together with a definition of the coordinate system used and the relevant
kinematic variables, can be found in Ref.~\cite{CMS:2008xjf}.

Events of interest are selected using a two-tiered trigger system.
The first level (L1), composed of custom hardware processors,
uses information from the calorimeters and muon detectors to select events at a rate of around 100\unit{kHz} within a
fixed latency of about 4\mus~\cite{CMS:2020cmk}.
The second level, known as the high-level trigger,
consists of a farm of processors running a version
of the full event reconstruction software optimized for
fast processing, and reduces the event rate to around
1\unit{kHz} before data storage~\cite{CMS:2016ngn}.

\section{Data samples and signal simulation}
\label{sec:samples}

This analysis uses data from proton-proton collisions at $\sqrt{s}=13\TeV$ collected by the CMS experiment during 2016--2018
with a combination of single-, double-, and triple-lepton triggers.
The minimum lepton transverse momentum (\pt) requirements are chosen such that all events are within the fully efficient regions of the triggers. The trigger efficiencies, calculated using independent missing transverse energy triggers, are higher than 95\% across the full \pt spectrum with uncertainties, systematic and statistical, smaller than 2\%.

The analysis aims to study dimension-six EFT effects on processes in which top quarks are produced in association with additional charged leptons. Processes that lead to the same multilepton final-state signatures but are not impacted by these EFT operators are backgrounds for this analysis.
The expected background contributions are estimated using a combination of simulated samples and control samples in data, as described in Section~\ref{sec:backgrounds}.

\begin{figure}[htp!]
\centering
\includegraphics[width=0.23\textwidth]{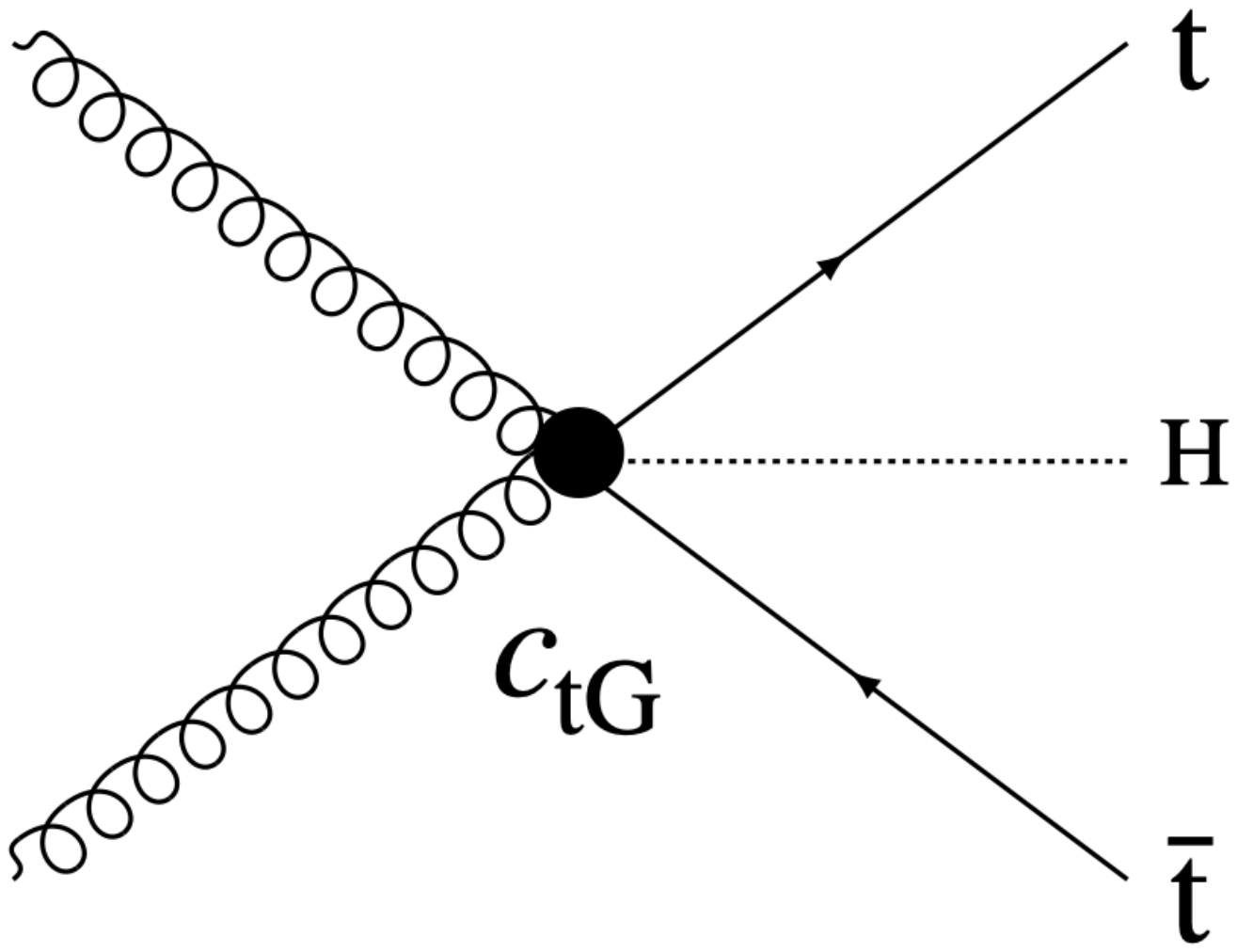}
\includegraphics[width=0.23\textwidth]{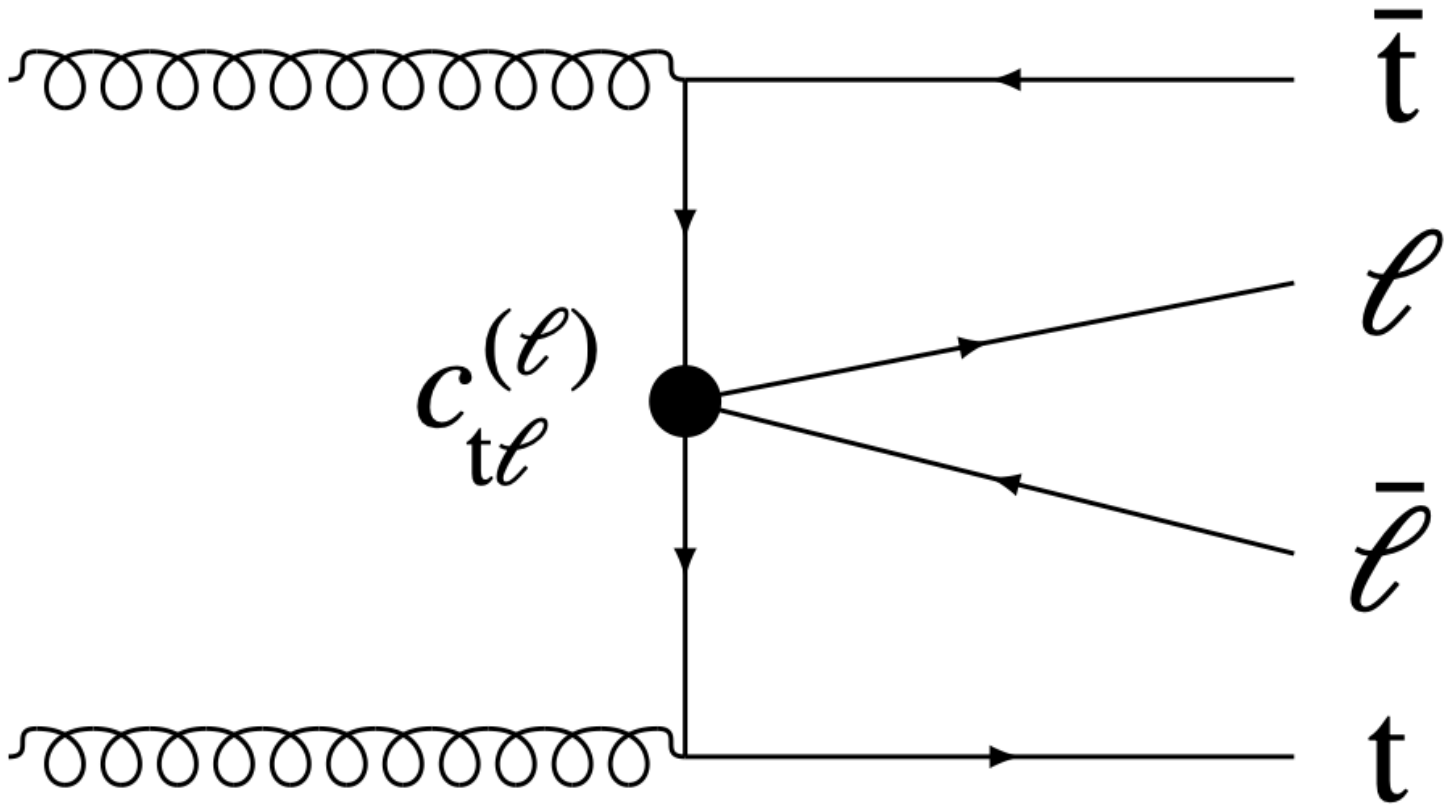}
\includegraphics[width=0.23\textwidth]{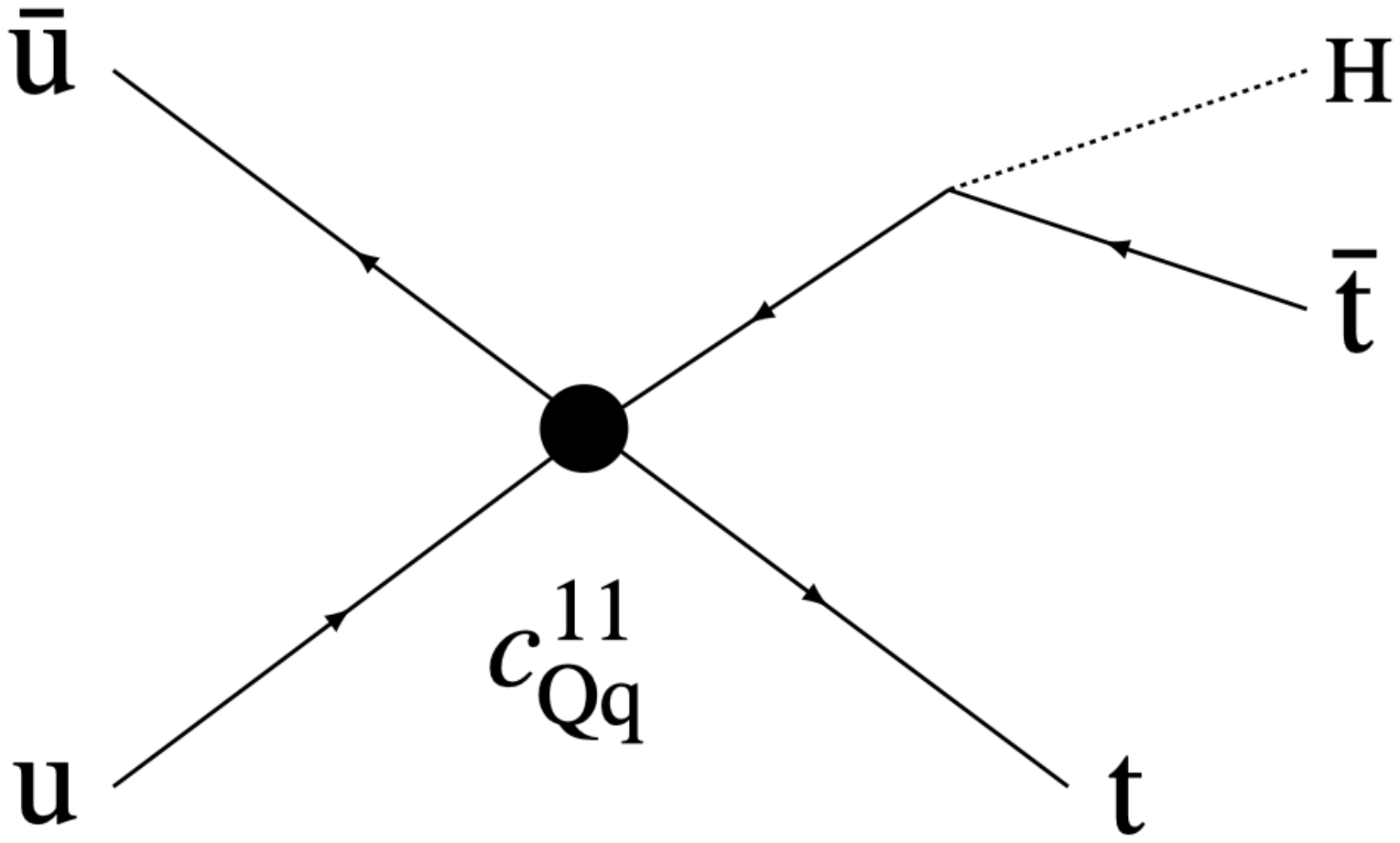}
\includegraphics[width=0.23\textwidth]{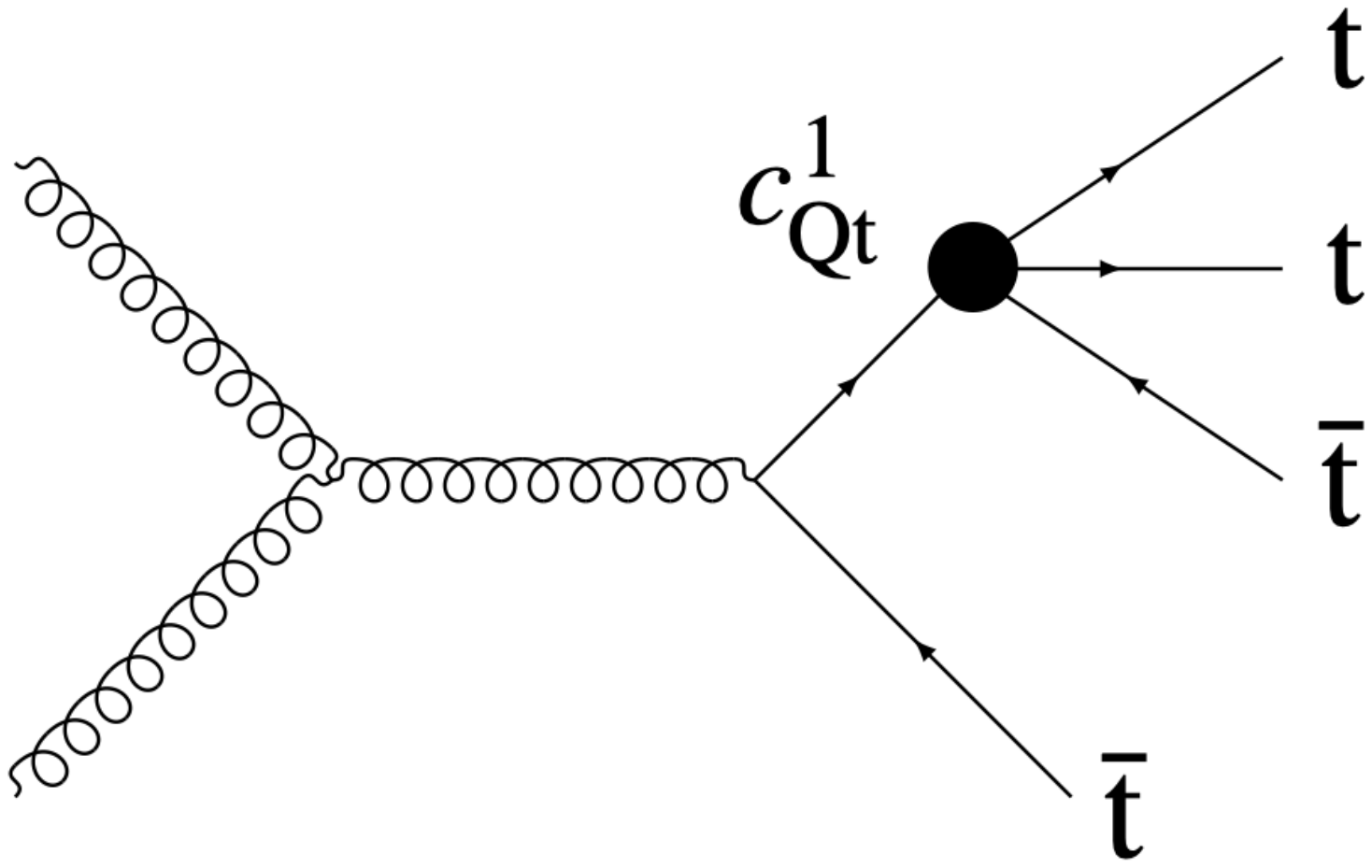}
\caption{Example Feynman diagrams illustrating Wilson coefficients from each of the categories listed in Table~\ref{tab:wc_lst}. From left to right, the diagrams show vertices associated with the \ctG, \ctl, \cQqOneOne, and \cQtOne.}
\label{fig:feynman_diagrams_categories_ex}
\end{figure}

The signal contribution is modeled at leading order (LO) using the \MGvATNLO event generator (version 2.6.5)~\cite{MadGraph5_aMCatNLO,Frederix:2012ps,Alwall:2007fs} with the \textsc{dim6top} model described in Ref.~\cite{AguilarSaavedra:2018nen} to incorporate the EFT effects.
Using the Warsaw basis~\cite{Grzadkowski:2010es} of gauge invariant dimension-six operators, this model focuses on operators involving one or more top quarks, providing tree-level modeling of their effects.
While the model described in Ref.~\cite{AguilarSaavedra:2018nen} allows for the EFT effects to vary independently for each generation of leptons, we impose the assumption that the EFT effects impact each lepton generation in the same way.
In this analysis, we aim to include all operators from Ref.~\cite{AguilarSaavedra:2018nen} that significantly impact processes in which one or more top quarks are produced in association with charged leptons; as listed in Table~\ref{tab:wc_lst}, this amounts to \nwc operators in total.
The \nwc operators fall into four main categories:\ operators involving four heavy quarks (\fourhq), operators involving two heavy quarks and two light quarks (\twohqtwolq), operators involving two heavy quarks and two leptons (\twohqtwolep), and operators involving two heavy quarks and bosons (\twohqV).
The leptons in the \twohqtwolep operators may be either charged or neutral, and the bosons in the \twohqV operators include both gauge bosons and the Higgs boson.
Figure~\ref{fig:feynman_diagrams_categories_ex} shows examples of how WCs from each of these categories can impact associated top processes.
The definitions of the operators associated with all of these WCs are provided in Ref.~\cite{AguilarSaavedra:2018nen}.
In order to allow \MGvATNLO to properly handle the emission of gluons from the vertices involving the \ctG WC (which impacts interactions involving top quarks, gluons, and the Higgs boson), an extra factor of the strong coupling $g_\mathrm{s}$ is applied to the \ctG coefficient, as explained in Ref.~\cite{Goldouzian:2020ekx}.
The simulations use the NNPDF3.1~\cite{NNPDF3} sets of parton distribution functions (PDFs).
Parton showering and hadronization are performed with \PYTHIA 8.240~\cite{pythia8} with the CP5 tune~\cite{Sirunyan:2019dfx}.
The decays of Higgs bosons and top quarks are handled with \PYTHIA.
Both the leptonic and hadronic top quark decays can contribute.
The top quark mass used in the simulation is 172.5\GeV.
The default input scheme in the \textsc{dim6top} model is used.
The \tllq and \tHq samples are produced using the four-flavor scheme, while the five-flavor scheme is used for the rest of the samples (\ttH, \ttll, \ttlnu, and \tttt).

\begin{table}[!htp]
\centering
\topcaption{List of Wilson coefficients (WCs) included in this analysis, grouped according to the categories of WCs defined in Ref.~\cite{AguilarSaavedra:2018nen}; the abbreviations for the WCs categories used in this paper are noted parenthetically. The definitions of the WCs and the definitions of the corresponding operators can be found in Table~1 of Ref.~\cite{AguilarSaavedra:2018nen}. An extra factor of the strong coupling is applied to the \ctG coefficient, as explained in the text.}
\renewcommand{\arraystretch}{1.6}
\begin{tabular}{ll}
    \hline
    Operator category & Wilson coefficients\\
    \hline
    Two-heavy               (\twohqV)      & \ctp, \cpQM, \cpQa, \cpt, \cptb, \ctW, \ctZ, \cbW, \ctG \\
    Two-heavy-two-lepton    (\twohqtwolep) & \cQla, \cQlM, \cQe, \ctl, \cte, \ctlS, \ctlT \\
    Two-heavy-two-light     (\twohqtwolq)  & \cQqOneThree, \cQqEightThree, \cQqOneOne, \cQqEightOne, \ctqOne, \ctqEight\\
    Four-heavy              (\fourhq)      & \cQQOne, \cQtOne, \cQtEight, \cttOne\\
    \hline
\end{tabular}
\label{tab:wc_lst}
\end{table}

Processes that are significantly impacted by these operators constitute the signal processes for this analysis:\ \ttH, \ttll, \ttlnu, \tllq, \tHq, and \tttt.
The \ttll and \tllq samples incorporate contributions from on- and off-shell \PZ bosons, contributions from virtual photons, and contributions in which the lepton pair is produced directly from a \twohqtwolep EFT vertex.
The \ttlnu process similarly includes lepton pairs produced from on-shell \PW bosons, as well as those from nonresonant processes, allowing the effects from \twohqtwolep operators to be incorporated in the sample.
The processes involving a Higgs boson enter our signal selection (defined in Section~\ref{sec:selection}) when the Higgs boson decays into one or more leptons.
All simulated signal processes are normalized with their respective cross sections which are given Table~\ref{tab:xsec}. The cross section computations may include quantum chromodynamics (QCD) and electroweak (EWK) corrections.

\begin{table}[!htp]
\centering
\topcaption{Theoretical cross sections at next-to-LO (NLO) used for normalization of simulated signal samples. The items are ordered by cross section. The uncertainties are broken into normalization components due to modeling the parton distribution functions (PDFs) and QCD order. Entries without a value are negligible.}
\renewcommand{\arraystretch}{1.4}
\cmsTable{\begin{tabular}{lrll}
    \hline
    Process & Cross section (pb) & Accuracy & Ref. \\
    \hline
    \ttH & $0.5071\pm2.4\%\PDF\,^{+7.6\%}_{-7.1\%}\QCD$ & NLO (QCD + EWK) & \cite{LHCHiggsCrossSectionWorkingGroup:2016ypw} \\
    \ttll ($\mll>10\GeV$) & $0.113\,^{+12\%}_{-10\%}\QCD$ & NLO (QCD + EWK) & \cite{LHCHiggsCrossSectionWorkingGroup:2016ypw} \\
    \multirow{2}{*}{\ttlnu} & \multirow{2}{*}{$0.235\,^{+10\%}_{-11\%}\QCD$} & NLO (QCD + EWK) (incl.\ $\alpS \alpha^4$ & \multirow{2}{*}{\cite{Frederix:2021agh}} \\[-5pt]
    & & terms and multijet merging) & \\
    \tllq ($\mll>30\GeV$) & $0.076\pm2.7\%\PDF\pm2.0\%\QCD$ & NLO QCD & \cite{MadGraph5_aMCatNLO,Frederix:2012ps,Alwall:2007fs} \\
    \tHq & $0.071\pm5.1\%\PDF\,^{+6.5\%}_{-15\%}\QCD$ & NLO QCD & \cite{LHCHiggsCrossSectionWorkingGroup:2016ypw} \\
    \tttt & $0.01337\pm6.9\%\PDF\,^{+3.6\%}_{-11\%}\QCD$ & NLO (QCD + EWK) + NLL' & \cite{vanBeekveld:2022hty} \\
    \hline
\end{tabular}}
\label{tab:xsec}
\end{table}

For each of the six signal processes, we account for diagrams with zero EFT vertices (\ie, the SM contribution) and diagrams with one EFT vertex (\ie, the new physics contribution). The amplitude for each process will thus depend linearly on the WCs, and the cross section will depend quadratically on the WCs.
With \nwc WCs, the dependence of the cross section on the WCs will therefore be given by a \nwc-dimensional quadratic function.
Since the weight of each generated event corresponds to the event's contribution to the inclusive cross section, each event weight will also depend quadratically on the \nwc WCs.
For each generated event, we determine the \nwc-dimensional quadratic parametrization using the \MGvATNLO event reweighting technique~\cite{Mattelaer:2016gcx}.
Once we have obtained the \nwc-dimensional quadratic parametrization for each event, we can find the dependence of any observable bin (\ie, distinct category of events defined by the properties of the final-state objects) on the WCs by summing the quadratic parametrizations for each of the events that passes the selection criteria for the given bin.
Since we are thus able to write the predicted yield of any observable bin as a function of the \nwc WCs, we can obtain detector-level predictions at any arbitrary point in the \nwc-dimensional EFT space.
This is the key enabling concept of this analysis, and it allows us to rigorously account for all EFT effects across all analysis bins simultaneously when performing the likelihood fitting with the statistical framework.
This approach was developed in Ref.~\cite{CMS:2020lrr}, which contains a more detailed description of the method of parametrizing the predicted yields in terms of the WCs.

Similar to Ref.~\cite{CMS:2020lrr}, we include an additional final-state parton in the matrix element generation for the \ttX processes using the MLM scheme~\cite{Alwall:2007fs}. The inclusion of the additional parton can improve the modeling at high jet multiplicities, and can also significantly impact the dependence of the \ttX processes on the WCs~\cite{Goldouzian:2020ekx}.
For the other processes (\tllq, \tHq, and \tttt), an additional jet is not included because of technical limitations, and an additional uncertainty is applied to account for this where relevant, as described in Section~\ref{sec:systematics}.

\section{Object reconstruction and identification}
\label{sec:reconstruction}

The global event reconstruction (also called particle-flow
event reconstruction~\cite{CMS:2017yfk}) aims to reconstruct
and identify each individual particle in an event, with an
optimized combination of all subdetector information.
In this process, the identification of the particle type
(photon (\PGg), electron, muon, charged hadron, neutral hadron)
plays an important role in the determination of the particle
direction and energy.
Photons 
are identified as ECAL energy clusters not
linked to the extrapolation of any charged particle
trajectory to the ECAL.
Electrons 
are
identified as a primary charged particle track and
potentially many ECAL energy clusters corresponding to this
track extrapolation to the ECAL and to possible
bremsstrahlung photons emitted along the way through the tracker material.
Muons are
identified as tracks in the central tracker consistent with
either a track or several hits in the muon system,
and associated with calorimeter deposits compatible with the
muon hypothesis. Charged hadrons are identified as charged
particle tracks neither identified as electrons, nor as muons.
Finally, neutral hadrons are identified as HCAL energy
clusters not linked to any charged hadron trajectory, or as a
combined ECAL and HCAL energy excess with respect to the
expected charged hadron energy deposit.

The energy of photons is obtained from the ECAL measurement.
The energy of electrons is determined from a combination of
the track momentum at the main interaction vertex,
the corresponding ECAL cluster energy, and the energy sum of
all bremsstrahlung photons attached to the track.
The energy of muons is obtained from the corresponding track momentum.
The energy of charged hadrons is determined from a
combination of the track momentum and the corresponding ECAL
and HCAL energies, corrected for the response function of
the calorimeters to hadronic showers.
Finally, the energy of neutral hadrons is obtained from the
corresponding corrected ECAL and HCAL energies.

The electron momentum is estimated by combining the energy
measurement in the ECAL with the momentum measurement in the tracker.
The momentum resolution for electrons with
$\pt\approx 45\GeV$ from $\PZ \to \Pe \Pe$ decays ranges
from 1.6 to 5\%.
It is generally better in the barrel region than in the endcaps,
and also depends on the bremsstrahlung energy emitted by
the electron as it traverses the material in front of the ECAL~\cite{CMS:2020uim,CMS-DP-2020-021}.
We require electrons to have pseudorapidity $\abseta < 2.5$.

Muons are measured in the pseudorapidity range
$\abseta < 2.4$, with detection planes made using three
technologies:\ drift tubes, cathode strip chambers, and
resistive plate chambers.
The single muon trigger efficiency exceeds 90\% over the
full $\eta$ range, and the efficiency to reconstruct and
identify muons is greater than 96\%.
Matching muons to tracks measured in the silicon tracker
results in a relative transverse momentum resolution,
for muons with \pt up to 100\GeV, of 1\% in the barrel
and 3\% in the endcaps.
The \pt resolution in the barrel is better than 7\% for
muons with \pt up to 1\TeV~\cite{CMS:2018rym}.

Reconstructed electrons and muons are required to satisfy
selection criteria aiming to select prompt leptons produced
in decays of the \PW or \PZ bosons, as well as those
that couple directly to the top quarks in the beyond-SM scenarios we
consider. This lepton selection, fully described in Ref.~\cite{CMS:2020mpn},
is performed by means of a multivariate discriminator~\cite{CMS-PAS-MUO-22-001} that takes as its input the
variables related to the lepton isolation and its impact parameter.
In addition, kinematic information of charged and neutral particles around
the lepton candidate is used by feeding it into the jet reconstruction and \PQb tagging
algorithms, described below. The score of the \PQb tagging algorithm is used as an input to the discriminator.

The hadronic jets are clustered using the infrared and collinear safe
anti-\kt algorithm~\cite{Cacciari:2008gp, Cacciari:2011ma}
with a distance parameter of 0.4. Jet momentum is determined
as the vectorial sum of all particle momenta in the jet, and
is found from simulation to be, on average, within 5 to 10\%
of the true momentum over the whole \pt spectrum and detector
acceptance. Additional proton-proton interactions within the
same or nearby bunch crossings (pileup) can contribute
additional tracks and calorimetric energy depositions to the
jet momentum. To mitigate this effect, charged particles
identified to be originating from pileup vertices are
discarded and an offset correction is applied to correct for
remaining contributions~\cite{CMS:2020ebo}. Jet energy corrections are derived
from simulation to bring the measured response of jets to
that of particle level jets on average. In situ measurements
of the momentum balance in dijet,
$\PGg + \text{jet}$, $\PZ + \text{jet}$,
and multijet events are used to account for any residual
differences in the jet energy scale (JES) between data and
simulation~\cite{CMS:2016lmd}. The jet energy resolution (JER)
amounts typically to 15--20\% at 30\GeV, 10\% at 100\GeV, and 5\% at 1\TeV~\cite{CMS:2016lmd}.
Additional selection criteria are applied to each jet to
remove jets potentially dominated by anomalous
contributions from various  subdetector components or reconstruction failures.

Jets originating from \PQb quark decays (\PQb jets) are identified using the
algorithm~\cite{BTV-16-002} known as \DeepJet~\cite{Bols:2020bkb,CMS-DP-2023-005},
which uses a deep neural network to classify \PQb jets with
different working points. The analysis uses
a medium working point which correctly identifies \PQb jets with an efficiency of about
70\%, and a loose working point with an efficiency of about 85\%.
The misidentification rate for gluon or light-flavor quark jets for these two working points is 1.0\% and 10\%, respectively.

\section{Event selection and categorization}
\label{sec:selection}

The analysis targets events in which top quarks are produced in association with additional charged leptons.
The event selection categories are defined primarily by the multiplicity of charged leptons, jets, and \PQb jets.
This event selection scheme aims to isolate subsamples of the broader multilepton data set into categories containing different admixtures of the contributing signal processes, resulting in 43 unique categories.
The events in each category are binned according to a kinematical variable, which will be described in Section~\ref{sec:eventSelection_diff_vars}.
While it is not possible to completely isolate any of these individual processes, the division of the sample into subsamples with different compositions increases the statistical power to determine which specific processes might be responsible for any deviation from SM expectations that may be observed in data.

\begin{figure}[!htp]
\centering
\includegraphics[width=0.8\textwidth]{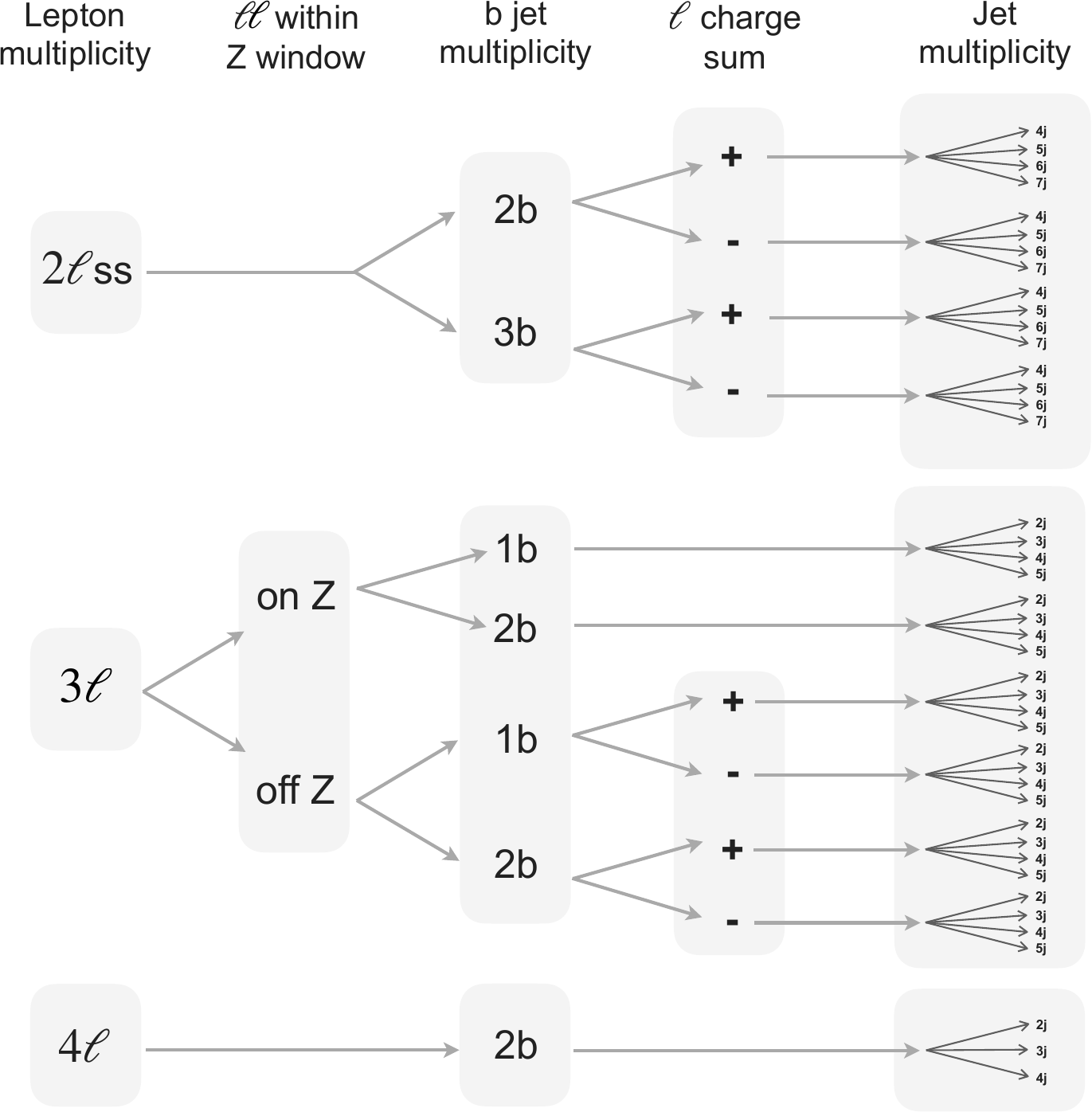}
\caption{Summary of the event selection categorization. The details for the selection requirements are described in Sections~\ref{sec:eventSelection_2lss}--\ref{sec:eventSelection_4l}.}
\label{fig:event_selection_summary}
\end{figure}

We require each event to have either two same-sign leptons (\twolss), three leptons (\threel), or four or more leptons (\fourl).
The \threel event category is further subdivided into events with and without on-shell \PZ boson decays (pair of same-flavor and opposite-sign leptons with an invariant mass within 10\GeV of the \PZ boson mass), referred to as the on-\PZ and off-\PZ subcategories, respectively.
The events in the \threeloffZ category and in the \twolss category are subdivided based on the sum of the charges of the leptons.
A schematic summary of the event selection categorization is shown in Fig.~\ref{fig:event_selection_summary}.
The selected events in all categories must have at least two jets with $\pt > 30\GeV$ and $\abseta<2.4$.
Events containing a pair of leptons with an invariant mass of less than 12\GeV are rejected to avoid backgrounds from light resonances. We use the \DeepJet algorithm
to impose the further requirement of one or more \PQb-tagged jets,
depending on the lepton multiplicity.
The \PGt leptons only enter our event selections via their $\PGt \to \Pe$ and $\PGt \to \PGm$ decay modes.
Table~\ref{tab:Categories} provides a summary of the requirements for each event category, which are detailed in the following sections.

\begin{table}[!htp]
\centering
\topcaption{Object requirements for the 43 event selection categories. Requirements separated by commas indicate a division into subcategories. The kinematical variable that is used in the event category is also listed. Section~\ref{sec:eventSelection_diff_vars} provides further details regarding the kinematical distributions.}
\label{tab:Categories}
\renewcommand{\arraystretch}{1.2}
\cmsTable{
\begin{tabular}{lcccccc}
    \hline
    Event category             & Leptons  & \mll & \PQb tags & Lepton charge sum & Jets & Kinematical variable \\
    \hline
    \twolsstwob           & 2        & No requirement                 & 2       & $>$0, $<$0     & 4, 5, 6, $\geq$7 & \ptljz \\
    \twolssthreeb         & 2        & No requirement                 & $\geq$3 & $>$0, $<$0     & 4, 5, 6, $\geq$7 & \ptljz \\
    \threeloffZoneb & 3        & $\mZmll > 10\GeV$    & 1       & $>$0, $<$0     & 2, 3, 4, $\geq$5 & \ptljz \\
    \threeloffZtwob & 3        & $\mZmll > 10\GeV$    & $\geq$2 & $>$0, $<$0     & 2, 3, 4, $\geq$5 & \ptljz \\
    \threelonZ  1\PQb & 3        & $\mZmll < 10\GeV$ & 1       & No requirement & 2, 3, 4, $\geq$5 & \ptZ \\
    \threelonZ  2\PQb & 3        & $\mZmll < 10\GeV$ & $\geq$2 & No requirement & 2, 3, 4, $\geq$5 & \ptZ or \ptljz\\
    \fourl             & $\geq$4   & No requirement                 & $\geq$2 & No requirement & 2, 3, $\geq$4   & \ptljz \\
    \hline
\end{tabular}}
\end{table}

\subsection{The \texorpdfstring{\twolss}{2lss} event category}
\label{sec:eventSelection_2lss}

The \twolss event category targets \ttH, \ttlnu, and \tttt events, where the events contain two leptons of the same charge which must pass the tight object selection criteria.
The higher (lower) \pt lepton must have $\pt > 25$ (15)\GeV.
The charge requirement significantly reduces $\ttbar+\text{jets}$ background
by leveraging the precision of the CMS detector to reliably reconstruct the electron and muon charges.
We require the uncertainty in the muon \pt to be smaller than 20\% and apply electron selection criteria, described as the ``selective algorithm'' in Ref.~\cite{CMS:2020uim}, that demand the consistency among three independent measurements of the electron charge, based on two different parametrizations of the electron track and the relative positions of the electron track and its energy deposit in the ECAL.
The \twolss events must have jet multiplicity of $\geq$4 with $\pt > 30\GeV$ and $\abseta<2.4$.
We also split the \twolss events based on total lepton charge because the \ttWp cross section in proton-proton collisions is roughly twice that of the \ttWm cross section.
The \twolss events are further subdivided based on \PQb jet multiplicity, which helps target \tttt events since events with
higher numbers of \PQb-tagged jets are enriched in \tttt events.

\subsection{The \texorpdfstring{\threel}{3l} event category}

The \threel event category targets \ttll, \tllq, \ttH, and \ttlnu events.
This category requires  exactly three leptons to pass the tight selection criteria.
The first, second, and third leptons must be above the \pt threshold of 25, 15, and 10\GeV, respectively.
In case the third lepton is an electron, the threshold is $\pt > 15\GeV$,
which suppresses the contributions from nonprompt electrons and helps stay above trigger thresholds.
We also require at least two jets with $\pt > 30\GeV$.
In the \threel event category, we separate events which contain a same-flavor opposite-sign pair of leptons with a mass within 10\GeV of \mZ (91.2\GeV) in order to isolate an enhanced contribution from the on-shell \PZ boson decay, primarily from the \ttZ process.
Events that do not lie within this region are further separated based on whether the sum of the lepton charges is positive or negative to again exploit the difference in cross section between \ttWp and \ttWm.
All \threel events are also categorized by the number of \PQb jets passing the medium \DeepJet working point:\ exactly one \PQb jet, or $\geq$2 \PQb jets.
Requiring one \PQb jet enhances \tllq events, while
requiring $\geq$2 \PQb jets helps to separate \tllq and \ttll events.

\subsection{The \texorpdfstring{\fourl}{4l} event category}
\label{sec:eventSelection_4l}

The \fourl event category targets \ttll and \ttH events, requiring at least four leptons passing the specific selection criteria.
The first through fourth leptons must be above the \pt threshold of 25, 15, 10, and 10\GeV, respectively.
If the last two leptons are electrons, the requirement becomes $\pt > 15\GeV$ for both because of the same reasons described in the \threel event category. The events must have at least two jets with $\pt > 30\GeV$ and $\abseta < 2.4$. At least two of these jets must be \PQb jets, where one of them is required to pass the \DeepJet  medium working point, while the second is allowed to pass the loose working point.

\subsection{Kinematical variables}
\label{sec:eventSelection_diff_vars}

The selections described in Sections \ref{sec:eventSelection_2lss}--\ref{sec:eventSelection_4l} result in 43 unique categories of events.
In order to gain additional sensitivity to EFT effects, the events in each of the 43 categories are binned according to a kinematical variable.

\begin{figure}[!htp]
\centering
\includegraphics[width=0.5\textwidth]{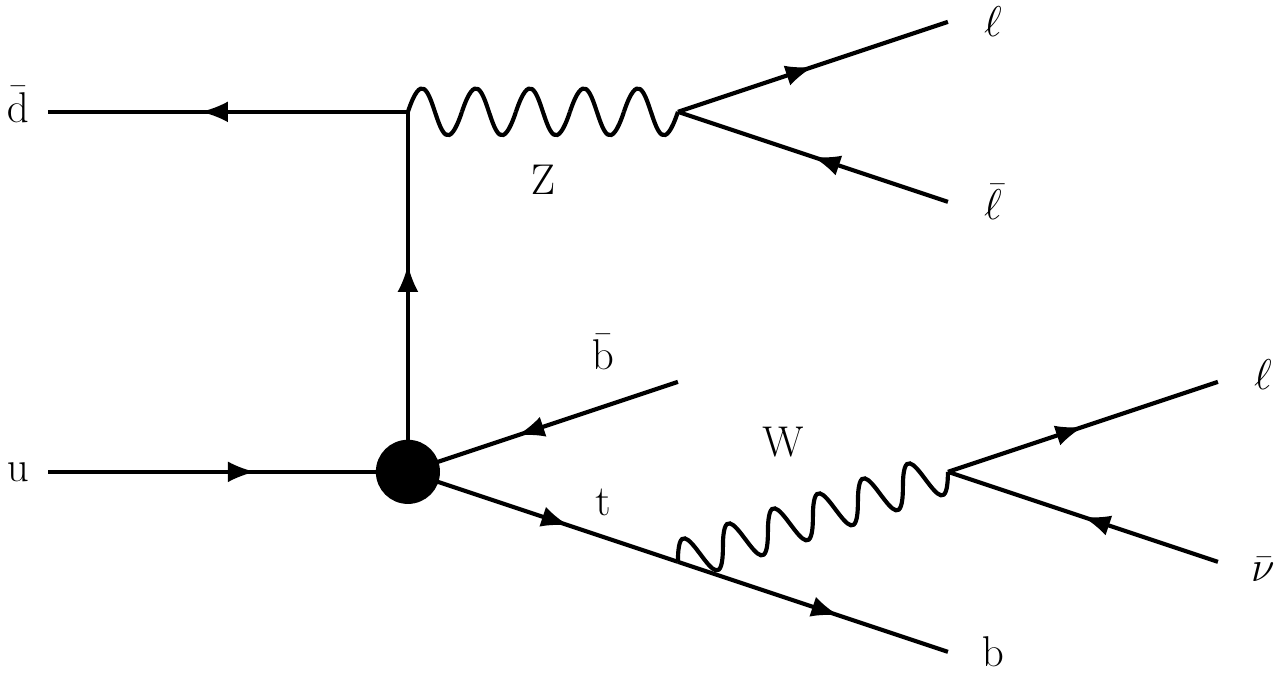}
\caption{Example Feynman diagram illustrating how the \cQqOneThree and \cQqEightThree WCs can impact the processes in \threel categories with two \PQb quarks and an on-shell \PZ boson. These two WCs are a part of the \twohqtwolq group, but unlike the other WCs in this group, these two WCs are associated with operators that have vertices involving a top and bottom quark pair, as pictured in the figure.}
\label{fig:2hq2lq_cqQ83_cqQ13_tZq}
\end{figure}

{\tolerance=800
For most of the event categories, we use the variable \ptljz.
To form this variable, we sum vectorially the momenta of all possible pairs of objects in the collections of leptons and jets and select the combination with the largest \pt. The value of \ptljz is the \pt of that combination.
Thus, this variable may represent the transverse momenta of two leptons, two jets, or a lepton and a jet.
Expressed mathematically, the \ptljz variable can be described as follows: 
\begin{linenomath}
\begin{equation}
\label{eq:ptljz_def}
   \ptljz = \max\left( \max \left[ \pt(\ell,\ell')\right], \, \max \left[ \pt(\PQj,\PQj') \right], \, \max \left[ \pt(\ell,\PQj) \right] \right),
\end{equation}
\end{linenomath}
where $\max[\pt(\ell,\ell')]$ indicates the \pt of the pair of unique leptons with the largest \pt, $\max[\pt(\PQj,\PQj')]$ indicates the \pt of the pair of unique jets with the largest \pt, and $\max[\pt(\ell,\PQj)]$ indicates the \pt of the lepton-jet pair with the largest \pt.
The \ptljz variable is useful because it combines sensitivity to a broad range of EFT effects that grow with energy with access to a combination of EFT operators involving jets and/or leptons.
The \ptljz variable thus provides broadly good sensitivity to most of the WCs included in this analysis, motivating its use in the majority of the event categories.
For most of the on-shell \PZ \threel categories, however, we do not use \ptljz.
Instead, we use a variable that aims to provide sensitivity to the EFT operators involving \PZ bosons, as these operators may modify the kinematical variables of the \PZ boson.
Denoted as \ptZ, this variable is defined as the \pt of the same-flavor opposite-sign lepton pair associated with the \PZ boson.
The \ptZ variable is used for all of the on-shell \PZ boson \threel event categories, except for the event categories with two or three jets and two \PQb jets.
As shown in Fig.~\ref{fig:2hq2lq_cqQ83_cqQ13_tZq},
events with \cQqOneThree and \cQqEightThree vertices can contribute to these final states, and in these cases the \PZ boson is not associated with the EFT vertex, so using \ptZ in these event categories decreases sensitivity to the \cQqOneThree and \cQqEightThree coefficients.
For this reason, we utilize \ptljz instead of \ptZ in these specific on-shell \PZ categories, since this more general variable provides good sensitivity to the \cQqOneThree and \cQqEightThree WCs.
\par}

In summary, the \ptZ variable is used for all of the on-shell \PZ categories, except for the two and three jet categories with two \PQb-tagged jets; the \ptZ variable is thus used in a total of six event categories. In the remaining 37 event categories, the \ptljz variable is used.
For the categories that are binned in \ptljz, four bins are used.
For the categories that are binned in \ptZ (which generally have larger numbers of selected events), five bins are used.
This results in 178 total analysis bins.
Binning the 43 analysis categories in terms of the \ptljz and \ptZ variables provides an improvement in sensitivity of up to a factor of about two, compared to the case where the 43 analysis categories are not further subdivided.

\section{Background estimation}
\label{sec:backgrounds}

In addition to contributions from signal processes, we expect contributions from other SM processes to the signal regions (SRs) defined above.
We distinguish between reducible and irreducible backgrounds. A background
is considered to be irreducible if all final state leptons are genuine prompt leptons.
The contribution from irreducible backgrounds is dominated by diboson (\WZ and \ZZ) production, but
we expect a subleading contribution from triboson and \tWZ production processes.
Although the \tWZ process in principle could be impacted by a subset of the WCs studied in this analysis, the analysis does not have strong sensitivity to the effects (and the predicted contribution is also small, making up only about 3\% of the total predicted yield); the process is therefore categorized as background.
We estimate the contribution of these processes
using simulations with the \MGvATNLO, \POWHEG~\cite{POWHEG1,POWHEG2,POWHEG3,POWHEG_VV,Frixione:2007nw}, and \MCFM~\cite{Campbell:1999ah} programs.
The \WZ production is normalized to the next-to-NLO
in perturbative QCD and NLO in EWK theory~\cite{Grazzini:2019jkl}.
For \ZZ production, the $\qqbar\to\ZZ$ samples are next-to-LO (NLO) in QCD and LO in electroweak~\cite{POWHEG1, POWHEG2}; the $\Pg\Pg\to\ZZ$ samples are LO, and their cross sections are scaled to match a cross section that is NLO in QCD~\cite{Caola:2015psa}.
The diboson contributions are modeled with \POWHEG, \MGvATNLO and \MCFM. The triboson and \tWZ contributions are modeled with \MGvATNLO.

We distinguish three types of reducible backgrounds. The leading contribution is due to processes with
nonprompt leptons (\eg, from the decay of \PQb hadrons) in the final state. We estimate the contribution from these events following the
misidentification probability (MP) method described in Ref.~\cite{CMS:2020mpn}. This method is performed by
selecting events passing all the criteria described in Section~\ref{sec:selection} with the exception
of those on leptons, which are required to pass a looser identification criteria, but to fail
the full set of requirements described in Section~\ref{sec:reconstruction}.
This region, denoted as the application region, is enriched in events
with nonprompt leptons, but resembling the kinematics of the events in the SR.

We obtain an estimation of the nonprompt-lepton contribution to the SR by weighting each event in the application region
by a weight $w$, defined as
\begin{linenomath}
\begin{equation}
    w = (-1)^{n+1}\prod_{i=1}^{n} \frac{f(\pt,\eta)_i}{1-f(\pt,\eta)_i},
\end{equation}
\end{linenomath}
where $n$ is the number of leptons in the event failing the lepton selection criteria and $f(\pt,\eta)_i$ is the MP, defined as the probability
for a nonprompt lepton passing the looser lepton selection to pass the required selection. This probability
is measured in a sample of multijet events collected by a set of low-\pt lepton triggers and is measured as a
function of the lepton \pt and $\eta$, separately for electrons and muons. The validity of the method has been checked using samples of
simulated \ttbar and multijet events, as well as dedicated control regions (CRs), as well as dedicated \twolss control regions (CRs), which are defined with lower jet multiplicities than the \twolss SRs in order to ensure orthogonality with the SR.

An additional contribution to reducible backgrounds in the \twolss SR is due to cases in which the charge
of one of the final state leptons is measured incorrectly. This contribution is
dominated by \ttbar events and is estimated by selecting events passing the same selection
as the \twolss SR, but inverting the same-sign requirement on the dilepton system. Events in this region are weighted
by the probability of each of the leptons to have their charge measured with the wrong sign. This probability
is negligible for muons; however the probability for electrons is larger and we estimate it using simulations. Additionally,
we compare the prediction from the simulation with the observation of a region enriched in $\PZ \to \Pe\Pe$,
where the charge of one of the electrons has been measured with the incorrect sign. The two
agree within the applied uncertainties (30\%), but we correct for the residual differences by a constant
scale factor (SF) in our prediction.

Finally, a small contribution arises from the conversion of photons interacting with the
detector material. This contribution is significantly suppressed by the electron reconstruction and
identification algorithms, since we require electron tracks  to have
hits in the innermost layers of the silicon tracker and we veto electron candidates
that are matched to a reconstructed photon conversion vertex. We estimate this contribution using \ttgamma simulation (similar to Ref.~\cite{CMS:2017moi}) at LO with \MGvATNLO.

\section{Statistical methods}
\label{sec:fitting}

A maximum likelihood fit is performed across all of the analysis bins, treating the number of observed events in each bin as an independent Poisson measurement~\cite{Verkerke:2003ir}.
The total yield in each bin is parametrized as a quadratic function of the \nwc WCs, which are the parameters of interest in the likelihood fit.
The yield in each bin also depends on the effects of the systematic uncertainties (as listed in Section~\ref{sec:systematics}), which are treated as nuisance parameters (NPs) in the likelihood fit.
The likelihood has the functional form
\begin{linenomath}
\begin{equation}
    L = \prod^{N_{\text{bins}}}_{i=1} P(n_i|\nu_i(c, \theta)) \prod^{N_{\mathrm{NP}}}_{j=1}p(\hat{\theta}_j|\theta_j),
\end{equation}
\end{linenomath}
where $P(n_i|\nu_i(c, \theta))$ is the probability of observing $n_i$ events in the $i$-th category given by the Poisson distribution,
and $p(\hat{\theta}_j|\theta_j)$ is the prior probability for the $j$-th NP evaluated at the maximum likelihood estimate $\hat{\theta}_j$.
As described in Section~\ref{sec:samples}, the number of expected events is parametrized as a quadratic function of the WCs. This can be expressed as
\begin{linenomath}
\begin{equation}
    \nu_i(c, \theta) = \mathrm{SM}_i(\theta) + \sum_{j} \frac{c_j}{\Lambda^2} l_{ij}(\theta) + \sum_{j} \frac{c_j^2}{\Lambda^4} q_{ij}(\theta) + \sum_{j \neq k} \frac{c_j c_k}{\Lambda^4} m_{ijk}(\theta),
\end{equation}
\end{linenomath}
where $c$ are the WCs, $\theta$ are the NPs, and $l_j$, $q_j$, and $m_{jk}$ respectively are the linear, quadratic, and cross-term structure constants of the parametrization.
In order to find the 1 and 2 standard deviation ($\sigma$) confidence intervals (CIs) for each WC, a scan is performed over each WC, profiling the other \nwcone WCs and the NPs.
The test statistic is then given by
\begin{linenomath}
\begin{equation}
    \Lambda_{p}(c_{i}) = -2\ln{\frac{L(\hat{c_{i}}, \hat{\theta})}{L(c_{i}, \hat{\hat{\theta}})}},
\end{equation}
\end{linenomath}
where $\hat{c_{i}}$ and $\hat{\theta}$ are the values which maximize the likelihood, and $\hat{\hat{\theta}}$ corresponds to all the profiled parameters which maximize the likelihood at a particular WC value.
The test statistic asymptotically follows a $\chi^2$ distribution where the degrees of freedom correspond to the number of free parameters.
The scan is performed over a discrete set of values for the selected WC; at each WC value, the likelihood fit finds the corresponding values of the other WCs and NPs which minimize the negative log-likelihood (NLL) function.
The 1 and 2$\sigma$ CIs are extracted at the points where the values of the test statistic curve are equal to the values of 1 and 4 respectively.
In order to explore the effects of each WC individually (without interference between WCs), a scan is also performed where the other \nwcone WCs are held to their SM values of zero (instead of profiled).
The 1 and 2$\sigma$ CIs for this scan are extracted in the same way as the profiled fit.
Simultaneous scans are also performed for a selected subset of WC pairs (denoted as 2D).
The 2D scans are performed similarly to the single WC scans;
we step through a discrete set of points for the pair of scanned WCs, with the other \nwctwo WCs either profiled or fixed to their SM values of zero.
The 68.3, 95.5, and 99.7\% 2D CIs are extracted where test statistic is equal to 2.30, 6.18, and 11.83, respectively.

\section{Systematic uncertainties}
\label{sec:systematics}

The systematic uncertainties for this analysis are split into two main categories:\ uncertainties which only affect
the rate of the signal and background processes, and ones which affect both the rate and shape of the
measured distributions. Shape uncertainties which are specified as fully correlated across
all distributions and data-taking years are expressed as a single NP per systematic term.
The sources of systematic uncertainties considered are the following:\ the integrated luminosity, the JES and JER, \PQb jet tagging SFs,
the theoretical cross section, the renormalization (\mur)
and factorization (\muf) scales, the parton shower, the additional radiation,
the electron and muon identification and isolation, the trigger efficiency, the pileup, the L1 ECAL trigger efficiency corrections, the misidentified-lepton rate,
the charge misreconstruction rate, and the mismodeling of the jet multiplicity in diboson events
A breakdown of the systematic uncertainties and their average impact on the nominal predicted yields can be found in Table~\ref{tab:systs}.
\begin{description}
\item[Integrated luminosity:]
The uncertainty in the integrated luminosity is estimated to be 1.6\%~\cite{CMS-LUM-17-003, CMS-PAS-LUM-17-004, CMS-PAS-LUM-18-002} for the 2016--2018 data set.
\item[Jet energy scale and resolution:]
These systematic uncertainties are evaluated by shifting the scale and resolution applied to the reconstructed jets by $\pm 1\sigma$ in bins of \pt and $\eta$.
The JES uncertainty is correlated across years, and is modeled with a total of five independent parameters. The JER uncertainty is uncorrelated across years, and is modeled with a total of four independent parameters.
The impact of the JES and JER uncertainties is on average 1\% of the nominal prediction across all analysis bins.
\item[\PQb jet tagging SFs:] The uncertainties resulting from the \PQb tagging efficiency and misidentification rate are assessed by varying, within their uncertainties, the \PQb tagging data-to-simulation SFs.
The SFs for the heavy-flavor (\PQb and \PQc quark) jets are varied together, and the SFs for the light-flavor (gluon and \PQu, \PQd, and \PQs quark) jets are also together, independently from the heavy-flavor SFs.
The uncertainties for both flavor components are split into a component that is correlated across all data-taking years and components that are uncorrelated across all data-taking years.
The impact of the \PQb jet tagging uncertainty is on average 1\% of the nominal prediction across all analysis bins.
\item[Theoretical cross section:] The predicted yields for all signal and background categories are
normalized to their theoretical cross section values, calculated at NLO precision or greater, with the exception of \ttgamma, which is LO.
The theoretical uncertainties on these calculations come from the PDF choice and the choice of the QCD scales (\mur and \muf).
The average uncertainty across all analysis bins due to the scales is 1--4\% of the total nominal predicted rate, while that from the PDFs is 1\%.
\item[Renormalization and factorization scales:]
Uncertainties in the matrix element generators due to the \mur and \muf scales are measured by shifting the \mur and \muf up (down) by a factor of 2 (0.5) independently, which allows the shape of the variations to be incorporated coherently across bins.
Since the inclusive cross sections for the simulated samples are normalized to NLO predictions and the uncertainty on this overall normalization is already accounted for with the theoretical cross section uncertainty discussed above, the \mur and \muf systematic uncertainties do not impact the inclusive cross section.%
\footnote{Because the precision of the \ttgamma cross section is LO, the \mur and \muf uncertainties impact both the inclusive cross section and the kinematic shape for this sample.}
Rather, these systematic uncertainties affect the kinematic shapes and, correspondingly, the acceptance.
The uncertainty due to  the choice of \mur and \muf is about 3\% of the nominal prediction across all analysis bins.
\item[Parton shower:]
The uncertainty due to initial- and final-state radiation (ISR and FSR) in the parton-shower simulation is estimated by varying the scale of each up (down) by a factor of 2 (0.5) respectively.
The uncertainty due to the FSR and ISR modeling is on average 1--2\% of the nominal prediction across all analysis bins.
\item[Additional radiation:] Because of parton-matching limitations in the \MGvATNLO generator,
additional partons cannot be included in the single \PQt (\tllq and \tHq) LO EFT samples.
Instead, the \tllq LO sample is compared to the NLO \tZq sample. In each jet bin, any discrepancy not covered by the
systematic uncertainties is ascribed as an additional radiation uncertainty. Since this is an uncertainty due
to issues in \MGvATNLO $t$-channel simulations, the same uncertainty is also applied to the \tHq sample.
The average change in yield due to the additional radiation is 7\% of the total nominal predicted rate.
\item[Electron and muon identification and isolation:]
The lepton SFs used to correct the efficiency on simulation to reproduce the efficiency in data are derived with a ``tag-and-probe'' method~\cite{CMS:2020uim,CMS:2018rym,Khachatryan:2010xn}.
The lepton identification, isolation, and tracking efficiency SF uncertainties are estimated and propagated to the final fitting variable distributions. The total uncertainty of the resulting SFs is the quadratic sum of the statistical and systematic uncertainties, for electrons and muons separately.
The uncertainty in the electron (muon) SF results in about 2 (1)\% variation of the nominal prediction across all analysis bins.
\item[Trigger efficiency:]
The impact due to the uncertainty in the trigger efficiency is estimated as well by varying the SFs within their uncertainties separately for each data-taking year and final state. This uncertainty is treated as uncorrelated between data-taking years.
The uncertainty is on average $\leq$1\% of the nominal prediction across all analysis bins.
\item[Pileup:]
Effects due to the uncertainty in the distribution of the number of pileup interactions are evaluated by varying the total inelastic proton-proton cross section used to calculate the number of pileup interactions in data by 4.6\% from its nominal value, which corresponds to a 1$\sigma$ variation~\cite{CMS:2018mlc}.
The uncertainty in each analysis category is on average 1\% of the nominal prediction across all analysis bins.
{\tolerance=800
\item[L1 ECAL trigger efficiency:]
To model the ECAL L1 trigger efficiency in 2016--2017~\cite{CMS:2020cmk}, a weight with its uncertainty is applied to the simulation.
The uncertainty in the predicted yields due to the L1 ECAL trigger efficiency is about 1\% of the total nominal predicted rate.
\par}
\item[Misidentified-lepton rate:]
The misidentification rates used to estimate the nonprompt-lepton background are affected by a statistical uncertainty associated with the number of events in the kinematic region used in the measurement as well as by uncertainties due to the different composition of this measurement region and the SR. Three sources of systematic uncertainty are considered. The first uncertainty stems from the statistical uncertainty associated with the multijet measurement region, while the second accounts for the uncertainty in the subtraction of the prompt-lepton contribution from the yield in this region. The effect of these two uncertainties is taken into account as variations of the MP map overall scale, as well as the dependences on \pt and $\eta$. In addition to  the measurement in the MP, the residual differences between the MP estimation with  multijet and \ttbar simulated samples is taken as an additional source of systematic uncertainty.
The uncertainty in the misidentified-lepton rate is on average of 30\% of the total expected misidentified leptons, and an average of 3\% of the nominal prediction across all analysis bins.
\item[Charge misreconstruction rate:]
An uncertainty of 30\% is assigned to the yield of the misrecon\-structed-charge background to account for the differences observed between the prediction and data in the charge misidentification CR, as noted in Section~\ref{sec:backgrounds}.
The uncertainty in each analysis category is on average 1\% of the nominal prediction across all analysis bins.
\item[Jet mismodeling:]
A discrepancy between the simulations and data was observed in some of the diboson CRs for high jet multiplicities.
This is due to a mismodeling in QCD radiation.
An additional uncertainty derived from the difference between the data and simulation in each jet bin in this CR is added to cover this discrepancy.
The uncertainty is treated as correlated across jet bins.
The uncertainty in the jet mismodeling is on average 7\% of the nominal prediction across all analysis bins.
\end{description}
For the majority of the WCs studied in this analysis, the precision of the result is dominated by the statistical uncertainty; the systematic uncertainties represent the dominant contribution for six of the WCs, including two WCs from the \twohqtwolq category (\cQqEightOne and \ctqEight) as well as four WCs from the \twohqV category (\ctG, \ctp, \cpQM, and \cpt).
For all of the WCs that are dominated by systematic uncertainties, the NLO cross section uncertainties (to which the LO samples are normalized) represent the leading sources of uncertainty.
While normalization uncertainties represent the leading systematic uncertainties, this does not imply that the EFT primarily impacts the normalization.
As discussed in Section~\ref{sec:eventSelection_diff_vars}, binning the events in each selection category according to a kinematical distribution significantly improves the sensitivity, confirming that many of the WCs indeed have strong impacts on the kinematical shapes.
Rather, the fact that the leading systematic uncertainties are normalization uncertainties is a reflection of the fact that EFT may lead to small deviations with respect to the SM. A precise modeling of the SM distribution is thus important for identifying potentially small deviations from the SM, not only in the normalization but also in shape.
Other systematic uncertainties that often have relatively large impacts for these WCs include various uncertainties related to the modeling (\eg, ISR, FSR, \mur, and the diboson jet mismodeling uncertainty). For the WCs that are dominated by systematic uncertainties, the uncertainty on the nonprompt-lepton contribution generally represents the leading experimental systematic uncertainty.

The shape variation due to the PDF uncertainty is measured by reweighting the spectra using 100 replica sets and variations in \alpS. The total uncertainty is then measured using the recommendation in PDF4LHC~\cite{Butterworth:2015oua}. This uncertainty had a negligible effect on the analysis, so it was not included.

\begin{table}[!htp]
\centering
\topcaption{Summary of systematic uncertainties along with the average change in the SM prediction yields}
\renewcommand{\arraystretch}{1.1}
\begin{tabular}{lr}
\hline
Systematic uncertainty & Average change in the yields \\
\hline
Integrated luminosity & 1.6\% \\
Jet energy scale and resolution & 1\% \\
\PQb jet tagging scale factors & 1\% \\
Theoretical cross section & 1--4\% (QCD) 1\% (PDF) \\
Renormalization and factorization scales & 3\% \\
Parton shower & 1--2\% \\
Additional radiation & 7\% \\
Electron and muon identification and isolation & 2\% (electron) 1\% (muon) \\
Trigger efficiency & $\leq$1\% \\
Pileup & 1\% \\
L1 prefiring & 1\% \\
Misidentified-lepton rate & 3\% \\
Charge misreconstruction rate & 1\% \\
Jet mismodeling & 7\% \\
\hline
\end{tabular}
\label{tab:systs}
\end{table}

\section{Results}
\label{sec:results}

\begin{figure}[!htp]
\centering
\includegraphics[width=0.90\textwidth]{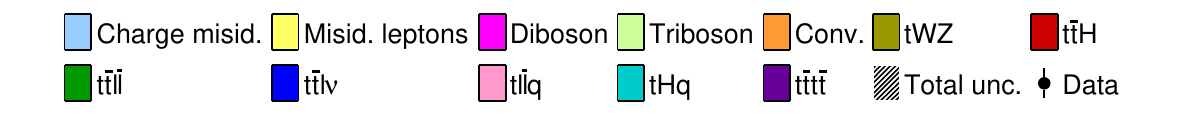}\\
\includegraphics[width=0.45\textwidth]{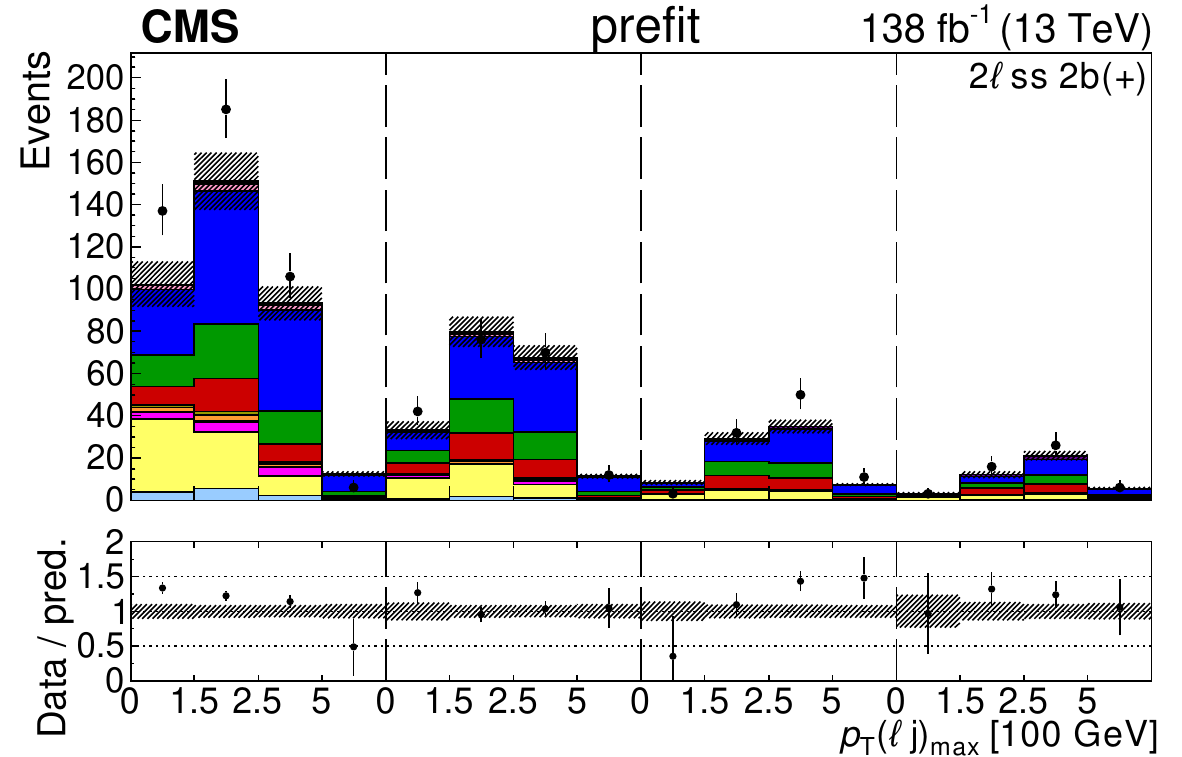}
\includegraphics[width=0.45\textwidth]{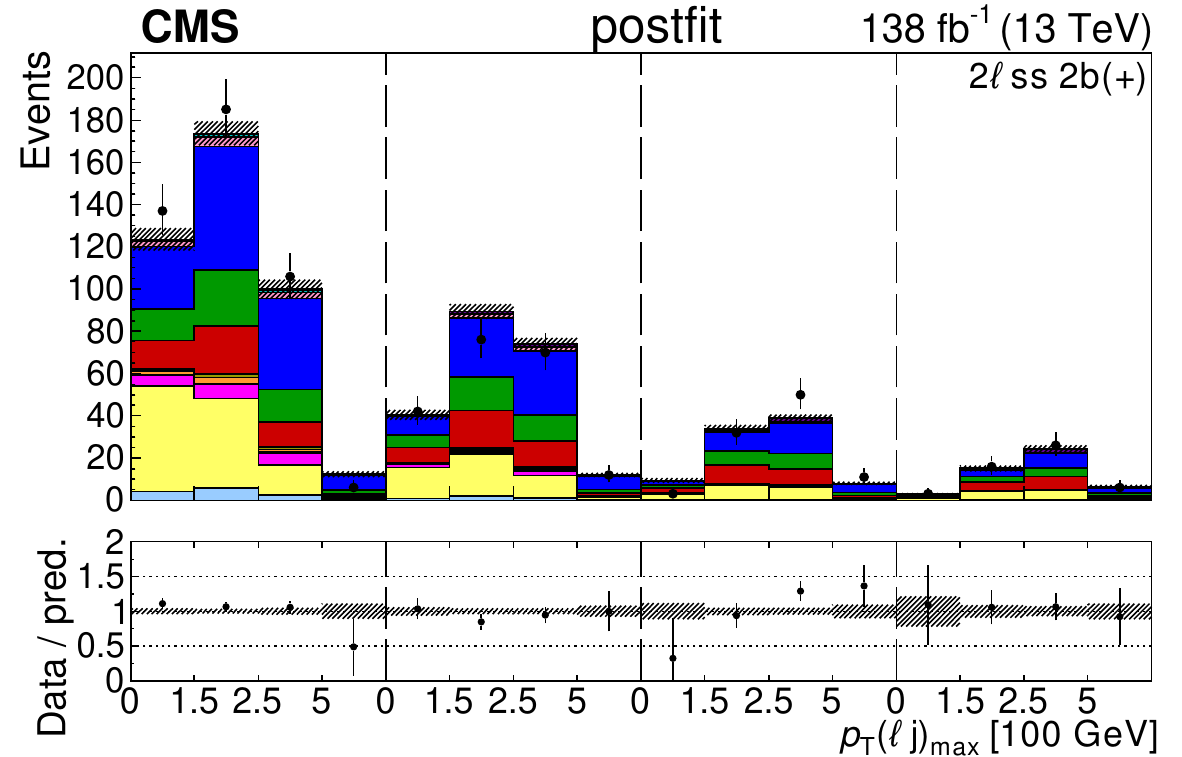}\\
\includegraphics[width=0.45\textwidth]{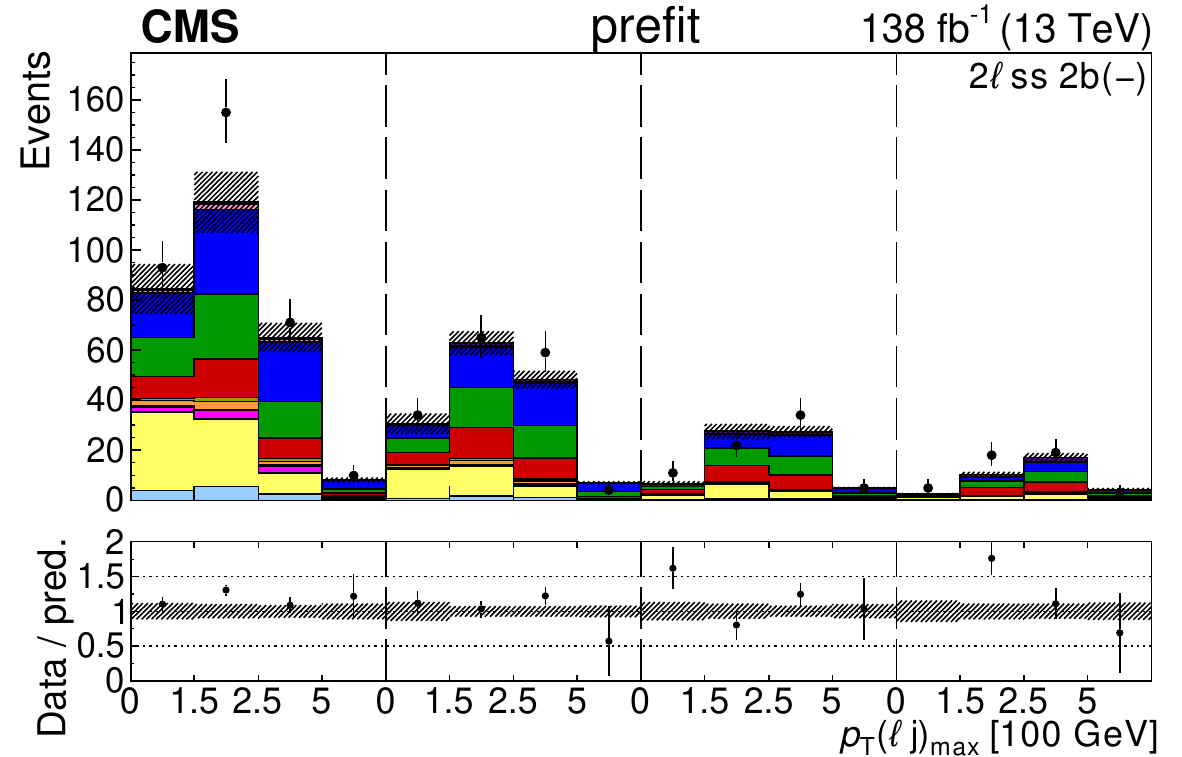}
\includegraphics[width=0.45\textwidth]{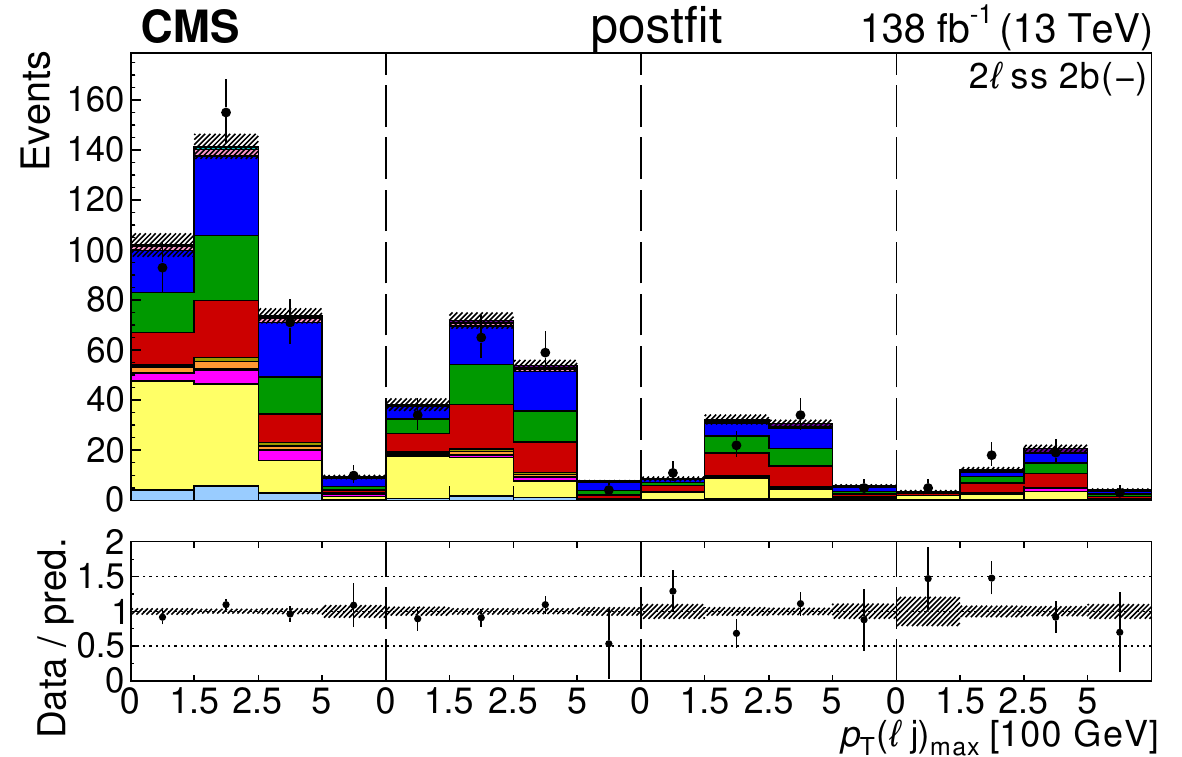}\\
\includegraphics[width=0.45\textwidth]{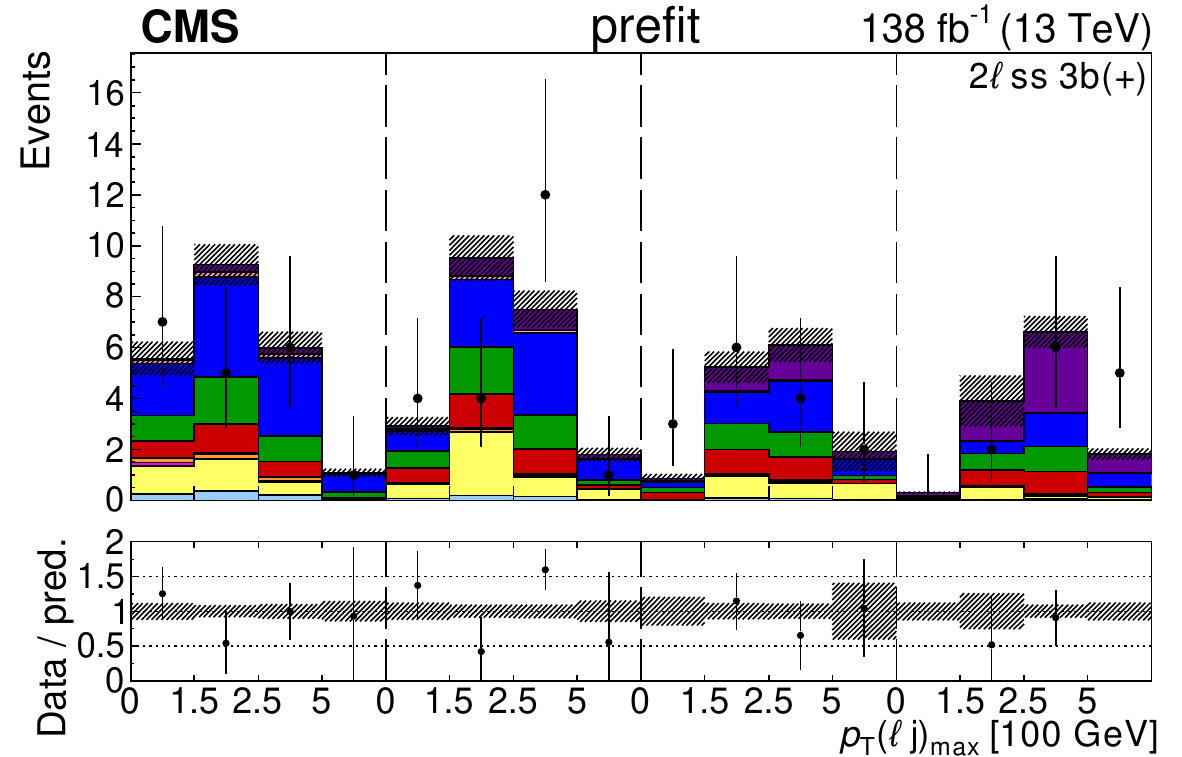}
\includegraphics[width=0.45\textwidth]{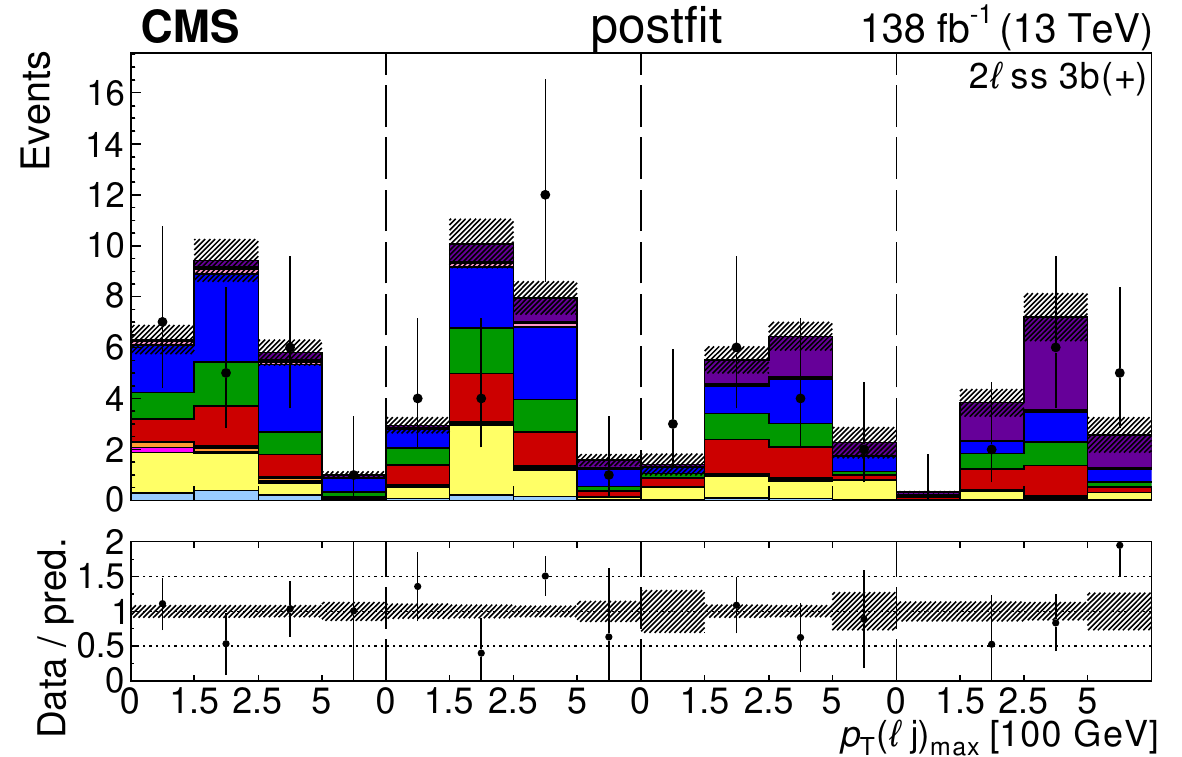}\\
\includegraphics[width=0.45\textwidth]{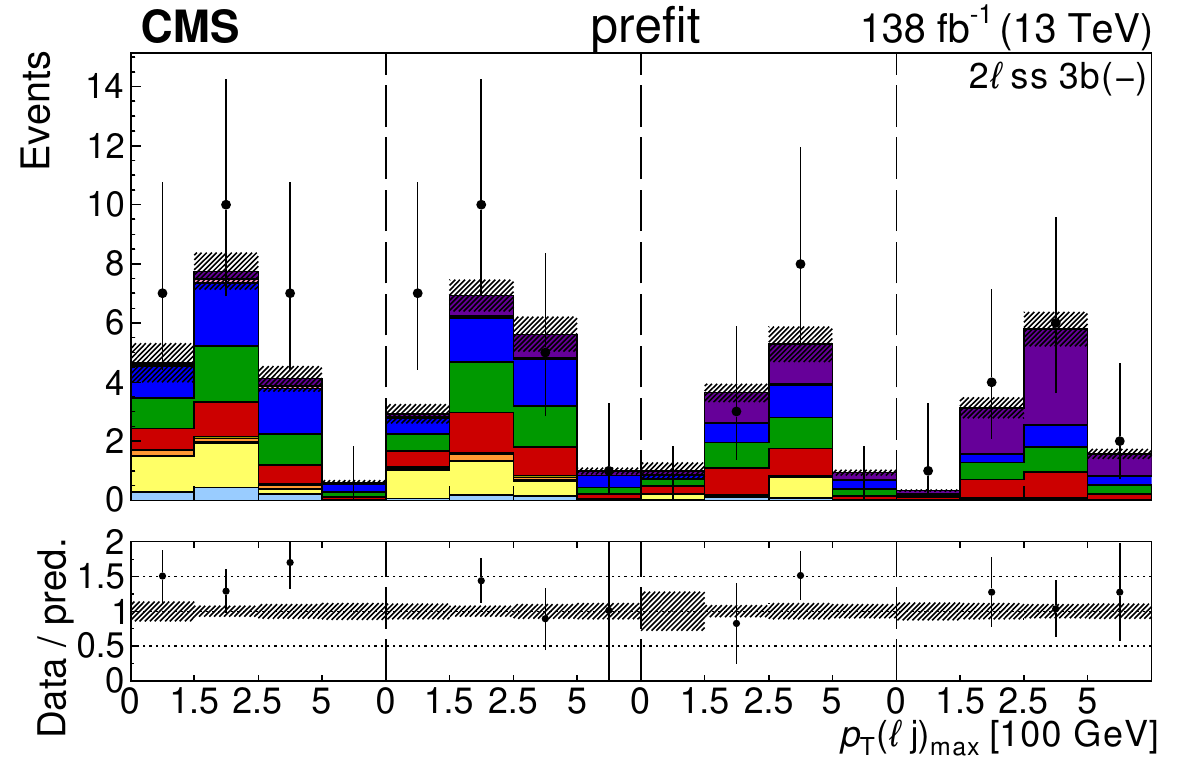}
\includegraphics[width=0.45\textwidth]{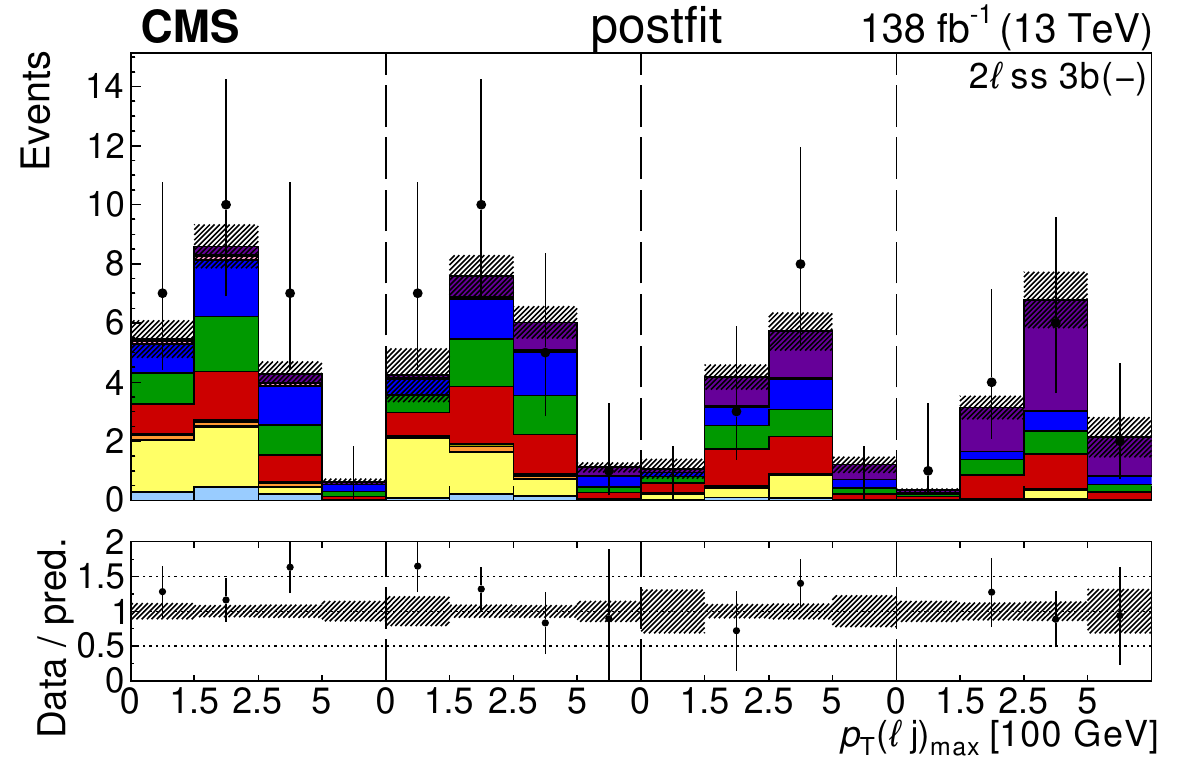}
\caption{%
    The categories shown in these plots are \twolsstwob and \twolssthreeb.
    The prefit plots for each category are shown on the left side while the postfit plots are shown on the right side.
    The differential distribution in the plots is \ptljz.
    The jet subcategories are arranged from low jet multiplicity to high jet multiplicity from left to right for each individual plot.
    For example, in the \mbox{\twolsstwob (+)} plot, the first four bins are the \ptljz variable for \mbox{\twolsstwob (+) 4\PQj},
    the next four bins are for \mbox{\twolsstwob (+) 5\PQj}, \etc
    The process labeled ``Conv.'' corresponds to the photon conversion background,
    ``Misid. leptons'' corresponds to misidentified leptons,
    and ``Charge misid.'' corresponds to leptons with a mismeasured charge.
}
\label{fig:subplots_part1}
\end{figure}

\begin{figure}[!htp]
\centering
\includegraphics[width=0.90\textwidth]{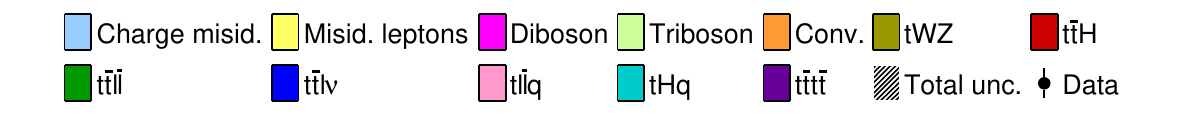}\\
\includegraphics[width=0.45\textwidth]{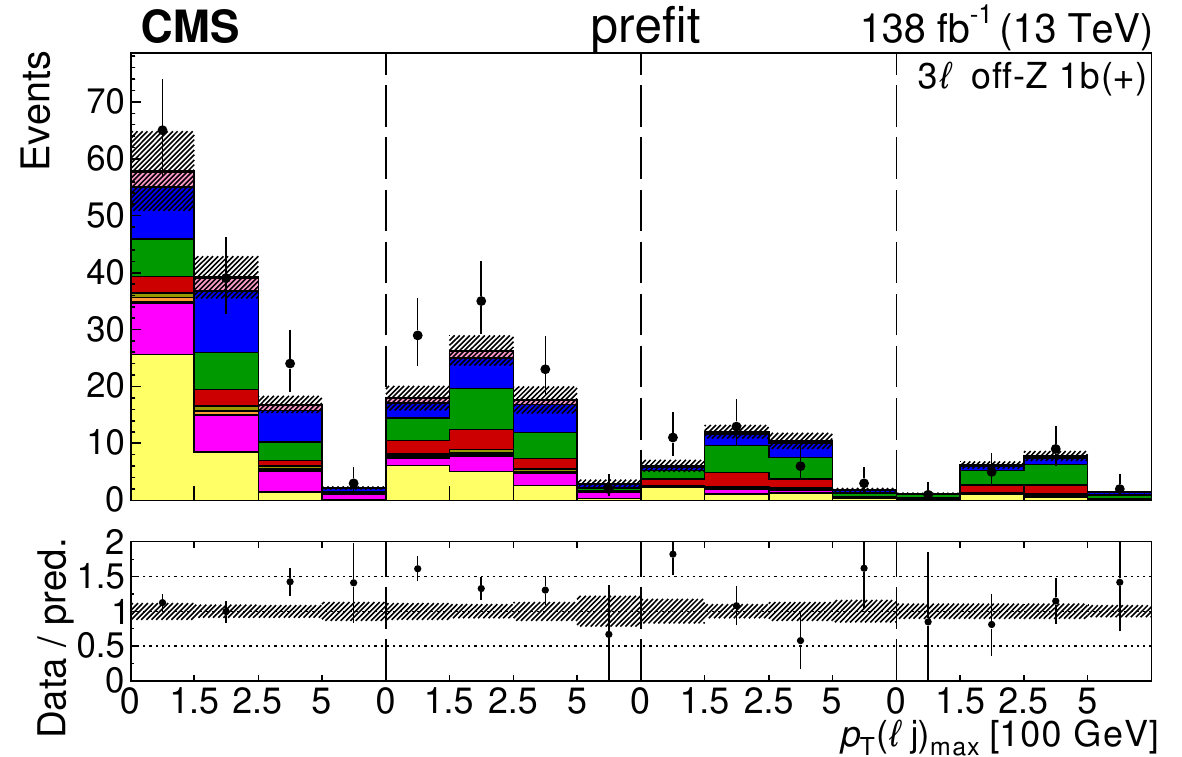}
\includegraphics[width=0.45\textwidth]{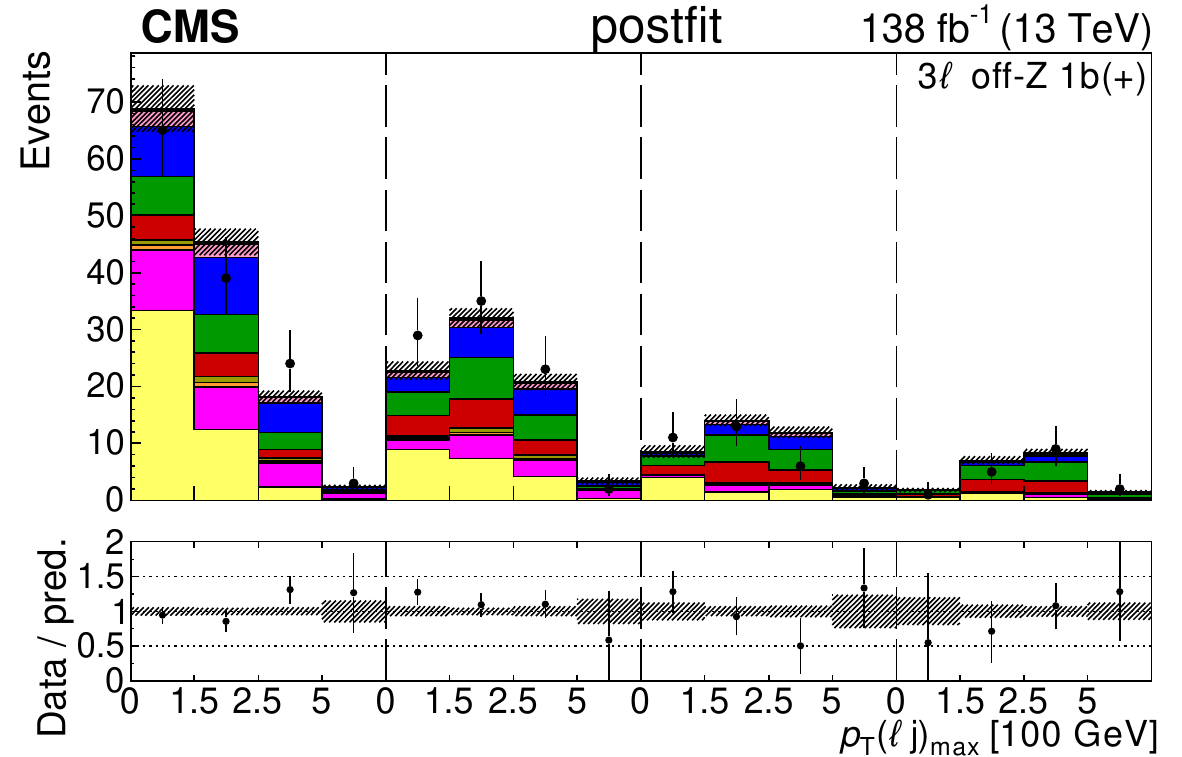}\\
\includegraphics[width=0.45\textwidth]{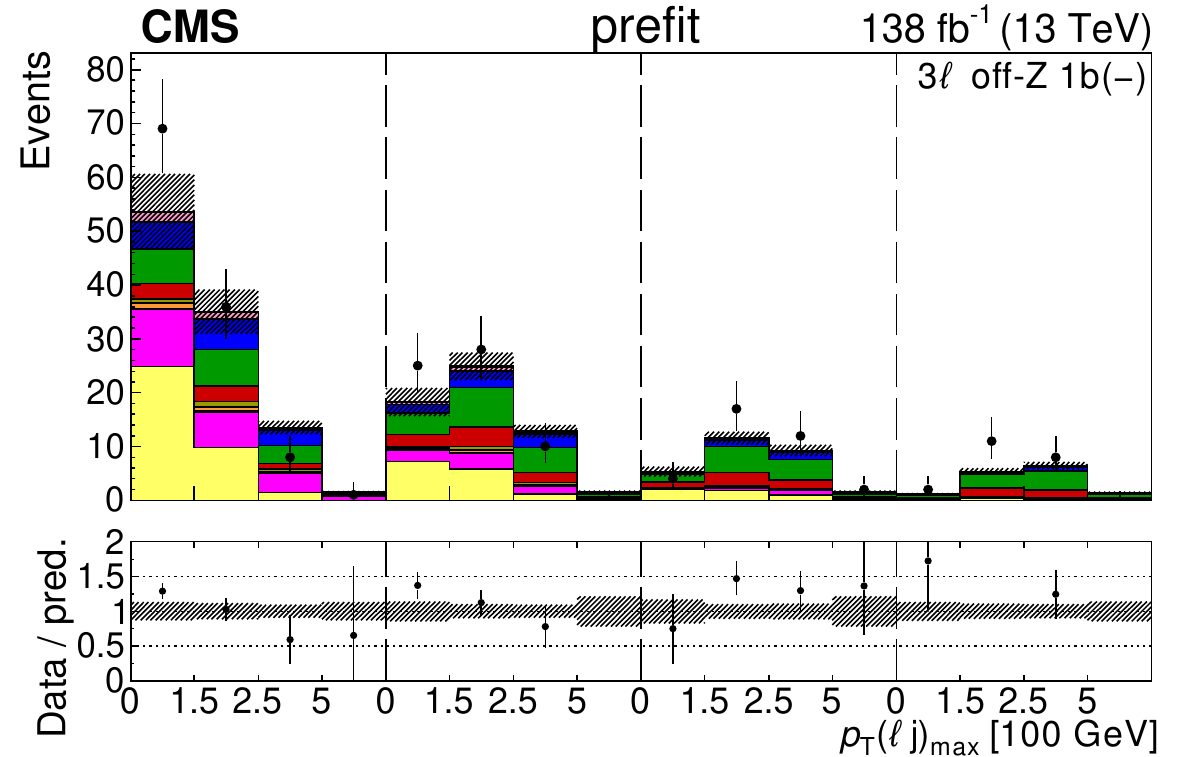}
\includegraphics[width=0.45\textwidth]{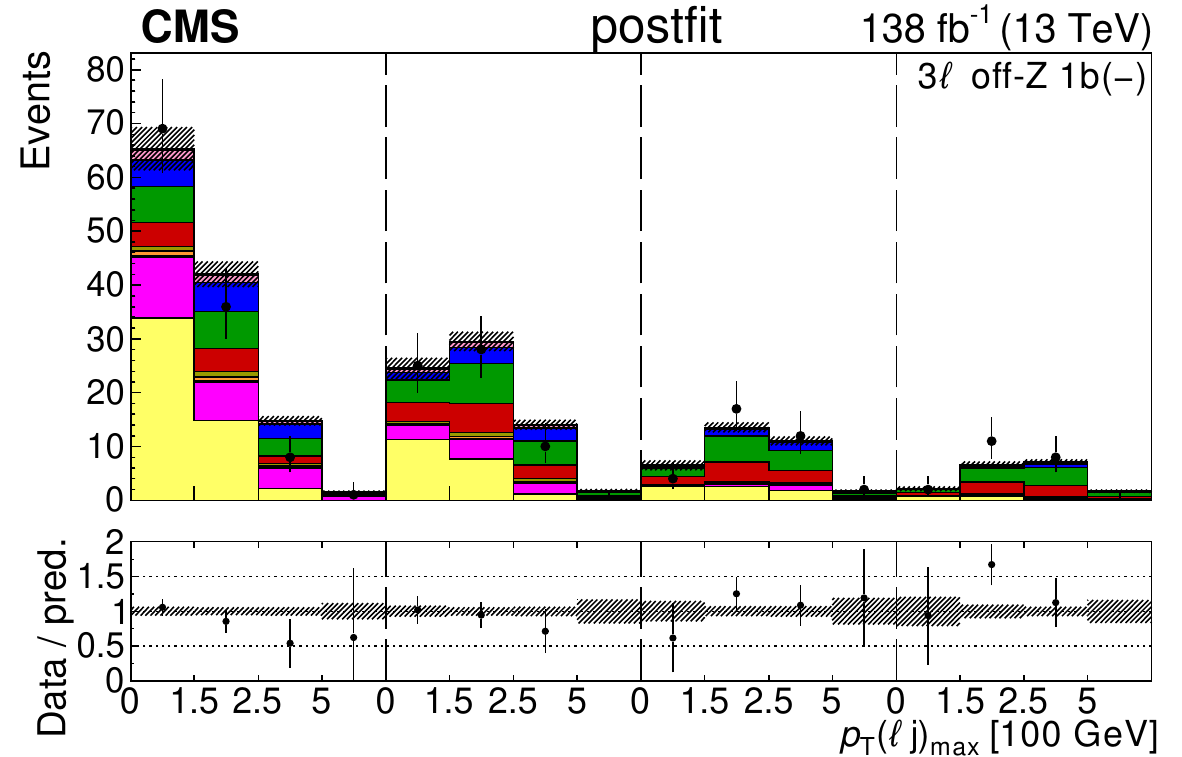}\\
\includegraphics[width=0.45\textwidth]{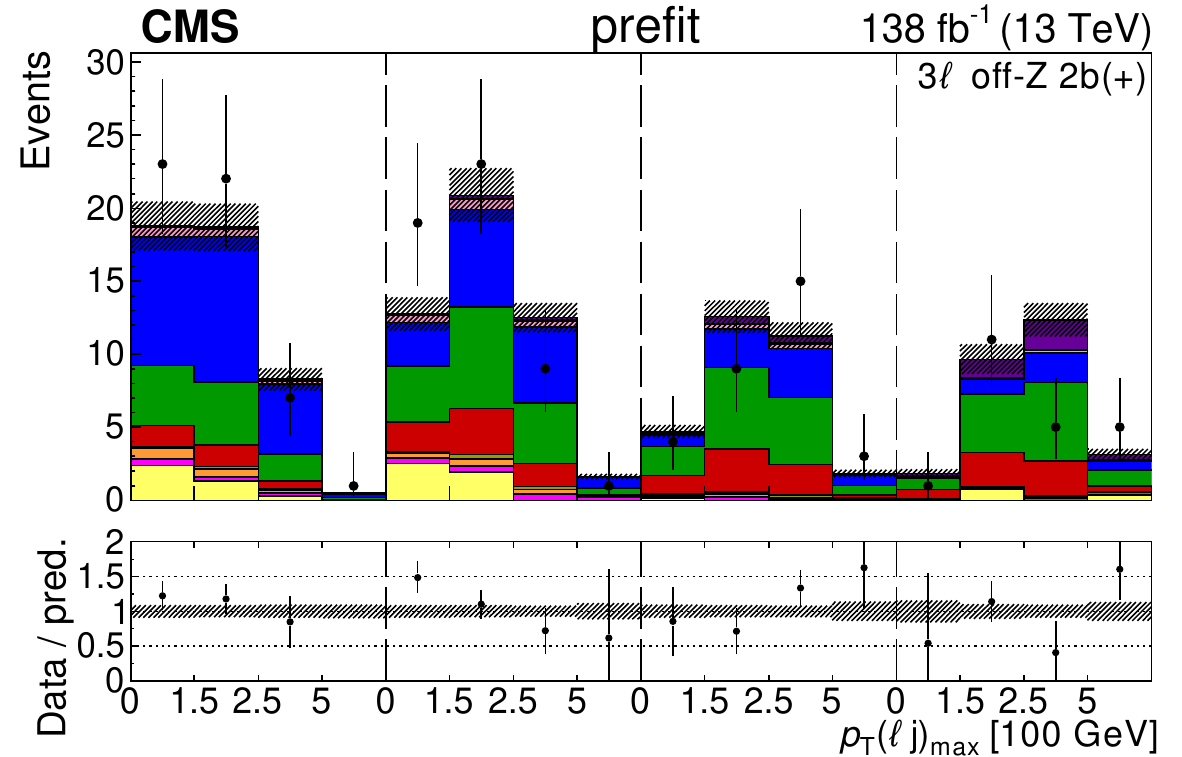}
\includegraphics[width=0.45\textwidth]{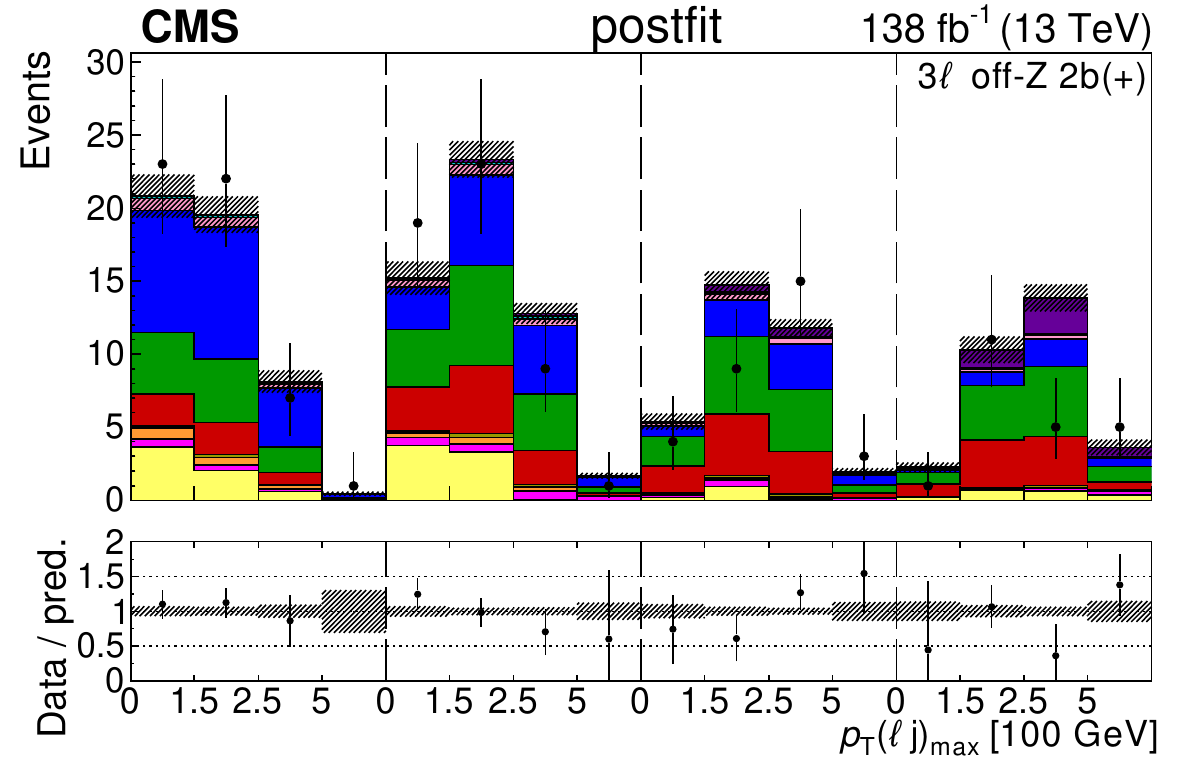}\\
\includegraphics[width=0.45\textwidth]{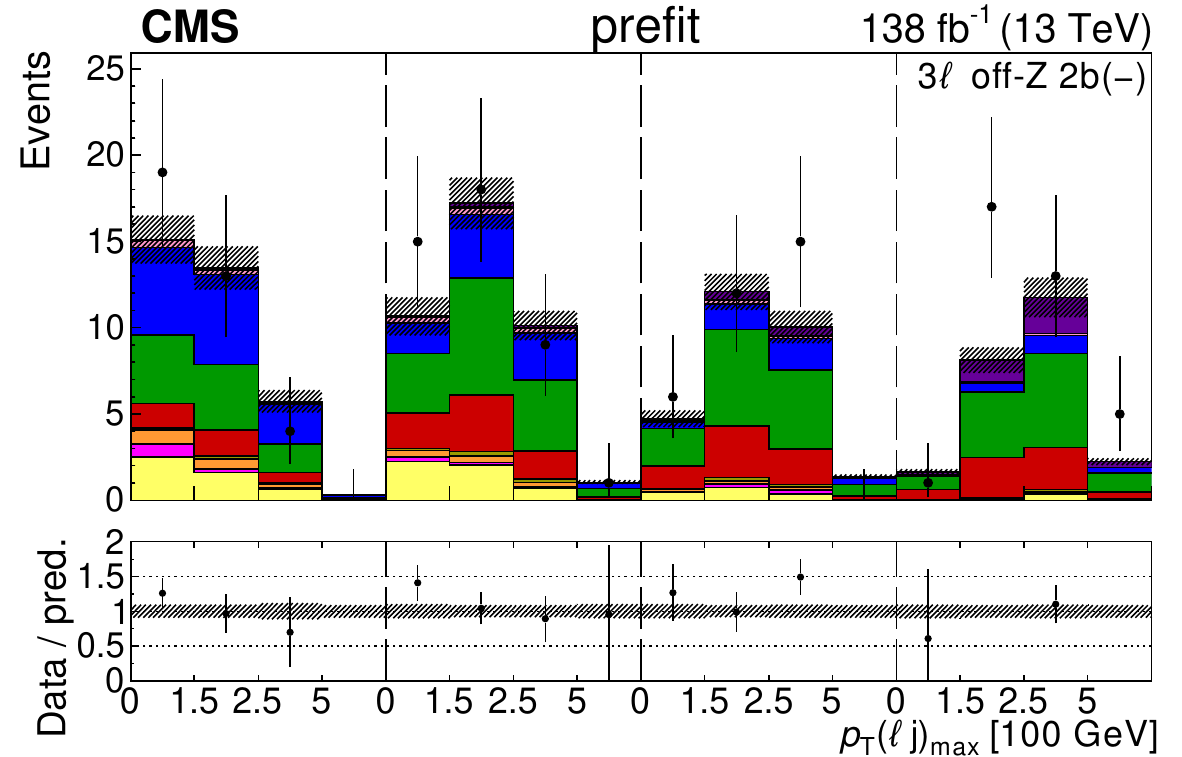}
\includegraphics[width=0.45\textwidth]{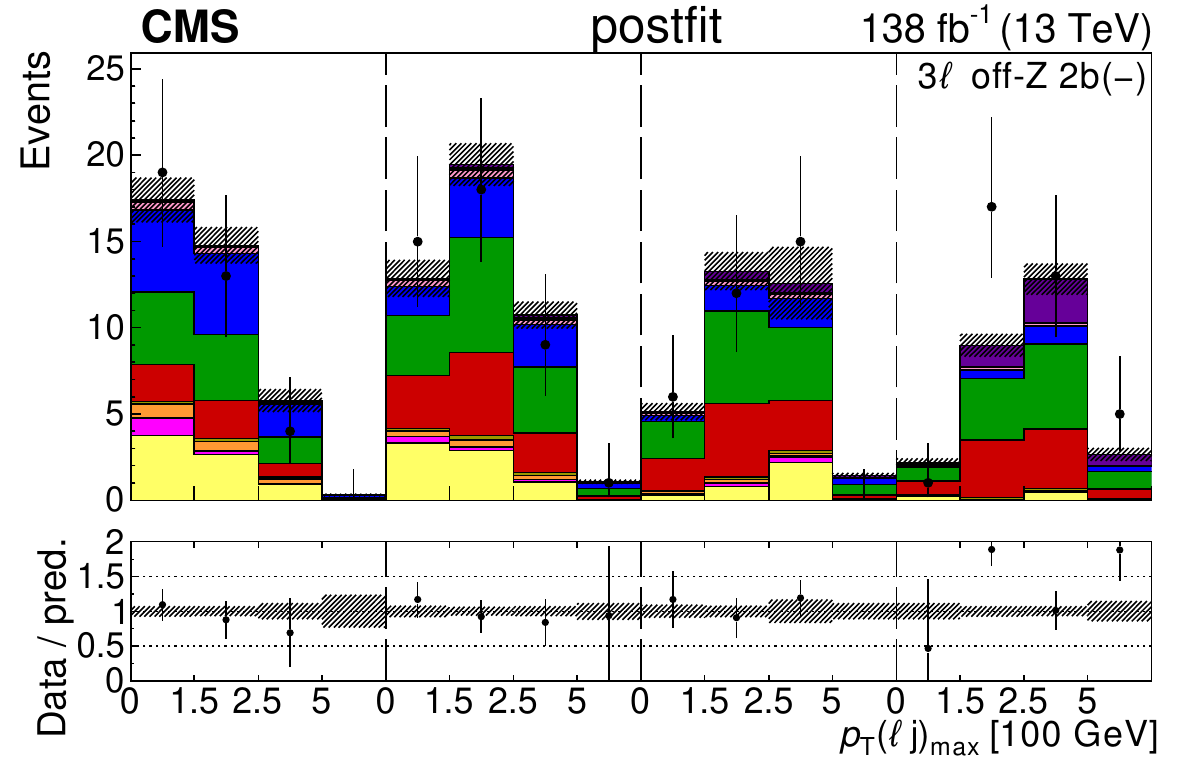}
\caption{%
    The categories shown in these plots are \threeloffZoneb and \threeloffZtwob.
    The prefit plots for each category are shown on the left side while the postfit plots are shown on the right side.
    The differential distribution in the plots is \ptljz.
    The jet subcategories are arranged from low jet multiplicity to high jet multiplicity from left to right for each individual plot.
    For example, in the \mbox{\threeloffZoneb (+)} plot, the first four bins are the \ptljz variable for \mbox{\threeloffZoneb (+) 2\PQj},
    the next four bins are for \mbox{\threeloffZoneb (+) 3\PQj}, \etc
}
\label{fig:subplots_part2}
\end{figure}

\begin{figure}[!htp]
\centering
\includegraphics[width=0.90\textwidth]{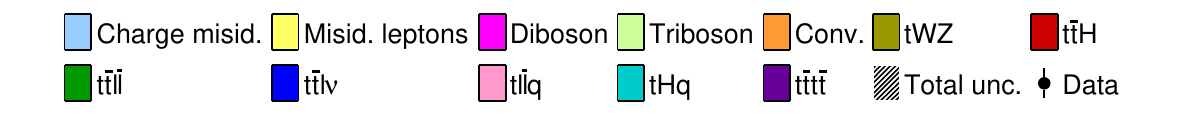}\\
\includegraphics[width=0.45\textwidth]{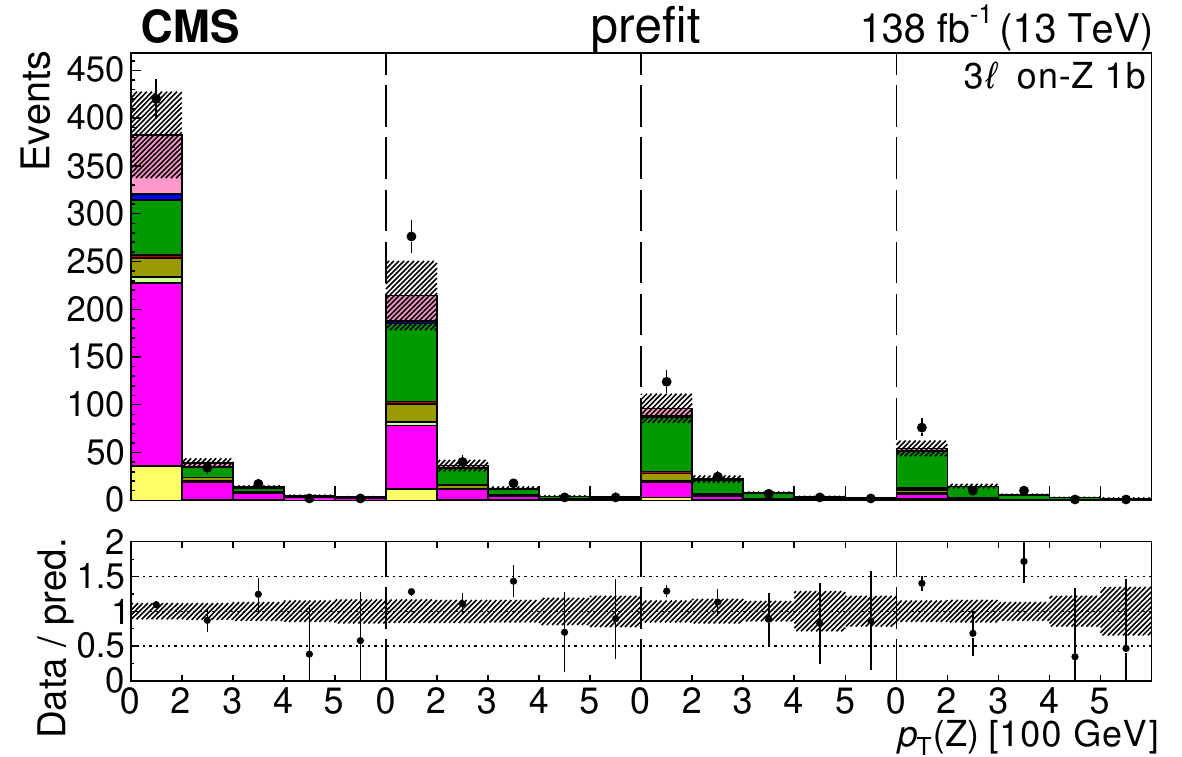}
\includegraphics[width=0.45\textwidth]{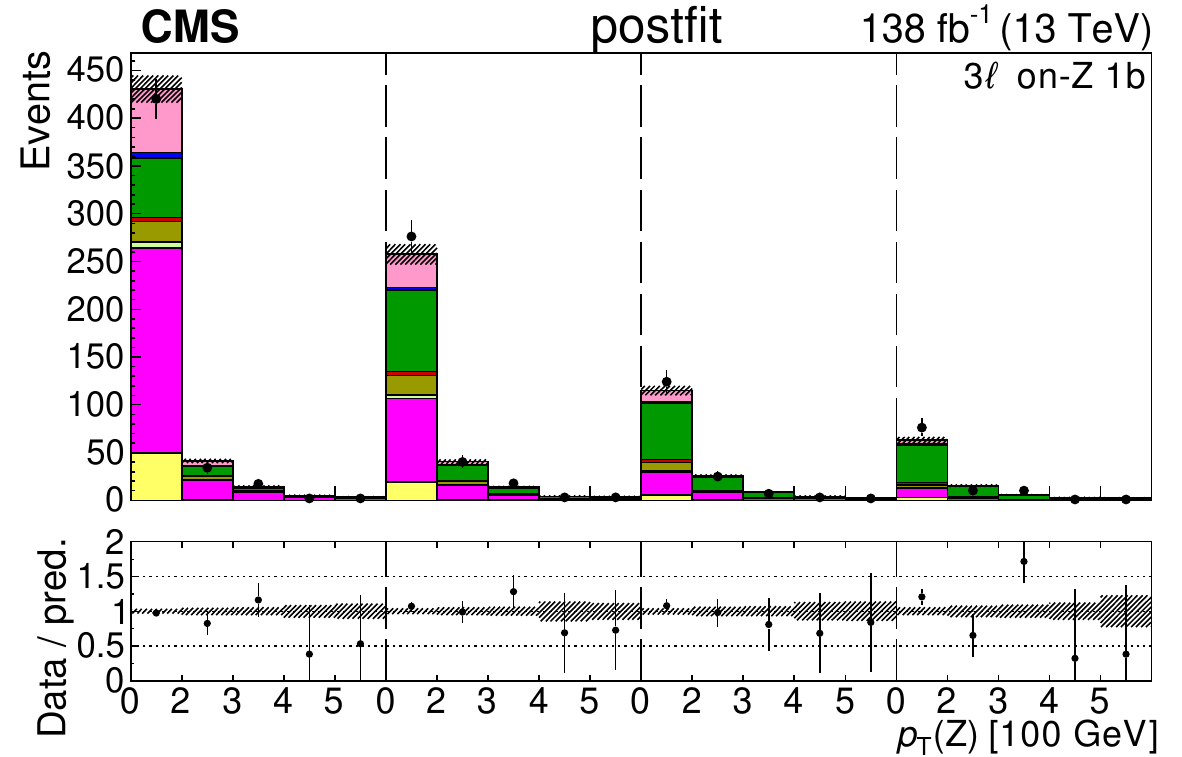}\\
\includegraphics[width=0.45\textwidth]{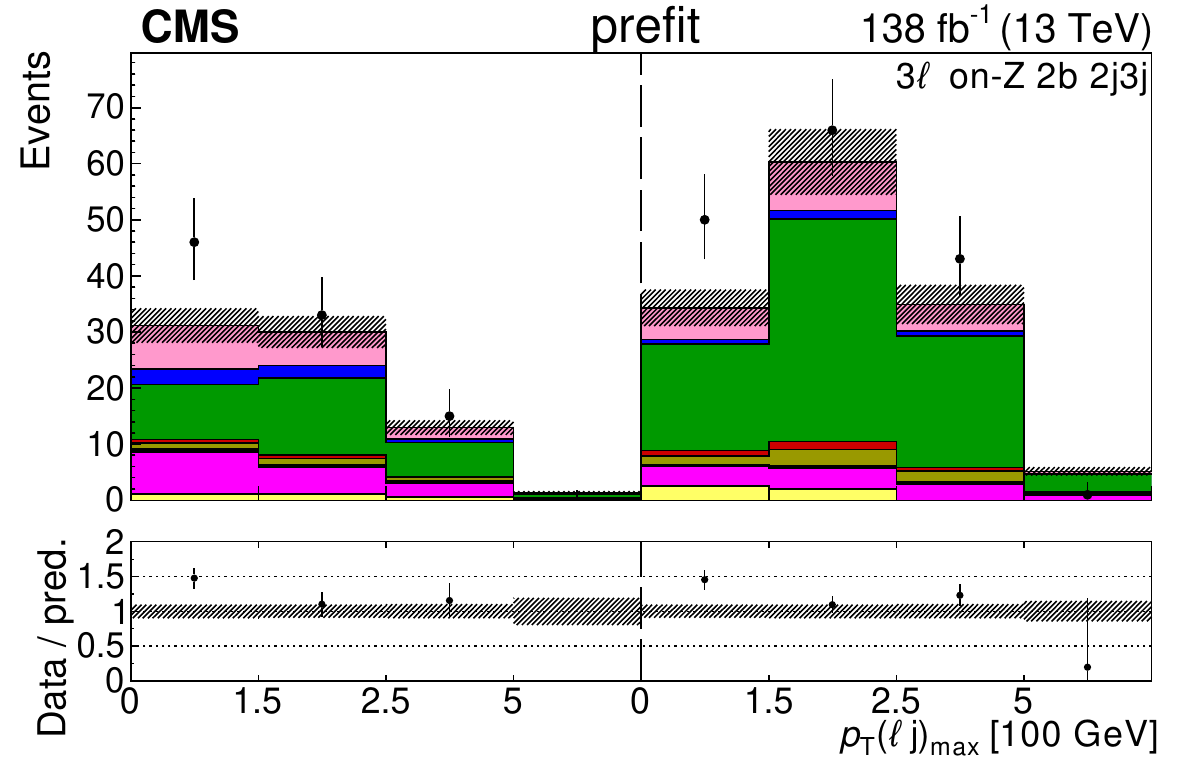}
\includegraphics[width=0.45\textwidth]{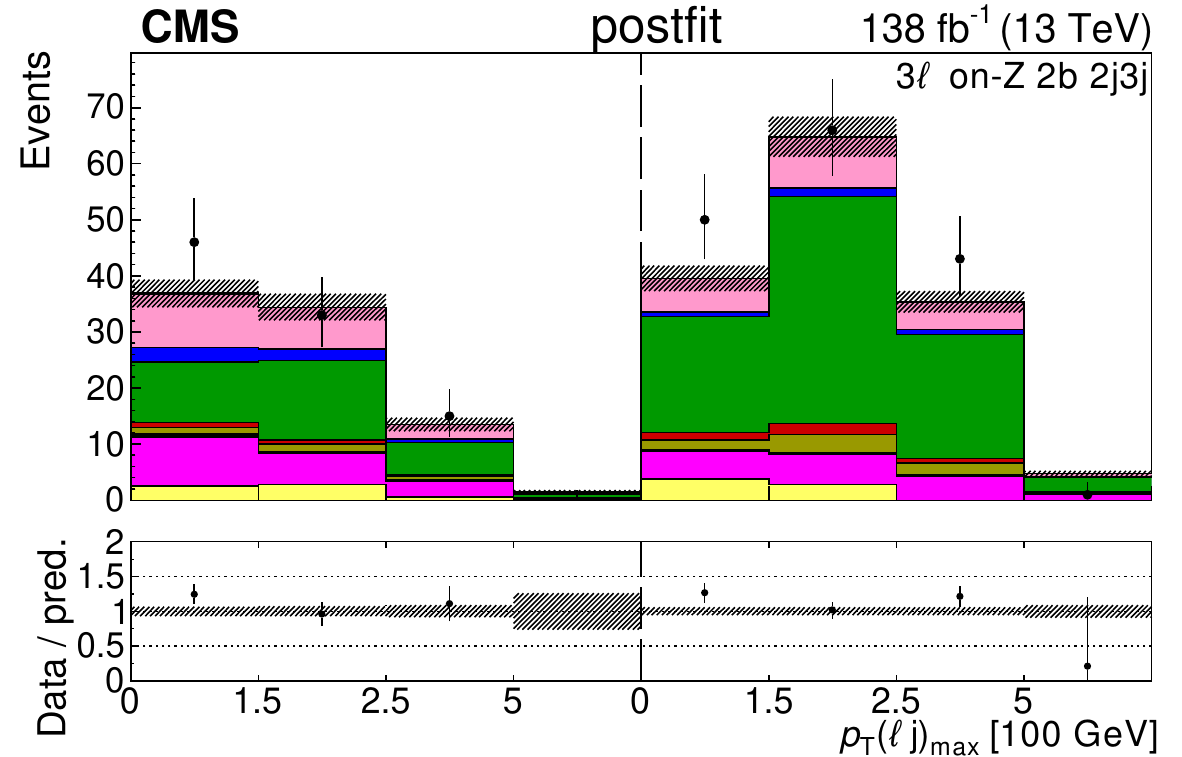}\\
\includegraphics[width=0.45\textwidth]{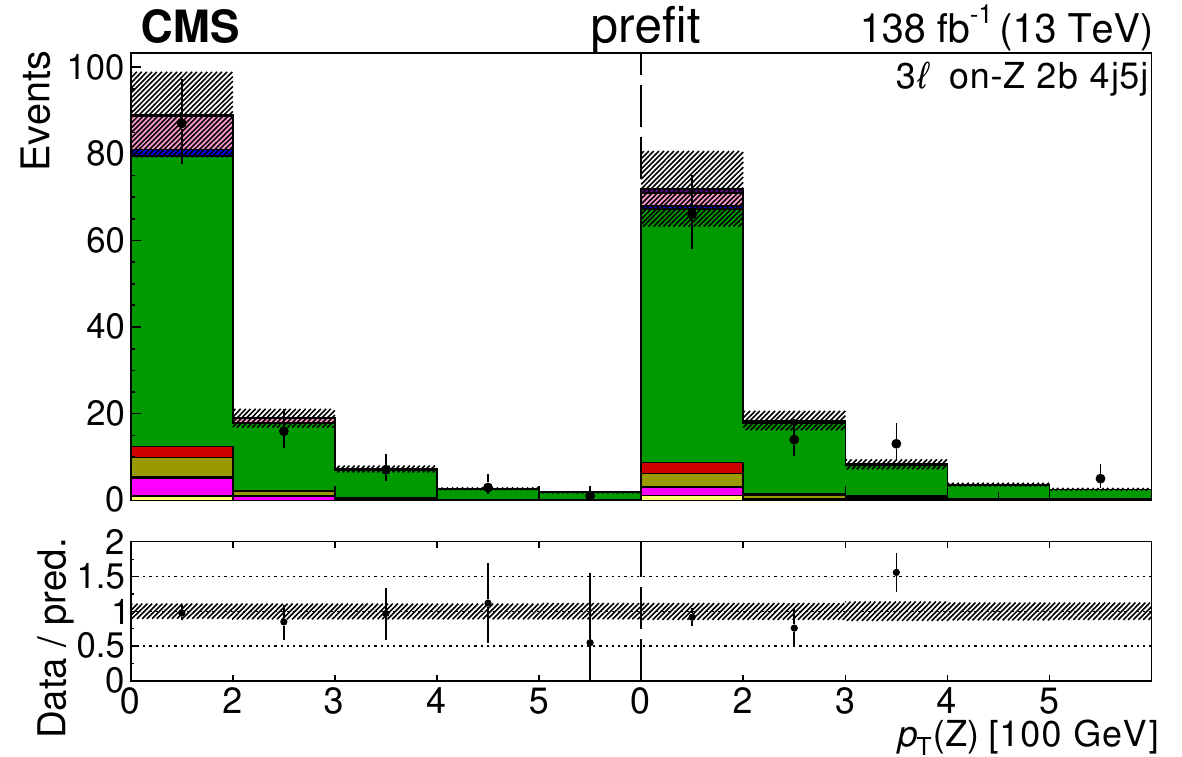}
\includegraphics[width=0.45\textwidth]{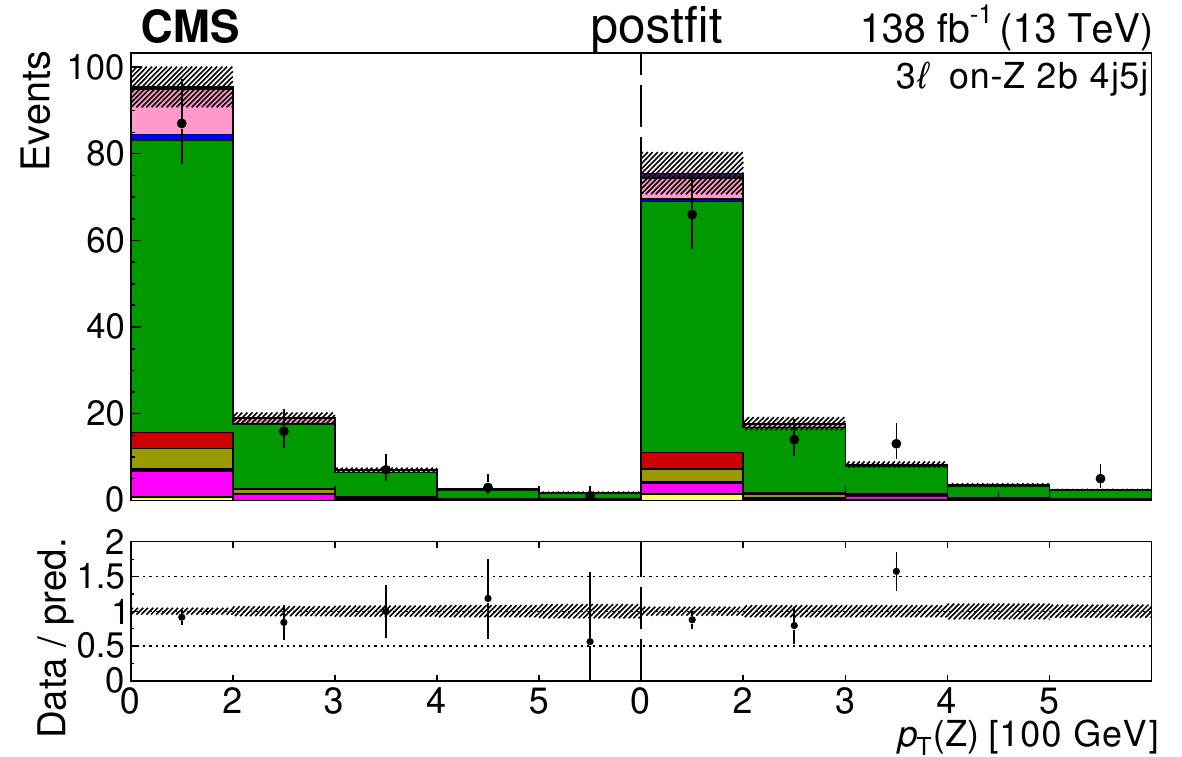}\\
\includegraphics[width=0.45\textwidth]{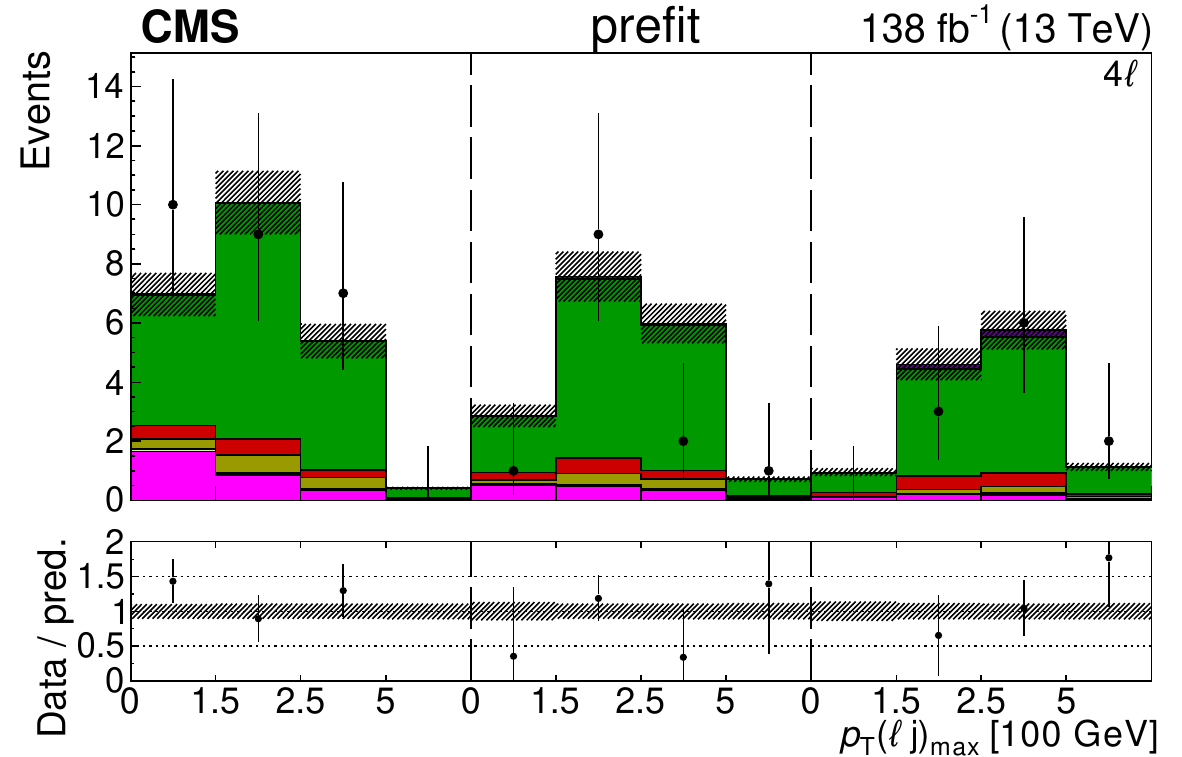}
\includegraphics[width=0.45\textwidth]{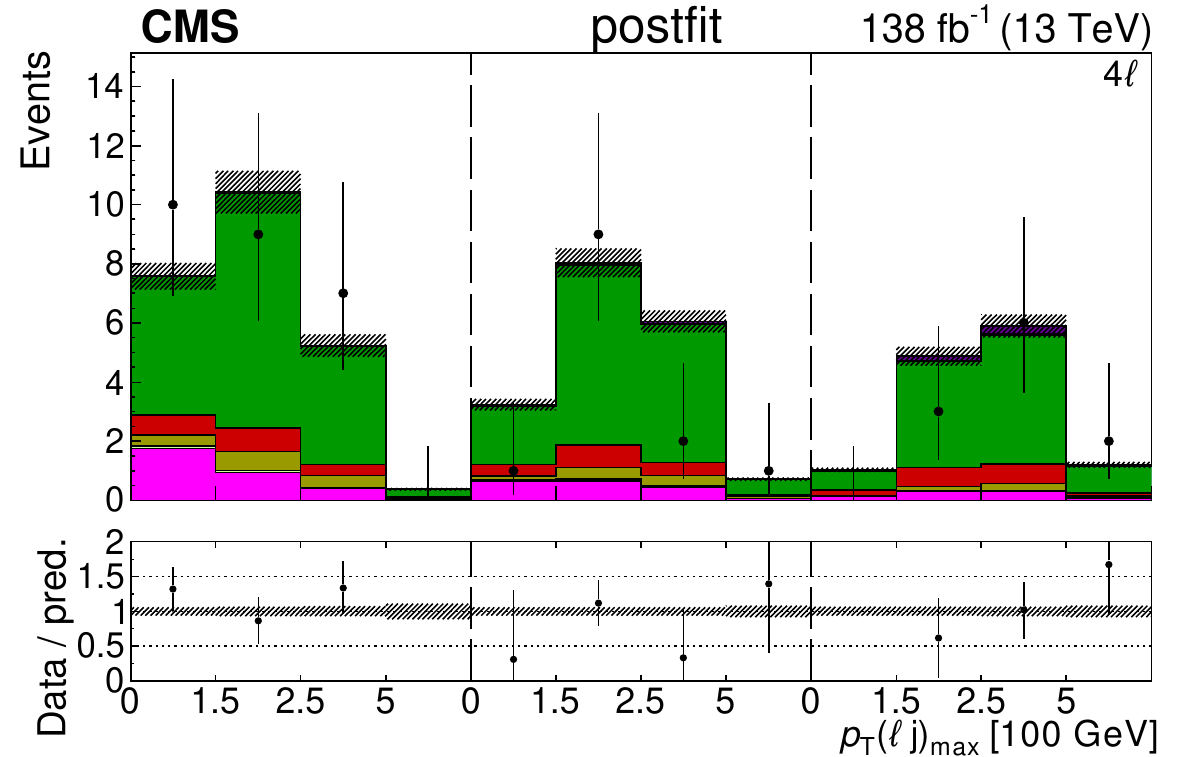}
\caption{%
    The categories shown in these plots are \threelonZoneb, \threelonZtwob, and \fourl.
    The prefit plots for each category are shown on the left side while the postfit plots are shown on the right side.
    The differential distribution is \ptZ in the plots of \threelonZoneb and \threelonZtwob (4\PQj and 5\PQj),
    and \ptljz in the plots of \threelonZtwob (2\PQj and 3\PQj) and \fourl.
    The jet subcategories are arranged from low jet multiplicity to high jet multiplicity from left to right for each individual plot.
    For example, in the \threelonZoneb plot, the first five bins are the \ptZ variable for \mbox{\threelonZtwob 2\PQj},
    the next five bins are for \mbox{\threelonZtwob 3\PQj}, \etc
}
\label{fig:subplots_part3}
\end{figure}

\begin{figure}[!htp]
\centering
\includegraphics[width=0.90\textwidth]{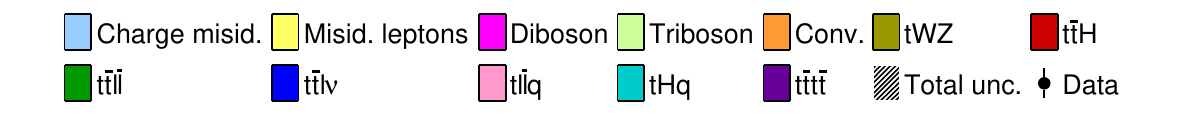}\\
\includegraphics[width=\textwidth]{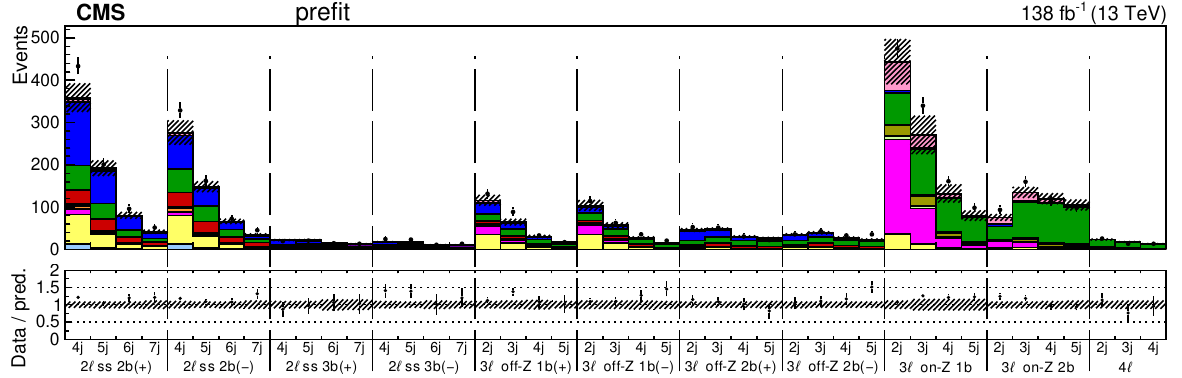} \\[1ex]
\includegraphics[width=\textwidth]{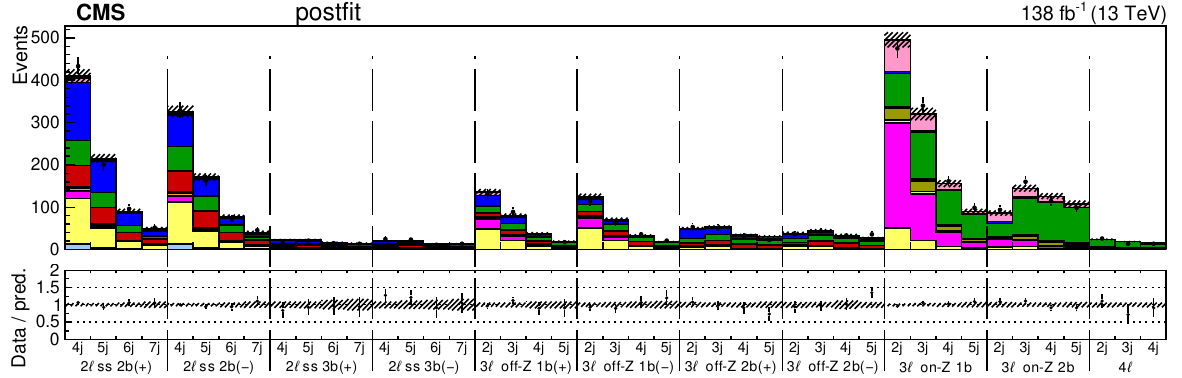}
\caption{%
    Observed data and expected yields in the prefit (upper) and postfit (lower) scenarios.
    All kinematic variables have been combined, resulting in distributions for the jet multiplicity only.
    The postfit values are obtained by simultaneously fitting all \nwc Wilson coefficients (WCs) and the nuisance parameters (NPs).
    The lower panel contains the ratios of the observed yields over the expected.
    The error bands are computed by propagating the uncertainties from the WCs and NPs.
}
\label{fig:prepost_njet}
\end{figure}

The number of observed events for all 178 bins is shown in Figs.~\ref{fig:subplots_part1}, \ref{fig:subplots_part2}, and~\ref{fig:subplots_part3} separated by the different signal categories,
with the expectation obtained by setting all WCs to their SM values of zero (prefit)
or simultaneously fitting the \nwc WCs and the NPs by minimizing the NLL (postfit).
To visualize the relative yields across the categories, Fig.~\ref{fig:prepost_njet} combines the bins of the kinematic variables,
resulting in a plot of jet multiplicity for each selection category.
The hatched regions in the stacked plot and shaded regions in the ratio plot correspond to the total systematic uncertainty.
We have quantified the level of agreement between data and the SM hypothesis using a saturated model. We obtain a p-value of 0.18, showing no significant discrepancies between the two.

\begin{figure}[!htp]
\centering
\includegraphics[width=0.85\textwidth]{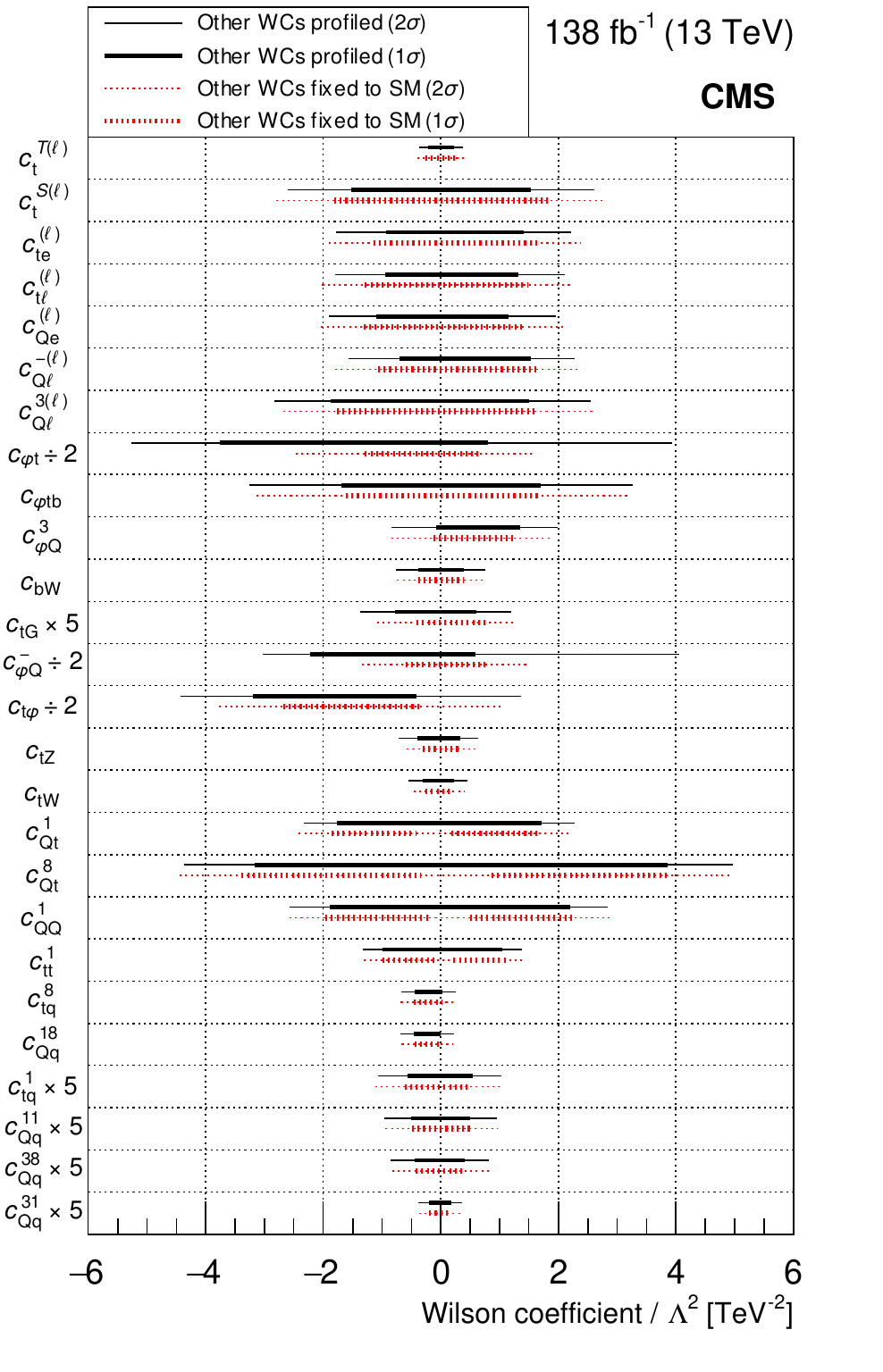}
\caption{Summary of CIs extracted from the likelihood fits described in Section~\ref{sec:fitting}. The WC 1$\sigma$ (thick line) and 2$\sigma$ (thin line) CIs are shown for the case where the other WCs are profiled (in solid black), and the case where the other WCs are fixed to their SM values of zero (in dashed red). To make the figure more readable, the intervals for \ctp,  \cpt, and \cpQM were scaled by 0.5, and the intervals for \ctG, \ctqOne, \cQqOneOne, \cQqEightThree, and \cQqOneThree were scaled by 5.}
\label{fig:ci_summary}
\end{figure}

The 1 and 2$\sigma$ CIs extracted from the likelihood fits described in Section~\ref{sec:fitting} are shown in Fig.~\ref{fig:ci_summary}.
The solid black (dashed red) bars show the results of the fits in which the other \nwcone WCs are profiled (fixed to their SM values of zero).
The CIs for all \nwc WCs are consistent with the SM.
The 1 and 2$\sigma$ CIs for each WC are listed in Tables~\ref{tab:ci_1sigma} and \ref{tab:ci_2sigma}, respectively.
We note that, as mentioned in Section~\ref{sec:samples}, the definition of the operator associated with \ctG here includes an explicit factor of the strong coupling constant, which should be accounted for when comparing to results extracted based on other conventions.

\begin{table}[!htp]
\centering
\topcaption{The 1$\sigma$ uncertainty intervals extracted from the likelihood fits described in Section~\ref{sec:fitting}. The intervals are shown for the case where the other Wilson coefficients (WCs) are profiled, and the case where the other WCs are fixed to their SM values of zero.}
\renewcommand{\arraystretch}{1.5}
\begin{tabular}{ccc}
    \hline
    WC/$\Lambda^2$ [{\perTeVsqns}] & 1$\sigma$ CI (other WCs profiled) & 1$\sigma$ CI (other WCs fixed to SM) \\
    \hline
    WC category \twohqtwolep & & \\[-3pt]
    \ctlT             &   [$-0.22$, 0.22]  &   [$-0.26$, 0.26]                   \\
    \ctlS             &   [$-1.52$, 1.53]  &   [$-1.81$, 1.82]                   \\
    \cte              &   [$-0.93$, 1.41]  &   [$-1.15$, 1.68]                   \\
    \ctl              &   [$-0.95$, 1.32]  &   [$-1.29$, 1.47]                   \\
    \cQe              &   [$-1.10$, 1.16]  &   [$-1.32$, 1.40]                   \\
    \cQlM             &   [$-0.71$, 1.53]  &   [$-1.07$, 1.64]                   \\
    \cQla             &   [$-1.87$, 1.50]  &   [$-1.76$, 1.63]                   \\
    WC category \twohqV & & \\[-3pt]
    \cpt              &   [$-7.52$, 1.59]  &   [$-2.59$, 1.32]                   \\
    \cptb             &   [$-1.69$, 1.70]  &   [$-1.61$, 1.67]                   \\
    \cpQa             &   [$-0.09$, 1.35]  &   [$-0.13$, 1.25]                   \\
    \cbW              &   [$-0.39$, 0.40]  &   [$-0.39$, 0.39]                   \\
    \ctG              &   [$-0.15$, 0.12]  &   [$-0.08$, 0.15]                   \\
    \cpQM             &   [$-4.44$, 1.17]  &   [$-1.20$, 1.57]                   \\
    \ctp              &   [$-6.40$, $-0.82$] &   [$-5.37$, $-0.64$]                  \\
    \ctZ              &   [$-0.40$, 0.32]  &   [$-0.31$, 0.32]                   \\
    \ctW              &   [$-0.31$, 0.22]  &   [$-0.26$, 0.21]                   \\
    WC category \fourhq & & \\[-3pt]
    \cQtOne           &   [$-1.77$, 1.71]  &   [$-1.86$, $-0.41$]${}\cup{}$[0.19, 1.70] \\
    \cQtEight         &   [$-3.17$, 3.86]  &   [$-3.39$, $-0.34$]${}\cup{}$[0.86, 3.87] \\
    \cQQOne           &   [$-1.89$, 2.19]  &   [$-1.96$, $-0.16$]${}\cup{}$[0.49, 2.25] \\
    \cttOne           &   [$-0.99$, 1.05]  &   [$-1.00$, $-0.08$]${}\cup{}$[0.21, 1.11] \\
    WC category \twohqtwolq & & \\[-3pt]
    \ctqEight         &   [$-0.45$, 0.03]  &   [$-0.46$, 0.02]                   \\
    \cQqEightOne      &   [$-0.47$, $-0.00$] &   [$-0.45$, 0.00]                   \\
    \ctqOne           &   [$-0.11$, 0.11]  &   [$-0.12$, 0.10]                   \\
    \cQqOneOne        &   [$-0.10$, 0.10]  &   [$-0.10$, 0.10]                   \\
    \cQqEightThree    &   [$-0.09$, 0.08]  &   [$-0.09$, 0.08]                   \\
    \cQqOneThree      &   [$-0.04$, 0.03]  &   [$-0.04$, 0.03]                   \\
    \hline
\end{tabular}
\label{tab:ci_1sigma}
\end{table}

\begin{table}[!htp]
\centering
\topcaption{The 2$\sigma$ uncertainty intervals extracted from the likelihood fits described in Section~\ref{sec:fitting}. The intervals are shown for the case where the other Wilson coefficients (WCs) are profiled, and the case where the other WCs are fixed to their SM values of zero.}
\renewcommand{\arraystretch}{1.5}
\begin{tabular}{ccc}
    \hline
    WC/$\Lambda^2$ [{\perTeVsqns}] & 2$\sigma$ CI (other WCs profiled) & 2$\sigma$ CI (other WCs fixed to SM) \\
    \hline
    WC category \twohqtwolep & & \\[-3pt]
    \ctlT             &  [$-0.37$, 0.37]    &  [$-0.40$, 0.40]   \\
    \ctlS             &  [$-2.60$, 2.62]    &  [$-2.80$, 2.80]   \\
    \cte              &  [$-1.78$, 2.21]    &  [$-1.91$, 2.39]   \\
    \ctl              &  [$-1.80$, 2.11]    &  [$-2.02$, 2.20]   \\
    \cQe              &  [$-1.91$, 1.96]    &  [$-2.04$, 2.12]   \\
    \cQlM             &  [$-1.58$, 2.28]    &  [$-1.80$, 2.33]   \\
    \cQla             &  [$-2.84$, 2.55]    &  [$-2.69$, 2.58]   \\
    WC category \twohqV & & \\[-3pt]
    \cpt              &  [$-10.52$, 7.87]   &  [$-4.93$, 3.18]   \\
    \cptb             &  [$-3.25$, 3.26]    &  [$-3.14$, 3.18]   \\
    \cpQa             &  [$-0.84$, 2.00]    &  [$-0.85$, 1.89]   \\
    \cbW              &  [$-0.76$, 0.76]    &  [$-0.75$, 0.75]   \\
    \ctG              &  [$-0.28$, 0.24]    &  [$-0.22$, 0.25]   \\
    \cpQM             &  [$-6.06$, 8.12]    &  [$-2.68$, 2.94]   \\
    \ctp              &  [$-8.85$, 2.75]    &  [$-7.54$, 2.11]   \\
    \ctZ              &  [$-0.71$, 0.64]    &  [$-0.58$, 0.59]   \\
    \ctW              &  [$-0.55$, 0.46]    &  [$-0.47$, 0.41]   \\
    WC category \fourhq & & \\[-3pt]
    \cQtOne           &  [$-2.34$, 2.27]    &  [$-2.41$, 2.22]   \\
    \cQtEight         &  [$-4.37$, 4.97]    &  [$-4.45$, 4.96]   \\
    \cQQOne           &  [$-2.56$, 2.84]    &  [$-2.57$, 2.89]   \\
    \cttOne           &  [$-1.33$, 1.38]    &  [$-1.31$, 1.43]   \\
    WC category \twohqtwolq & & \\[-3pt]
    \ctqEight         &  [$-0.68$, 0.25]    &  [$-0.68$, 0.24]   \\
    \cQqEightOne      &  [$-0.68$, 0.22]    &  [$-0.67$, 0.21]   \\
    \ctqOne           &  [$-0.21$, 0.21]    &  [$-0.22$, 0.20]   \\
    \cQqOneOne        &  [$-0.19$, 0.19]    &  [$-0.19$, 0.20]   \\
    \cQqEightThree    &  [$-0.17$, 0.16]    &  [$-0.17$, 0.16]   \\
    \cQqOneThree      &  [$-0.08$, 0.07]    &  [$-0.08$, 0.07]   \\
\hline
\end{tabular}
\label{tab:ci_2sigma}
\end{table}

The disjoint 1$\sigma$ intervals that appear in some of the individual scans (\ie, the scans in which the other \nwcone WCs are fixed to zero) shown in Fig.~\ref{fig:ci_summary} are a result of the quadratic nature of the EFT parametrization.
In principle, this inherent degeneracy would apply for all WCs; however, the degeneracy can be broken when contributions from multiple processes in multiple bins result in one of the two minima having significantly better agreement with the observed data.
The individual scans over the \fourhq WCs are the only cases with a double minima that is sufficiently degenerate to lead to disjoint CIs.
These double minima disappear when profiling over the other 25 WCs, since the interferences with the other WCs can compensate for one another's effects within the range between the two minima, resulting in a single long, flat minimum instead of two disjoint minima.
Double minima in the individual scans (even relatively shallow double minima that do not manifest in disjoint CIs) can broaden the CI interval and sometimes lead to individual CIs that are wider than the corresponding profiled CIs, as described in Ref.~\cite{CMS:2020lrr}.
For many of the WCs, the CIs obtained from the profiled and individual scans are similar; for some of the WCs, this results from the fact that there are not significant correlations between the given WC and the other WCs.
However, in other cases there are non-trivial correlations among several of the WCs, and this will be explored further in Section~\ref{sec:results_2dscans}.

Performing the likelihood fit over all \nwc WCs simultaneously, this work supersedes Ref.~\cite{CMS:2020lrr} as the most global detector-level EFT analysis to date.
Not only does this work incorporate 10 additional WCs that were not studied in Ref.~\cite{CMS:2020lrr}, but it also obtains significantly improved constraints on the WCs.
For the WCs that are common between the analyses, the 2$\sigma$ profiled CIs generally improve by factors of approximately 2 to 6, depending on the WC.
The differential approach leveraged by this analysis provides the majority of the improvement, though the larger data set also helps to increase the sensitivity.

\subsection{Two-dimensional correlations among WCs}
\label{sec:results_2dscans}

\begin{figure}[!htp]
\centering
\includegraphics[width=0.45\textwidth]{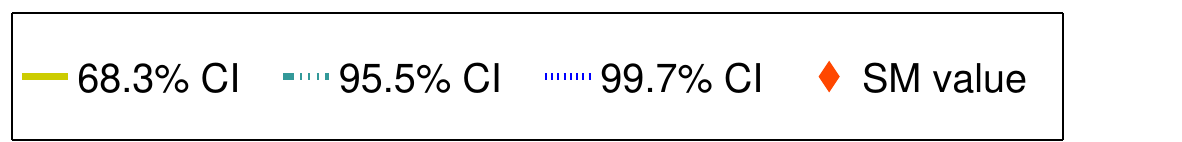}\\
\includegraphics[width=0.45\textwidth]{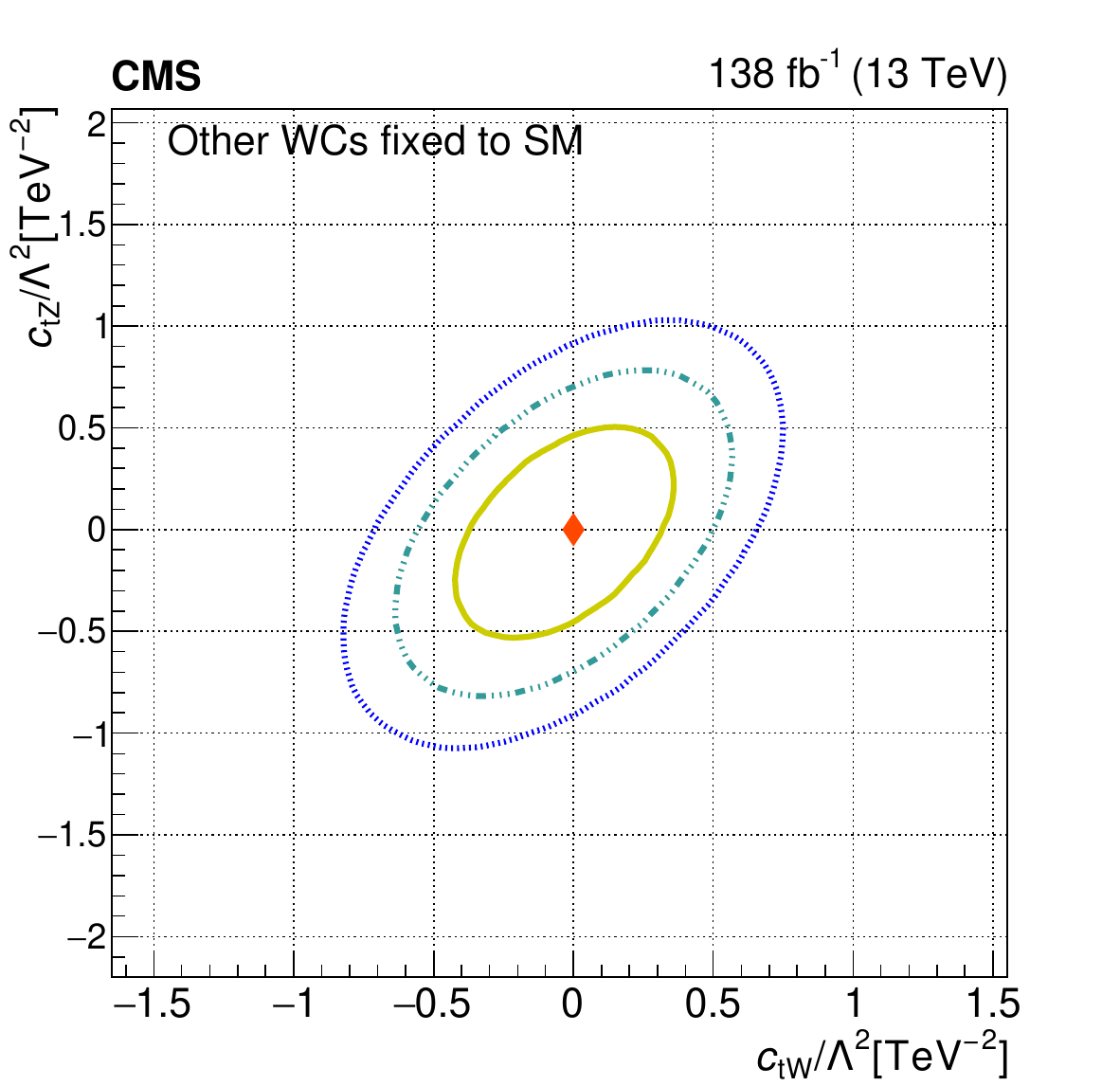}
\includegraphics[width=0.45\textwidth]{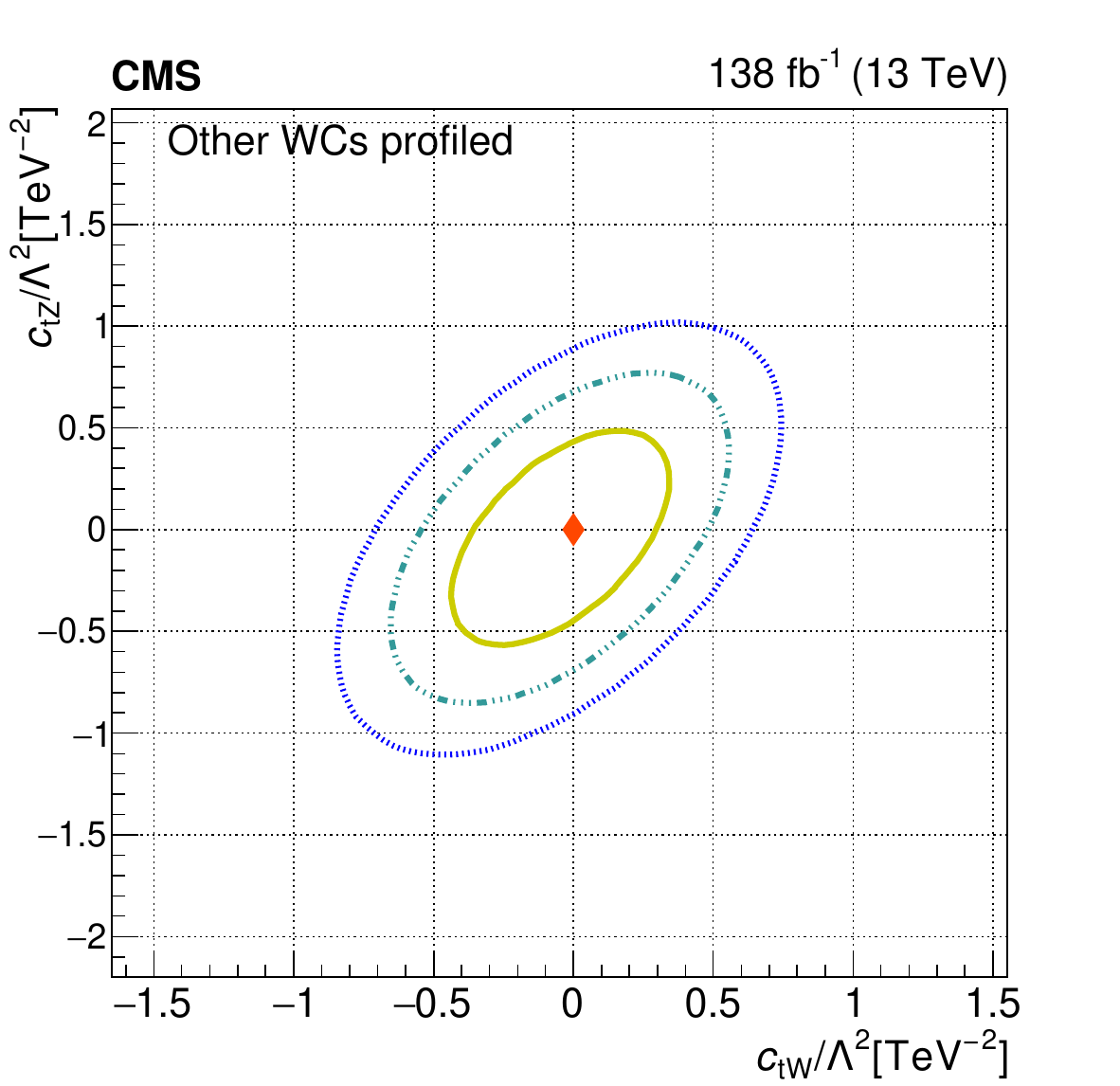}
\caption{The observed 68.3, 95.5, and 99.7\% confidence level contours of a 2D scan for \ctW and \ctZ with the other WCs fixed to their SM values (left), and profiled (right). Diamond markers show the SM prediction.}
\label{fig:Contours_ctW_ctZ}
\end{figure}

\begin{figure}[!htp]
\centering
\includegraphics[width=0.45\textwidth]{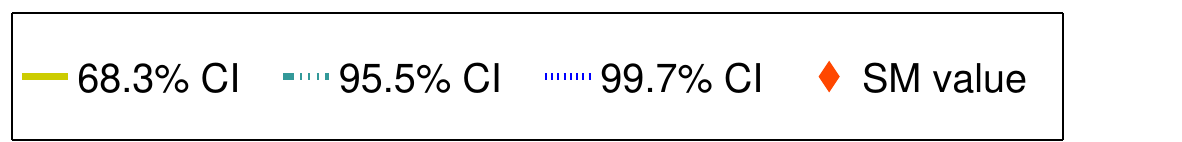}\\
\includegraphics[width=0.45\textwidth]{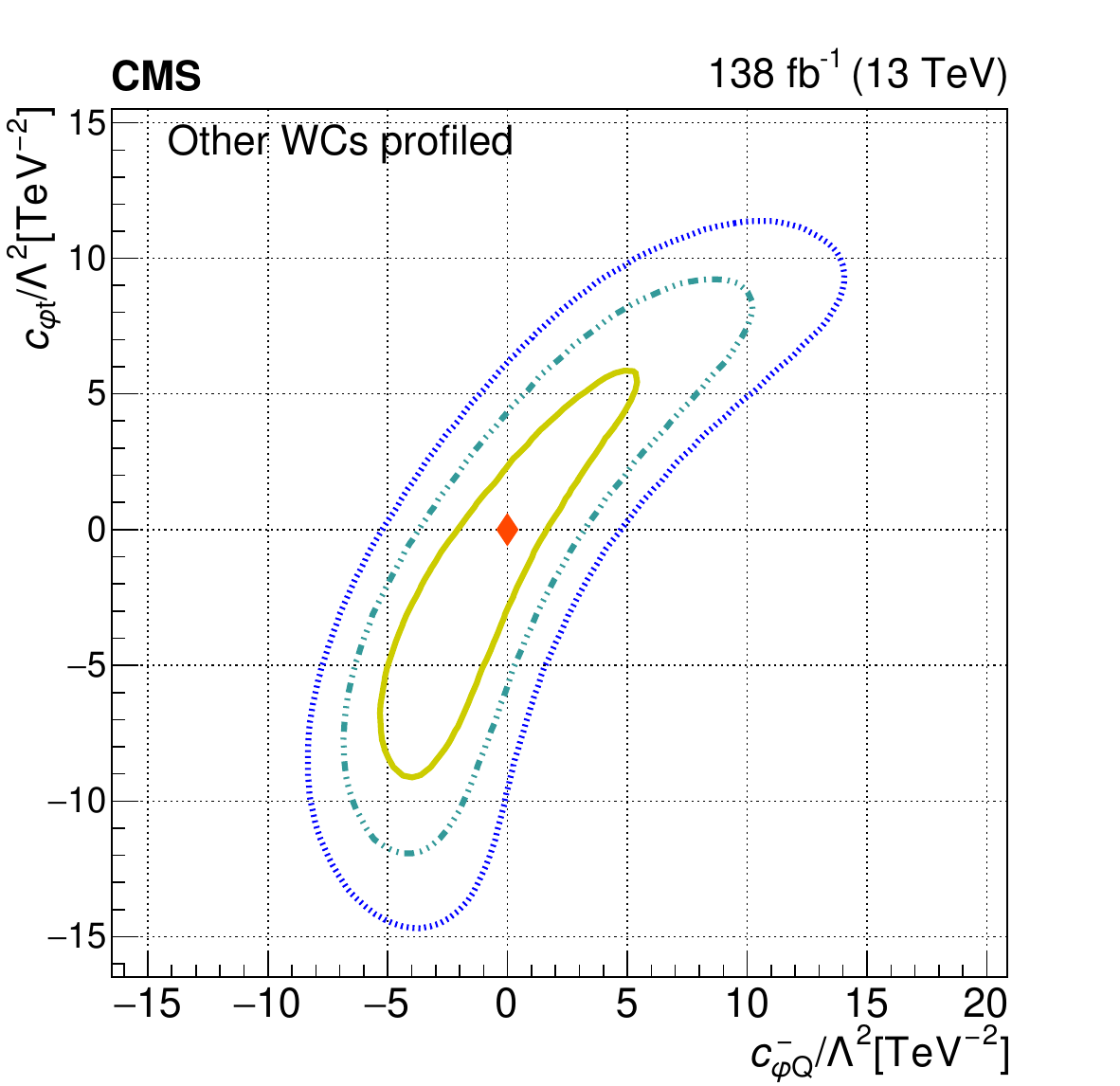}
\includegraphics[width=0.45\textwidth]{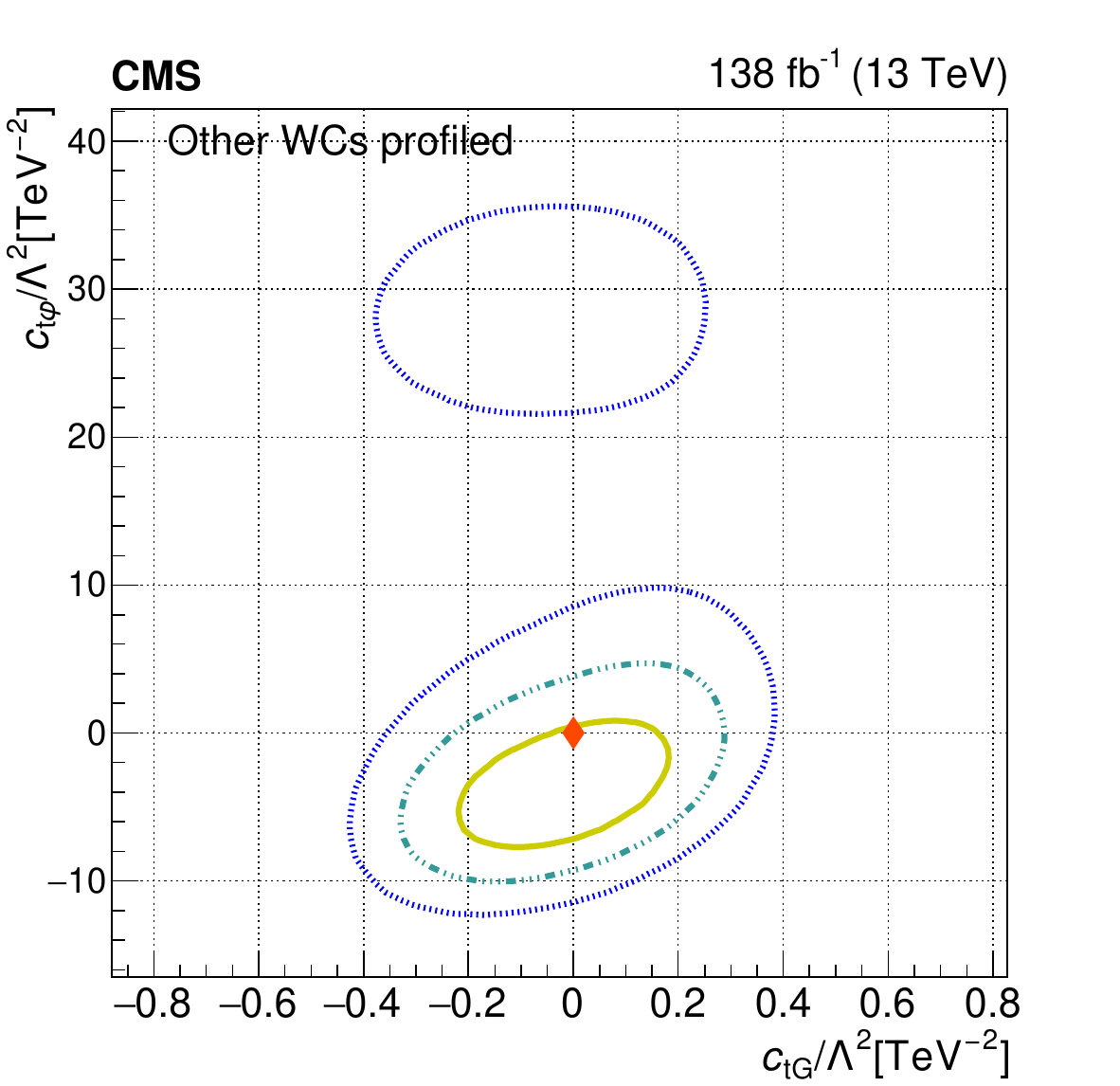}
\caption{The observed 68.3, 95.5, and 99.7\% confidence level contours of a 2D scan with the other WCs profiled, for \cpQM and \cpt (left), and for \ctG and \ctp (right). Diamond markers show the SM prediction.}
\label{fig:Contours_cpQMcpt_ctGctp}
\end{figure}

\begin{figure}[!ht]
\centering
\includegraphics[width=0.45\textwidth]{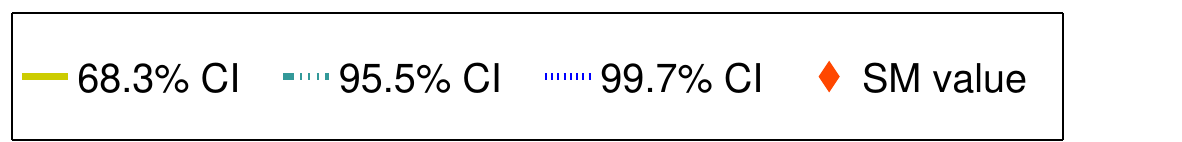}\\
\includegraphics[width=0.45\textwidth]{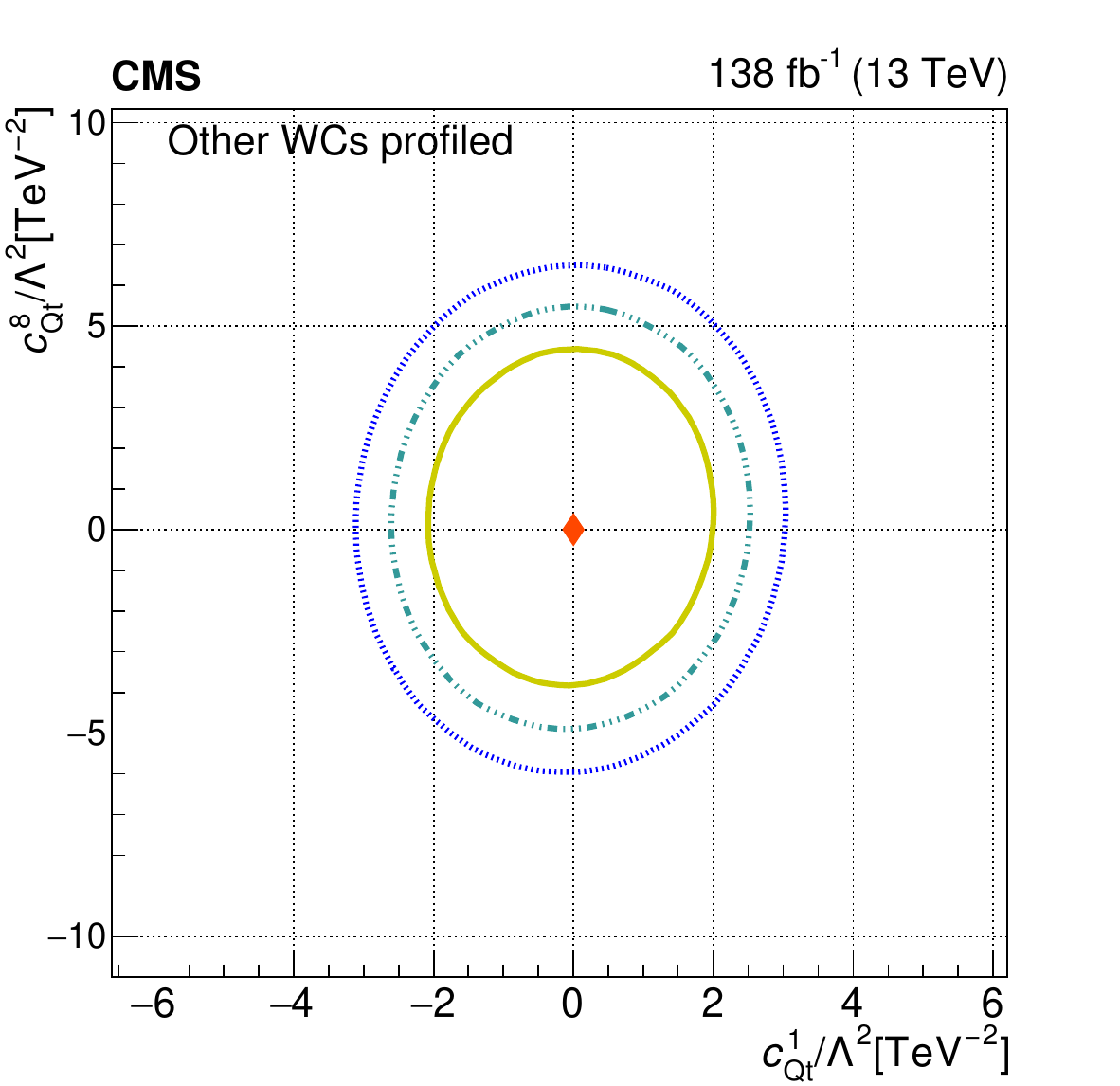}
\includegraphics[width=0.45\textwidth]{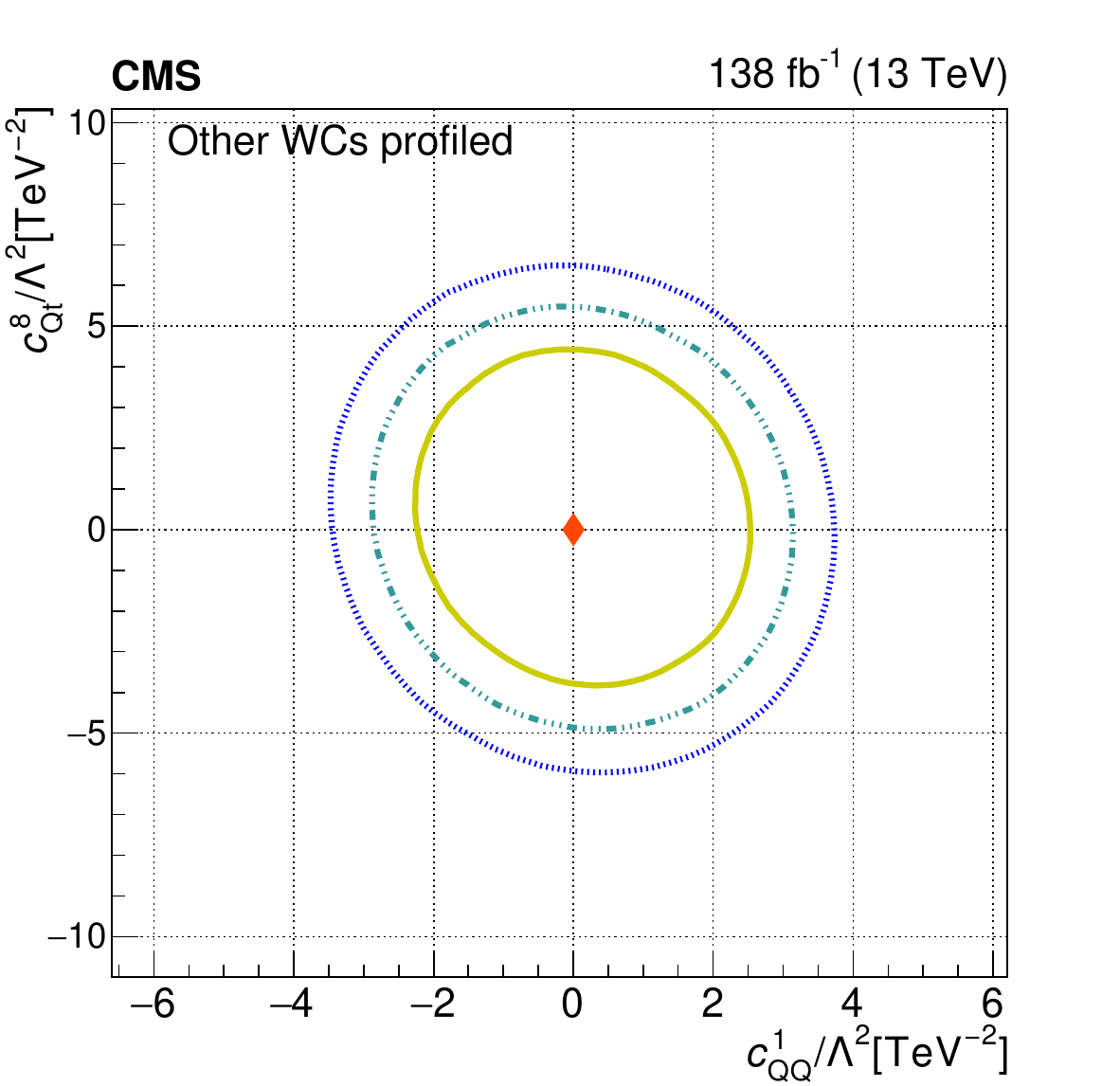}\\
\includegraphics[width=0.45\textwidth]{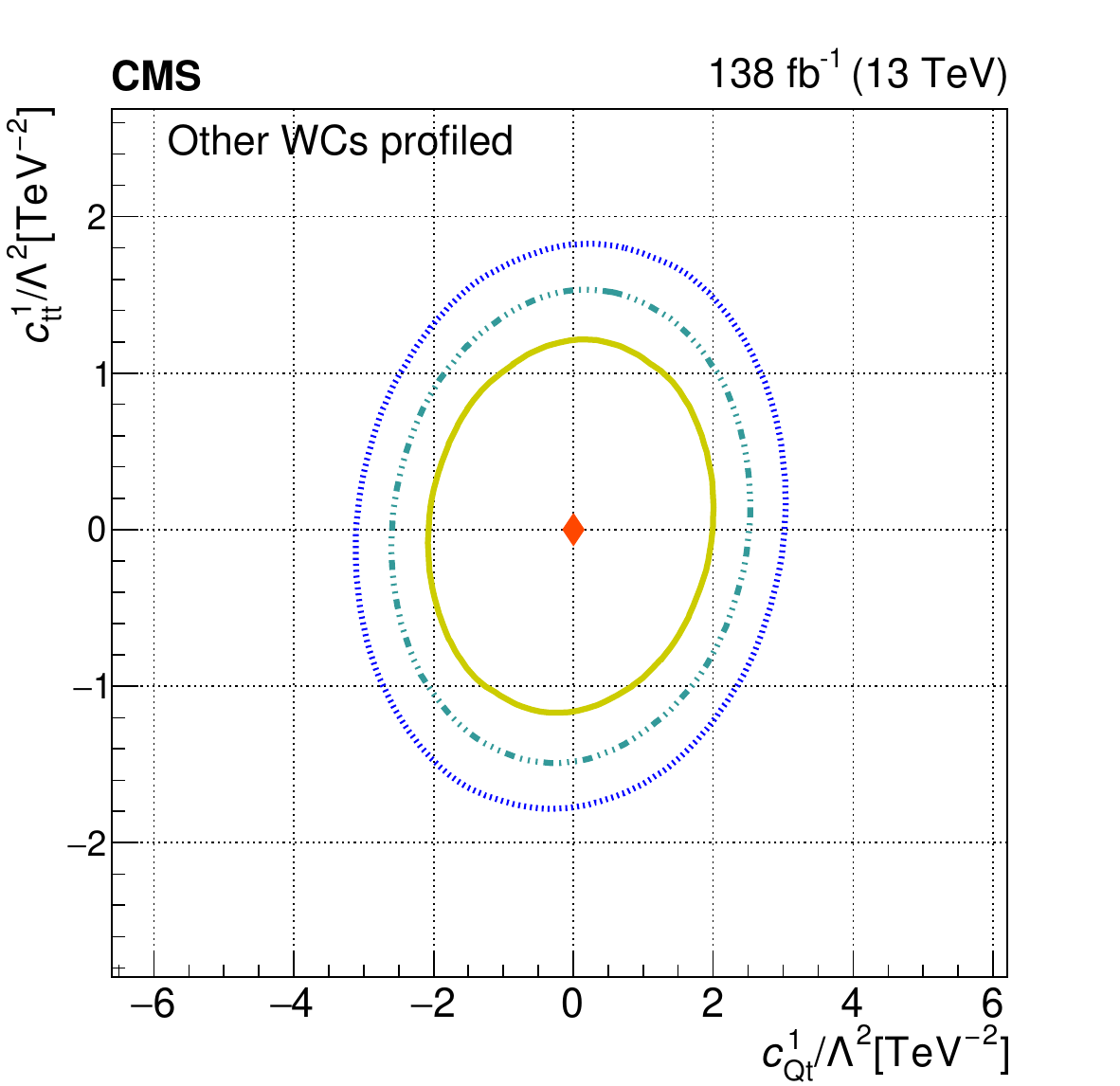}
\includegraphics[width=0.45\textwidth]{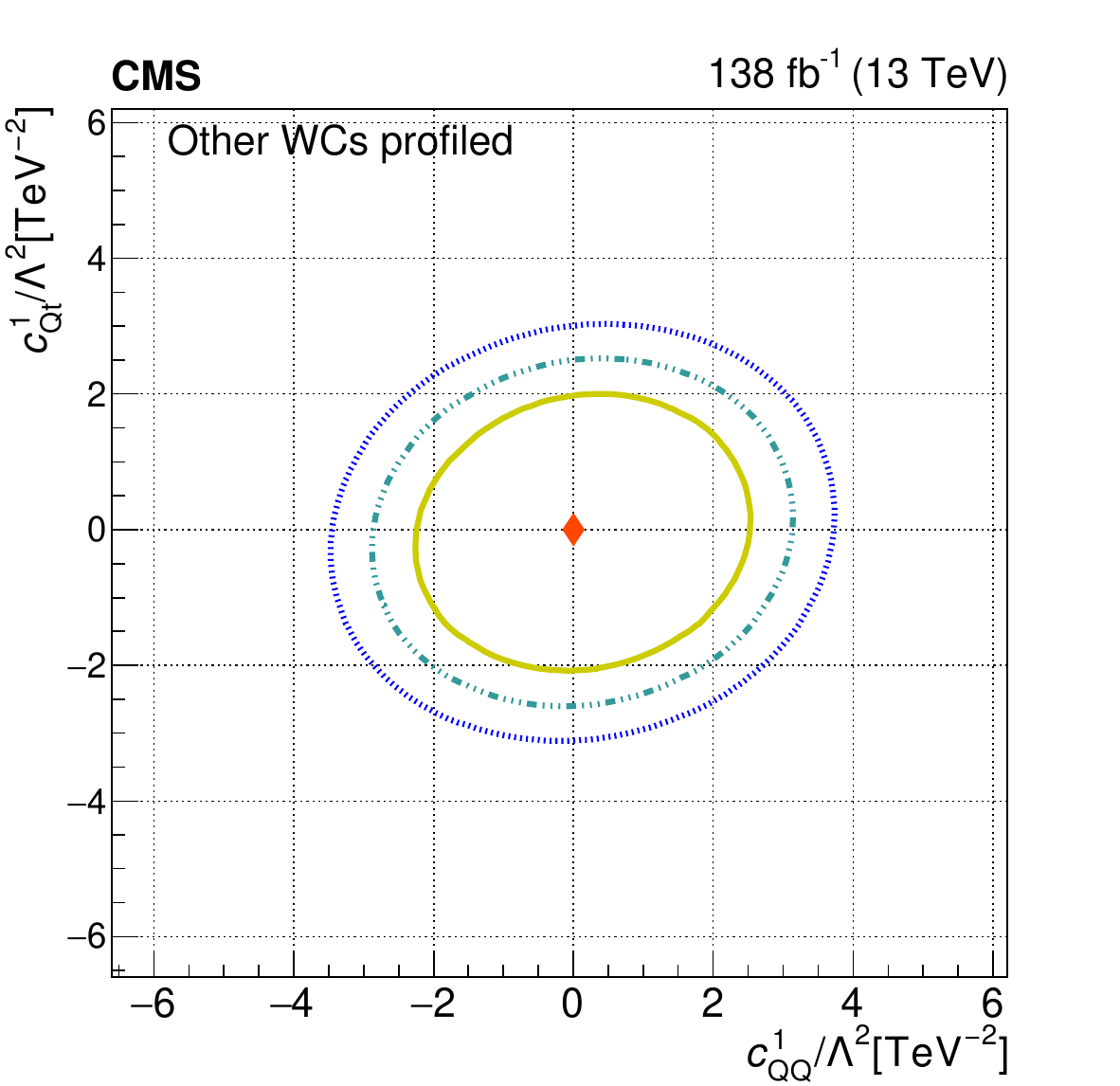}
\caption{The observed 68.3, 95.5, and 99.7\% confidence level contours of a 2D scan with the other WCs profiled, for \cQtOne and \cQtEight (upper left), for \cQQOne and \cQtEight (upper right), for \cQtOne and \cttOne (lower left), and for \cQQOne and \cQtOne (lower right). Diamond markers show the SM prediction.}
\label{fig:Contours_4heavy}
\end{figure}

To explore correlations among the WCs in the \nwc-dimensional fit, this section presents 2D scans for several pairs of WCs with nonnegligible correlations.
These pairs include several from the \twohqV category of WCs (shown in Figs.~\ref{fig:Contours_ctW_ctZ}--\ref{fig:Contours_cpQMcpt_ctGctp}) and several from the \fourhq category of WCs (shown in Fig.~\ref{fig:Contours_4heavy}).
In most cases, the 2D scans in which the other 24 WCs are profiled are very similar to the 2D scans in which the other WCs are fixed to zero, indicating that while the correlations between the given pair of WCs are important, the correlations with the other \nwctwo are less significant.
For example, Fig.~\ref{fig:Contours_ctW_ctZ} shows a 2D scan over the \ctZ and \ctW WCs with the other 24 WCs fixed to their SM values of zero (on the left) or profiled (on the right); while a strong correlation between \ctZ and \ctW is evident, the other 24 WCs are not significantly correlated with either of these two WCs.
Fig.~\ref{fig:Contours_cpQMcpt_ctGctp} shows correlations between other \twohqV WCs. Two disjoint contours of the 2$\sigma$ CI are visible in the 2D scan over \ctp and \ctG in the right-hand side of this figure; the confidence intervals shown in Fig.~\ref{fig:ci_summary} do not have two disjoint intervals because the minimum is not deep enough to cross the 2$\sigma$ threshold in the one-dimensional scan.
Pairs of WCs from the \fourhq category are shown in Fig.~\ref{fig:Contours_4heavy}.
Near the SM, where their interference terms compensate for one another as discussed above, these four WCs have significant correlations with each other.
However, farther from the SM at the 1 and 2$\sigma$ limits, the quadratic terms dominate over the linear terms, so the effects of all WCs monotonically increase the yields; for this reason, the WCs can no longer compensate for each other, so they become uncorrelated in the fit.
This is why the contour plots in Fig.~\ref{fig:Contours_4heavy} show minimal correlations between the pairs of WCs.

\subsection{Interpretation of sensitivity}
\label{sec:results_interp}

In this section, we discuss the sensitivity to the WCs (which is quantified by the CIs), focusing on the likelihood fits with other WCs profiled.
While all 178 analysis bins contribute to the sensitivity to the \nwc WCs, the relative contribution of each bin varies by WC.
Organizing the WCs based primarily on the interactions they modify and the processes they most strongly impact, the WCs may be classified into seven main groups, summarized in Table~\ref{tab:wc_sensitivity_groups}.
The WCs that belong to each grouping are listed in the center column of Table~\ref{tab:wc_sensitivity_groups}, while the right-hand column notes the category (or categories) of analysis bins that generally provide the leading sensitivity to the WCs in the given group.
It should be emphasized that the categories of bins listed in the right-hand column of Table~\ref{tab:wc_sensitivity_groups} represent a simplified picture of the interpretation of the sensitivity:\ while there are indeed some cases where the majority of the sensitivity to a WC is derived from a relatively clear subset of the analysis bins, the sensitivity to many of the WCs is provided by a diverse combination of bins across all selection categories.
Furthermore, when characterizing relevant bins, it is also important to keep in mind interference and correlations among WCs.
The following subsections will step through each of the groups of WCs outlined in Table~\ref{tab:wc_sensitivity_groups}, discussing the subsets of bins that provide the leading contributions to the sensitivity and discussing non-trivial correlations where relevant.

\begin{table}[!htp]
\centering
\topcaption{Summary of categories that provide leading contributions to the sensitivity for subsets of the Wilson coefficients (WCs).}
\renewcommand{\arraystretch}{1.6}
\begin{tabular}{ p{4.6cm} p{3.5cm} p{3cm}}
\hline
Grouping of WCs & WCs & Lead categories \\
\hline
\twohqtwolep & \cQla, \cQlM, \cQe, \ctl, & \threeloffZ \\
& \cte, \ctlS, \ctlT & \\
\fourhq & \cQQOne, \cQtOne, \cQtEight, \cttOne & \twolss \\
\twohqtwolq ``\ttlnu-like'' & \cQqOneOne, \cQqEightOne, \ctqOne, \ctqEight & \twolss \\
\twohqtwolq ``\tllq-like'' & \cQqOneThree, \cQqEightThree & \threelonZ \\
\twohqV ``\ttll-like'' & \ctZ, \cpt, \cpQM & \threelonZ and \twolss \\
\twohqV ``\tXq-like'' & \cpQa, \cptb, \cbW & \threelonZ \\
\twohqV (significant impacts & \ctG, \ctp, \ctW &  \threel and \twolss \\[-6pt]
on many processes) & & \\
\hline
\end{tabular}
\label{tab:wc_sensitivity_groups}
\end{table}

\subsubsection{The WCs from the \texorpdfstring{\twohqtwolep}{2hq2lep} category of operators}

Beginning with the WCs in the \twohqtwolep group, the \threeloffZ channels provide the majority of the sensitivity for these WCs, which are associated with four-fermion vertices that produce pairs of leptons without an intermediate \PZ boson.
To quantify the contributions of the off-\PZ channels, a fit is performed with only this subset of bins included. The resulting 2$\sigma$ profiled CIs show that the expected sensitivity is only degraded by about 5--7\% compared to the results when all bins are included.

\subsubsection{The WCs from the \texorpdfstring{\fourhq}{4hq} category of operators}

The next group of WCs are those associated with the \fourhq operators. The sensitivity to these WCs is provided primarily by the \twolss bins, with leading contributions from the bins requiring at least three \PQb-tagged jets.
Since the \tttt process contributes significantly to these bins and the four-heavy WCs strongly affect the \tttt process (both the total cross section and shape of the kinematic distributions), it is expected that these bins would contribute significantly to the sensitivity.
To obtain a quantitative characterization of the sensitivity provided by the \twolss bins, we performed a fit with only these bins included. The resulting 2$\sigma$ CIs are only degraded by about 4--6\% (with respect to a fit with all bins included), showing that the \twolss bins indeed represent the dominant source of sensitivity to the four-heavy WCs.

\subsubsection{The WCs from the \texorpdfstring{\twohqtwolq}{2hq2lq} category of operators}

The next set of WCs are those associated with the \twohqtwolq category of operators.
Four of these WCs (\cQqOneOne, \ctqOne, \cQqEightOne, and \ctqEight) primarily affect the \ttlnu process, so  bins populated significantly by \ttlnu are expected to provide important contributions to the sensitivity to these WCs.
Performing a fit with only the \twolss bins included, the expected 2$\sigma$ CIs are degraded by only about 6--15\%.
The \twolss bins thus provide the primary source of sensitivity for these WCs, though other bins (\eg, from the off-\PZ channels) also contribute to the sensitivity.

The remaining two WCs from the \twohqtwolq group (\cQqOneThree and \cQqEightThree) are distinct from the other \twohqtwolq WCs in that they feature $\PQt\PQb\PQq\PQq^\prime$  vertices.
These vertices allow \cQqOneThree and \cQqEightThree to significantly impact the \tllq process in the \threelonZ bins with two \PQb-tagged jets and low jet multiplicity (as discussed in Section~\ref{sec:eventSelection_diff_vars}).
The on-\PZ bins thus contribute significant sensitivity to these WCs.
While the \twolss and off-\PZ categories also contribute to the sensitivity to these WCs, the \threelonZ bins provide the leading contribution. The expected 2$\sigma$ CIs for these WCs each widen by more than 30\% when the \threelonZ bins are excluded from the fit.

\subsubsection{The WCs from the \texorpdfstring{\twohqV}{2hqV} category of operators}

The final set of WCs are those associated with the \twohqV category of operators.
These nine WCs impact a broad range of processes, leading to diverse effects across the full set of 178 analysis bins and making it challenging to definitively characterize subsets of bins that provide dominant contributions to the sensitivity.
However, the WCs can be classified into three main groups (as listed in Table~\ref{tab:wc_sensitivity_groups}) based on the  processes they impact most significantly.

The \ctZ, \cpQM, and \cpt WCs feature \ttZ EFT vertices and primarily affect the \ttll process; the on-\PZ bins are thus important for these WCs.
However, these WCs also impact other processes (\eg, \tttt), meaning other categories of bins can also provide important sensitivity.
Furthermore, the \ttll process also significantly populates the \twolss bins (making up about 20\% of the total expected yield), so the \ttll effects can also be relevant in the \twolss bins.
Thus, the \threelonZ bins and \twolss bins are important for these WCs.
The \threeloffZ bins provide a smaller (though nonzero) contribution to the sensitivity. Performing a fit with these bins excluded results in an approximately 6\% degradation of the expected 2$\sigma$ CIs for each of these three WCs.

Next, let us consider \cpQa, \cptb, and \cbW.
These WCs primarily impact \tllq and \tHq, and their sensitivity arises from multiple categories of analysis bins.
The \threelonZ bins represent the leading (though not overwhelmingly dominant) contribution.
Performing a fit with only the \threelonZ bins included, the expected 2$\sigma$ CIs for these WCs widen by  about 2--13\% (depending on the WC) compared to a fit with all bins included.

The final three WCs from the \twohqV group are \ctG, \ctp, and \ctW.
Impacting multiple processes, these WCs  gain sensitivity from the full spectrum of analysis bins.
For example, \ctG impacts \ttH (so the \twolss and \threeloffZ bins are important as \ttH significantly populates these bins) but also strongly impacts \ttll (so the on-\PZ and \twolss bins also play an important role).
The \ctp WC significantly impacts \ttH, \tHq, and \tttt; most of the analysis bins provide sensitivity to this WC, though the on-\PZ bins provide only minor contributions (dropping the on-\PZ bins only results in about a 5\% effect on the expected 2$\sigma$ profiled CIs for \ctp).
Finally, the \ctW WC impacts all signal processes and derives important sensitivity from many of the analysis bins.
Further complicating the picture, \ctW has significant interference with \ctZ, and the two WCs have a strong linear correlation in the profiled fit (as shown in Fig.~\ref{fig:Contours_ctW_ctZ}).
Thus, when we consider the 2$\sigma$ profiled CIs for \ctW, it is important to recall that the \ctZ operator is also set to a nonzero value, so bins that are affected by \ctZ can also be important when considering the sensitivity to \ctW.

\section{Summary}
\label{sec:summary}

A search for new physics in the production of one or more top quarks with additional leptons,
jets, and \PQb jets in the context of effective field theory (EFT) has been performed.
Events from proton-proton collisions with a center-of-mass energy of 13\TeV corresponding
to an integrated luminosity of 138\fbinv are used.
EFT effects are incorporated into the event weights of the simulated samples,
allowing detector-level predictions that account for correlations and interference effects among EFT operators and between EFT operators and standard model (SM) processes.

The Wilson coefficients (WCs) corresponding to \nwc EFT operators were simultaneously fit to the data.
Confidence intervals were extracted for the WCs either individually or in pairs by scanning the likelihood with the other WCs either profiled or fixed at their SM values of zero.
In all cases, the data are found to be consistent with the SM expectations.

\begin{acknowledgments}
We congratulate our colleagues in the CERN accelerator departments for the excellent performance of the LHC and thank the technical and administrative staffs at CERN and at other CMS institutes for their contributions to the success of the CMS effort. In addition, we gratefully acknowledge the computing centers and personnel of the Worldwide LHC Computing Grid and other centers for delivering so effectively the computing infrastructure essential to our analyses. Finally, we acknowledge the enduring support for the construction and operation of the LHC, the CMS detector, and the supporting computing infrastructure provided by the following funding agencies: SC (Armenia), BMBWF and FWF (Austria); FNRS and FWO (Belgium); CNPq, CAPES, FAPERJ, FAPERGS, and FAPESP (Brazil); MES and BNSF (Bulgaria); CERN; CAS, MoST, and NSFC (China); MINCIENCIAS (Colombia); MSES and CSF (Croatia); RIF (Cyprus); SENESCYT (Ecuador); MoER, ERC PUT and ERDF (Estonia); Academy of Finland, MEC, and HIP (Finland); CEA and CNRS/IN2P3 (France); BMBF, DFG, and HGF (Germany); GSRI (Greece); NKFIH (Hungary); DAE and DST (India); IPM (Iran); SFI (Ireland); INFN (Italy); MSIP and NRF (Republic of Korea); MES (Latvia); LAS (Lithuania); MOE and UM (Malaysia); BUAP, CINVESTAV, CONACYT, LNS, SEP, and UASLP-FAI (Mexico); MOS (Montenegro); MBIE (New Zealand); PAEC (Pakistan); MES and NSC (Poland); FCT (Portugal); MESTD (Serbia); MCIN/AEI and PCTI (Spain); MOSTR (Sri Lanka); Swiss Funding Agencies (Switzerland); MST (Taipei); MHESI and NSTDA (Thailand); TUBITAK and TENMAK (Turkey); NASU (Ukraine); STFC (United Kingdom); DOE and NSF (USA).

\hyphenation{Rachada-pisek} Individuals have received support from the Marie-Curie program and the European Research Council and Horizon 2020 Grant, contract Nos.\ 675440, 724704, 752730, 758316, 765710, 824093, 884104, and COST Action CA16108 (European Union); the Leventis Foundation; the Alfred P.\ Sloan Foundation; the Alexander von Humboldt Foundation; the Science Committee, project no. 22rl-037 (Armenia); the Belgian Federal Science Policy Office; the Fonds pour la Formation \`a la Recherche dans l'Industrie et dans l'Agriculture (FRIA-Belgium); the Agentschap voor Innovatie door Wetenschap en Technologie (IWT-Belgium); the F.R.S.-FNRS and FWO (Belgium) under the ``Excellence of Science -- EOS" -- be.h project n.\ 30820817; the Beijing Municipal Science \& Technology Commission, No. Z191100007219010; the Ministry of Education, Youth and Sports (MEYS) of the Czech Republic; the Shota Rustaveli National Science Foundation, grant FR-22-985 (Georgia); the Deutsche Forschungsgemeinschaft (DFG), under Germany's Excellence Strategy -- EXC 2121 ``Quantum Universe" -- 390833306, and under project number 400140256 - GRK2497; the Hellenic Foundation for Research and Innovation (HFRI), Project Number 2288 (Greece); the Hungarian Academy of Sciences, the New National Excellence Program - \'UNKP, the NKFIH research grants K 124845, K 124850, K 128713, K 128786, K 129058, K 131991, K 133046, K 138136, K 143460, K 143477, 2020-2.2.1-ED-2021-00181, and TKP2021-NKTA-64 (Hungary); the Council of Science and Industrial Research, India; the Latvian Council of Science; the Ministry of Education and Science, project no. 2022/WK/14, and the National Science Center, contracts Opus 2021/41/B/ST2/01369 and 2021/43/B/ST2/01552 (Poland); the Funda\c{c}\~ao para a Ci\^encia e a Tecnologia, grant CEECIND/01334/2018 (Portugal); the National Priorities Research Program by Qatar National Research Fund; MCIN/AEI/10.13039/501100011033, ERDF ``a way of making Europe", and the Programa Estatal de Fomento de la Investigaci{\'o}n Cient{\'i}fica y T{\'e}cnica de Excelencia Mar\'{\i}a de Maeztu, grant MDM-2017-0765 and Programa Severo Ochoa del Principado de Asturias (Spain); the Chulalongkorn Academic into Its 2nd Century Project Advancement Project, and the National Science, Research and Innovation Fund via the Program Management Unit for Human Resources \& Institutional Development, Research and Innovation, grant B05F650021 (Thailand); the Kavli Foundation; the Nvidia Corporation; the SuperMicro Corporation; the Welch Foundation, contract C-1845; and the Weston Havens Foundation (USA).
\end{acknowledgments}

\bibliography{auto_generated}
\cleardoublepage \appendix\section{The CMS Collaboration \label{app:collab}}\begin{sloppypar}\hyphenpenalty=5000\widowpenalty=500\clubpenalty=5000
\cmsinstitute{Yerevan Physics Institute, Yerevan, Armenia}
{\tolerance=6000
A.~Hayrapetyan, A.~Tumasyan\cmsAuthorMark{1}\cmsorcid{0009-0000-0684-6742}
\par}
\cmsinstitute{Institut f\"{u}r Hochenergiephysik, Vienna, Austria}
{\tolerance=6000
W.~Adam\cmsorcid{0000-0001-9099-4341}, J.W.~Andrejkovic, T.~Bergauer\cmsorcid{0000-0002-5786-0293}, S.~Chatterjee\cmsorcid{0000-0003-2660-0349}, K.~Damanakis\cmsorcid{0000-0001-5389-2872}, M.~Dragicevic\cmsorcid{0000-0003-1967-6783}, A.~Escalante~Del~Valle\cmsorcid{0000-0002-9702-6359}, P.S.~Hussain\cmsorcid{0000-0002-4825-5278}, M.~Jeitler\cmsAuthorMark{2}\cmsorcid{0000-0002-5141-9560}, N.~Krammer\cmsorcid{0000-0002-0548-0985}, D.~Liko\cmsorcid{0000-0002-3380-473X}, I.~Mikulec\cmsorcid{0000-0003-0385-2746}, J.~Schieck\cmsAuthorMark{2}\cmsorcid{0000-0002-1058-8093}, R.~Sch\"{o}fbeck\cmsorcid{0000-0002-2332-8784}, D.~Schwarz\cmsorcid{0000-0002-3821-7331}, M.~Sonawane\cmsorcid{0000-0003-0510-7010}, S.~Templ\cmsorcid{0000-0003-3137-5692}, W.~Waltenberger\cmsorcid{0000-0002-6215-7228}, C.-E.~Wulz\cmsAuthorMark{2}\cmsorcid{0000-0001-9226-5812}
% ...
\par}
\vskip\cmsinstskip
\dag:~Deceased\\
$^{1}$Also at Yerevan State University, Yerevan, Armenia\\
$^{2}$Also at TU Wien, Vienna, Austria\\
% ...
\end{sloppypar}
\end{document}